\documentclass[a4paper,11pt]{article}
\usepackage{jheppub} 
\usepackage[nameinlink,capitalize]{cleveref}
\usepackage{subcaption, tikz, tikz-cd}
\usepackage{soul}

\arxivnumber{2410.11034} 

\title{\boldmath Lattice Chern-Simons-Maxwell Theory and its Chirality}

\author[a,b]{Ze-An Xu}
\author[b]{and Jing-Yuan Chen}
\affiliation[a]{Department of Physics, Peking University,\\
Beijing, 100871, China}
\affiliation[b]{Institute for Advanced Study, Tsinghua University,\\ Beijing, 100084, China}

\abstract{We define and solve the $\text{U(1)}$ Chern-Simons-Maxwell theory on spacetime lattice, with an emphasis on the chirality of the theory. Realizing Chern-Simons theory on lattice has been a problem of interest for decades, and over the years it has gradually become clear that there are two key points: 1) Some non-topological term, such as a Maxwell term, is necessary---this is true even in the continuum, but more manifestly on the lattice; 2) the $\text{U(1)}$ gauge field should be implemented in the Villainized form to retain its topological properties. Putting the two ideas together seriously, we show all interesting properties of a chiral Chern-Simons theory are reproduced in an explicitly regularized manner on the lattice. These include the bosonic and fermionic level quantization, the bulk and chiral edge spectrum, the Wilson loop flux attachment (with point-split framing or geometric framing depending on the Maxwell coupling), the Wilson loop spin, the ground state degeneracy, and, most non-trivially, the chiral gravitational anomaly.
}

\begin{document}
\maketitle
\flushbottom
\newpage
\section{Introduction}
\label{introduction}
Chern-Simons (CS) theory is an earliest example of topological quantum field theory \cite{Schwarz:1978cn}. It is fascinating in its theoretical appeals \cite{Schwarz:1978cn, Witten:1988hf} and has found important application in quantum Hall physics \cite{Wen:1995qn}. Since its advent in late 1970s, there have been many efforts trying to realize it on the lattice. However, even for the abelian $\text{U(1)}$ CS,
\begin{align}
    S=\frac{k}{4\pi} \int_{3d} A \wedge dA,
    \label{cont_U1CS}
\end{align}
which is the simplest, non-interacting case, numerous early attempts towards a lattice realization have been unsuccessful, showing the problem is more non-trivial than it might seem. Being able to achieve a lattice realization will be very helpful for understanding of the subtleties in such a chiral topological field theory,
\footnote{And very illuminating for pedagogical purposes.}
manifesting the origin of its chiral edge mode and chiral gravitational anomaly. Over time, it has gradually become clear that there are two key points.

\textbf{1)} \ \ The first is that the lattice realization cannot be purely topological. It must involve some non-topological term, such as a Maxwell term, therefore we really should aim at realizing lattice CS-Maxwell theory. This is necessary for the lattice realization to have the correct local dynamics.

At the technical level, early attempts to naively ``discretize'' $A \wedge dA$ onto the lattice have found undesired extra zero modes, making the partition function not well-defined (as we will see in more details in the main text). In \cite{Berruto:2000dp} it was systematically analyzed that this problem is not associated with any particular way to discretize $A \wedge dA$---as long as the lattice action is local, gauge invariant and time-reversal odd, the problem will exist; but adding some time-reversal even, non-topological term, such as a Maxwell term, will lift this problem.

As the subject of topological order further developed, the problem has been understood with a more general argument/lore that roughly proceeds as the following. The $\text{U(1)}$ CS in continuum hosts gapless chiral edge mode which is robust with respect to boundary conditions. Suppose there is a purely topological lattice implementation, such that when we perform coarse-graining renormalization, the form of the lattice theory remains exactly unchanged (or at least remains exactly local, with any parameter staying bounded). We can start with a lattice system with boundary (such that the theory is implemented on the boundary cells in the same way as in the bulk cells
\footnote{This is often seen as a version of lattice Neumann boundary condition.}---assuming the robustness of the chiral edge mode with respect to boundary conditions), and then eventually coarse-grain the lattice until the space(time) consists of only a few lattice cells---but now it would obviously be impossible to host a gapless chiral edge mode, in contradiction to what we want.
\footnote{{Using a rigorous approach different from the lore above, a theorem estabilishes that a commuting projector Hamiltonian---the common systematic way to construct topological lattice theories---indeed cannot host a non-trivial thermal Hall conductivity which is a manifestation of chirality} \cite{Kapustin:2019zcq}.}
By contrast, those doubled $\text{U(1)}$ CS which can host gapped edges have indeed been realized onto the lattice purely topologically \cite{Chen:2019mjw}.

It might seem bizarre that the $\text{U(1)}$ CS theory is topological in the continuum but cannot be made topological on the lattice. In fact, closer scrutiny shows that even in the continuum, the $\text{U(1)}$ CS theory is not purely topological. To define the phase of the continuum path integral, some regulator must be introduced \cite{Witten:1988hf}, and it is not hard to see the regulator can be physically interpreted as a tiny Maxwell term (or Yang-Mills term if non-abelian) \cite{Bar-Natan:1991fix}. Moreover, in the presence of spacetime boundary, the regulator plays the role of the $i\epsilon$ prescription in the Green's function of the gapless chiral edge mode. So even in the continuum, the CS theory is in fact not purely topological---a conceptual point that is important but, unfortunately, not as widely appreciated as it should be.

\textbf{2)} \ \ The second key point is, to correctly capture the global aspects of the theory, the $\text{U(1)}$ gauge field must be implemented in the Villainized form.

The method of Villainization was originally invented to describe and analyze vortices in the XY model (lattice $S^1$ non-linear sigma model) \cite{Berezinsky:1970fr}, and to facilitate Monte-Carlo simulation of the theory \cite{Villain:1974ir}. Later it has been adapted to $\text{U(1)}$ gauge theory \cite{Einhorn:1977jm, Einhorn:1977qv}. In the recent years, the Villainization method has been re-emphasized as its nature and importance has become better understood \cite{Sulejmanpasic:2019ytl,Chen:2019mjw,Gorantla:2021svj}. It is the natural way to manifestly realize the topological fact $\pi_1(\text{U(1)})=\mathbb{Z}$ on the lattice.
\footnote{Or more generally, $\pi_1$ physics of generic target spaces or gauge groups \cite{Chen:2024ddr}.}
Since this fact is particularly important for $\text{U(1)}$ CS---the CS level quantization and ground state degeneracy are consequences of this fact---the Villainized form is necessary for an adequate lattice realization \cite{Chen:2019mjw,Jacobson:2023cmr}.

Very recently, one of the authors showed \cite{Chen:2024ddr} Villainization is the most elementary example of a much broader theme: We should, in general, refine the traditional lattice theories via category theory, so that the refined lattice theories more closely and more naturally represent the desired continuum theories, especially the topological aspects. This broader perspective can be applied to more generic cases, including a proposal for non-abelian lattice CS-Yang-Mills theory. The current problem of abelian lattice CS-Maxwell theory, being solvable as we will see, serves as a basic anchor point for this broader theme.

\

While this two key ideas have been gradually understood by (perhaps somewhat different groups of) people, in the past the two ideas have not been seriously put together in order to define and solve the lattice CS-Maxwell theory. This is our goal of this paper. We will show all the key properties of the desired CS theory in the continuum are reproduced, including the bosonic and fermionic level quantization, the bulk and chiral edge spectrum, the Wilson loop flux attachment (with point-split framing or geometric framing depending on the Maxwell coupling), the Wilson loop spin, the ground state degeneracy, and, most non-trivially, the chiral gravitational anomaly.

At this point we want to point out the relation between some previous works and our work:

In \cite{DeMarco:2019pqv}, a lattice bosonic CS-Maxwell theory is introduced, but with the action having discontinuities in the $\text{U(1)}$ gauge field. It is not hard to see the implementation of $\text{U(1)}$ gauge field in \cite{DeMarco:2019pqv} can be re-interpreted as starting with Villainization as in this paper, but always taking the saddle point approximation for the integer-valued Villain field (rather than allowing it fluctuate as in actual Villainized $\text{U(1)}$ gauge field). The cross-over between saddle points leads to discontinuity.

In \cite{Jacobson:2023cmr, Jacobson:2024hov}, Villainized $\text{U(1)}$ bosonic CS theory without Maxwell term has been proposed to implement the said topological properties due to $\pi_1(\text{U}(1))$, accompanied with an argument that the aforementioned extra zero modes \cite{Berruto:2000dp} might be a feature rather than a problem, in the sense that they restrict the form of Wilson loop observables. However, these extra zero modes do lead to divergence of the partition function and cannot be removed in a local manner \cite{Chen:2019mjw}, as we will review in the main text. Therefore, indeed, in the conclusion section of \cite{Jacobson:2023cmr}, the possibility of adding a Maxwell term has also been raised.

There is another line of development of lattice CS theory, by adding certain fine-tuned $B E_x$ and $B E_y$ terms (as opposed to the Maxwell $B^2$ and $E^2$ terms without any fine-tuning) in the Hamiltonian formalism to lift the undesired zero modes \cite{Eliezer:1991qh,Sun:2015hla}.
\footnote{We remark that these extra fine-tuned terms can be motivated from cup-1 product---a point that does not seem to have been mentioned in the previous literature. However, this is lengthy to explain and irrelevant to the main point of this paper, so we will leave this to future works.}
However, this approach does not seem to admit a description in terms of locally factorized Hilbert space subjected to locally implemented Gauss's law constraint, and does not seem to admit a spacetime lattice description; moreover, Villainization has not been done to implement the $\pi_1(\text{U(1)})=\mathbb{Z}$ topology. These aspects should be better understood.

One may also attempt to dynamically generate a lattice $\text{U(1)}$ CS theory by coupling a dynamical $\text{U(1)}$ gauge field to a massive lattice Dirac fermion in a Chern band. Yet such a theory is interacting in nature, and the generated dynamics is only guaranteed to appear as CS in the IR limit for weak gauge field configurations. For any careful analysis or application (such as coupling the CS theory to other sectors and analyzing the resulting dynamics), however, it is desired to have a lattice theory whose properties are controlled at the UV for generic dyanmical gauge field configurations. The massive Dirac fermion idea cannot serve the goal. 

\

This paper is organized as the following. In \cref{lattice_model} we define the theory and explain all the subtleties why it should be defined this way. In \cref{bulk_spectrum} and \cref{chiral_edge_mode} we solve for the bulk and the chiral edge spectrum in the Lorentzian signature. In \cref{ground_state_degeneracy} we compute the partition function in a Euclidean 3-torus to extract the ground state degeneracy. In \cref{gravitational_anomaly} we conduct the most non-trivial task of computing the chiral gravitational anomaly in the Euclidean signature, and giving it a UV complete physical interpretation. In \cref{Wilson_loop} we study the Wilson loop mutual statistic, self-statistics (with emphasis on the details on the Wilson loop  framing) and spin. Finally we give concluding remarks.

\

\emph{\textbf{Note}}: As this paper was being finalized, \cite{Peng:2024xbl} appeared, which also put the two aforementioned key ideas together and defined the same lattice CS-Maxwell theory as we do. Nonetheless, the emphases are very different between \cite{Peng:2024xbl} and this paper. \cite{Peng:2024xbl} focused on arriving at a Hamiltonian formulation, which we only discussed in \cref{hamiltonian_on_lattice}, {with some conceptual difference from} \cite{Peng:2024xbl}.
\footnote{{Our Hamiltonian formulation emphasizes manifest locality, at the price of having extra 1-form $\mathbb{Z}$ gauge constraint on each link, in addition to the usual 0-form gauge constraint. On the other hand, the Hamiltonian in }\cite{Peng:2024xbl} {fixes the 1-form $\mathbb{Z}$ gauge in a manner that depends on the global information of the space and the topological sector of the state, in the purview of future quantum computational implementation.}}
Our emphasis of this paper is to show the theory is indeed chiral, and for this purpose we presented detailed computations and physical interpretations for the chiral edge mode and the gravitational anomaly, which are not discussed in \cite{Peng:2024xbl}. Moreover, we also included an analysis of the Wilson loop framing, and a rigorous computation of the overall partition function.

\section{Lattice Model}
\label{lattice_model}
In the continuum, the $\text{U}(1)$ Chern-Simons-Maxwell action is written down as
\begin{equation}
    S=
    \int
    -\frac{1}{2e^2}\lVert F\rVert^2
    +\frac{k}{4\pi}A \wedge dA,
\end{equation}
where $\lVert F\rVert^2$ means $F^{\mu\nu}F_{\mu\nu}/2$, which is not necessarily positive in Lorentzian signature.
In the continuum, it is often said that a purely Chern-Simons (CS) theory can be defined without a Maxwell term. But this is not exactly true. A Maxwell term is needed for subtle regularization purpose \cite{Bar-Natan:1991fix}, as we will review later in this section. On the lattice, the necessity of the Maxwell term becomes more explicit. Now we will specify the lattice action on the lattice term by term.

\subsection{Lattice Path Integral Measure}
\label{lattice_path_integral_measure}
We will implement the $\text{U}(1)$ gauge field on the lattice by the Villainized degrees of freedom \cite{Einhorn:1977jm, Einhorn:1977qv}, which we will review now.

Naively, in a $\text{U}(1)$ lattice gauge theory, the dynamical degree of freedom is a lattice gauge field $e^{iA_l}\in \text{U}(1)$ on each link $l$, and the gauge flux (holonomy) around a plaquette $p$ is $e^{idA_p}\in \text{U}(1)$ where $dA_p$ is the lattice curl of $A_l$. The flux is invariant under the gauge transformation $e^{iA_l}\mapsto e^{id\phi_l} e^{iA_l}$, where $e^{i\phi_v}\in \text{U}(1)$ parameterizes the transformation on each vertex $v$, and $d\phi_l$ is the lattice derivative of $\phi_v$. 

There is a nice and conceptually important point of lattice gauge theory, that gauge fixing is not needed \cite{Wilson:1974sk}, because the gauge redundancy at each vertex contributes a finite constant factor to the partition function, and is thus equivalent to shifting the Lagrangian density by some finite local counter-term. Moreover, we do not need to demand observables to be gauge invariant, because any non-gauge invariant part of the observable will essentially vanish under the path integral, by Elitzur's theorem \cite{Elitzur:1975im}. Therefore, intrinsically, gauge redundancy does not, and should not, require any extra treatment.

\begin{figure}[htbp]
    \centering    \includegraphics[width=0.2\textwidth]{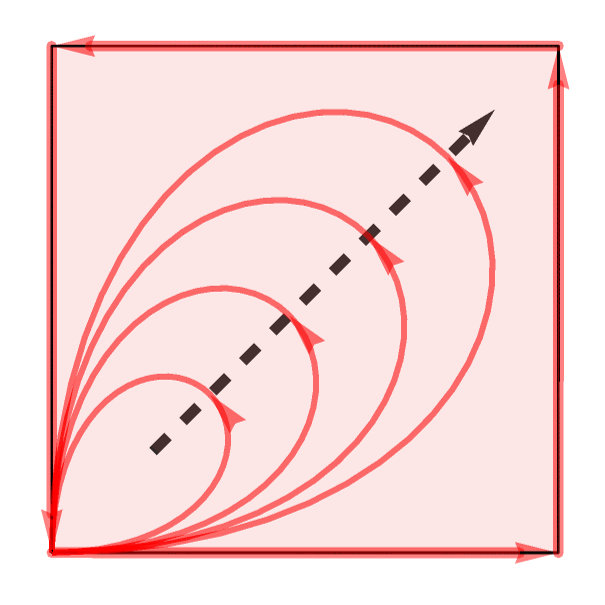}
    \caption{Villainization interpreted as extending the holonomy into the plaquette.}
    \label{extension}
\end{figure}

However, the naive lattice gauge theory does not allow topological configurations to be defined explicitly on the lattice, and this is why Villainization is needed. 
\footnote{Originally the Villainization method was introduced to explicitly describe vorticies in lattice $\text{U}(1)$ non-linear sigma model \cite{Berezinsky:1970fr}.}
Now we have the holonomy $e^{idA_p}\in\text{U}(1)$ around the plaquette. If we think of the lattice plaquette as being embedded in the continuum, let us consider the possible ways of how the holonomy around the plaquette can be extended into the inside of the plaquette. Consider the holonomy around a loop, which is gradually increasing its size, starting from residing at a single vertex and until going around the plaquette, as illustrated in \cref{extension}.
Over the process, the holonomy varies to form a continuous path in $\text{U}(1)$, starting at $1$ and ending at $e^{idA_p}$. While the starting and ending points are fixed, there are different possibilities of the interpolation, for instance the path may wind around $\text{U}(1)$ a few times before reaching $e^{idA_p}$. The path in $\text{U}(1)$ is characterized by a real number $F_p\in\mathbb{R}$ whose $\text{U}(1)$ part satisfies $e^{iF_p}=e^{idA_p}$ but whose $2\pi\mathbb{Z}$ part is unfixed by $e^{iA_l}$.  We will call $F_p\in\mathbb{R}$ the lattice gauge flux or lattice field strength, which can be viewed as a the integral of the continuum field strength over the plaquette. We can parameterize $F_p=dA_p-2\pi s_p$, and $s_p\in\mathbb{Z}$ is a new dynamical variable in the path integral. Here, we can either specify a $2\pi$ range for $A_l$, say $(-\pi, \pi]$, or we can leave the range of $A_l$ unspecified and note $A_l$ still essentially has $2\pi$ periodicity: $A_l\mapsto A_l+2\pi n_l \ (n_l\in\mathbb{Z})$ can be compensated by $s_p\mapsto s_p+2\pi dn_p$ to keep the physically meaningful $F_p$ invariant.

Now we are ready to see how Villainization allows topological configurations to be defined explicitly on the lattice. Let us consider a closed oriented 2d surface (possibly non-contractible) on the lattice formed by gluing plaquettes along edges. The total gauge flux through the surface is $\oint F := \sum_p F_p = -2\pi \sum_p s_p \in 2\pi\mathbb{Z}$. This is the Dirac quantization condition manifested on the lattice. Specially, if this closed surface is the boundary of a single cube $c$, the total flux $\sum_{p\in\partial c} F_p$ can be denoted as $dF_c$ (a lattice divergence), which is equal to $-2\pi ds_c \in 2\pi\mathbb{Z}$. This is the monopole charge located at the cube $c$. Therefore, viewed on the dual lattice (so that a plaquette $p$ corresponds to a dual link $l^\star$), $s_{l^\star}$ is like the Dirac string. We can control the occurrence of monopoles on the lattice by suitable path integral weights: If we want to forbid monopoles, we can introduce a $\text{U}(1)$ Lagrange multiplier field $\lambda_c$ with $e^{i\sum_c\lambda_c ds_c}$ in the path integral \cite{Sulejmanpasic:2019ytl,Chen:2019mjw}; or we can softly control the monopole fugacity by a suppression factor instead of a Lagrange multiplier. In this paper we will always forbid the monopoles with a Lagrange multiplier.

When monopoles are forbidden, up to the shift $A_l\mapsto A_l+2\pi n_l \ (n_l\in\mathbb{Z})$, $s_p\mapsto s_p+2\pi dn_p$, the $s$ configurations on the spacetime lattice $\mathcal{M}$ are classified by $H^2(\mathcal{M}; \mathbb{Z})$. This indeed agrees with the classification of $\text{U}(1)$ bundles in the continuum. Therefore Villainized  lattice $\text{U}(1)$ gauge field indeed reproduces the topological aspects of a continuum $\text{U}(1)$ gauge field.

In summary, we have two kinds of degrees of freedom: $A_l\in (-\pi, \pi]$ on each link and $s_p\in \mathbb{Z}$ on each plaquette; the field strength is $F_p=dA_p-2\pi s_p$ and the monopole density is $dF_c/2\pi=-ds_c$, which we will forbid by a Lagrange multiplier. The path integral measure is
\begin{equation}
    \left[\prod_{\text{link} \: l} \int_{-\pi}^\pi \frac{dA_l}{2\pi}\right] \left[\prod_{\text{plaq.} \: p} \sum_{s_p\in\mathbb{Z}} \right]
    \left[\prod_{\text{cube} \: c} \int_{-\pi}^\pi \frac{d\lambda_c}{2\pi} \: e^{i\lambda_c ds_c}\right] \ .
\end{equation}
It is often useful to think of $A_l\in\mathbb{R}$ instead, then there are two different kinds of gauge transformations: $A_l\mapsto A_l+d\phi_l$ for any $\phi_v\in\mathbb{R}$ on each vertex, and $A_l\mapsto A_l+2\pi n_l$, $s_p\mapsto s_p+dn_p$ for any $n_l\in\mathbb{Z}$ on each link. The latter kind of gauge transformation effectively restores the $2\pi$ periodicity of $A_l$ when $A_l$ is restricted to $(-\pi, \pi]$.

Before we move on, let us explain Villainization in more formal terms, which will be practically useful later. In the above we motivated Villainization starting with $e^{iA_l}\in \text{U}(1)$. Now, instead, let us start with a pure $\mathbb{R}$ gauge theory, with $A_l\in\mathbb{R}$. The $\mathbb{R}$ flux $dA_p$ is invariant under a 1-form $\mathbb{R}$ global symmetry, $A_l\mapsto A_l+\beta_l$ for any $\beta_l\in\mathbb{R}$ on the links satisfying $d\beta_p=0$.
\footnote{$\beta_l$ living on the link is what ``1-form'' means, and the condition $d\beta_p=0$ is what ``global'' means. This generalizes the ordinary (i.e. 0-form) global symmetry, where the transformation $\phi_v$ lives on vertices, and the ``global'' condition is $d\phi_l=0$ (so that $\phi_v$ is constant over each connected component of the spacetime).} 
Note this does not mean $\beta_l=d\phi_l$ when the spacetime is topologically non-trivial, i.e. a 1-form global symmetry is more than the usual gauge invariance. In order to reduce the $\mathbb{R}$ gauge theory to $\text{U}(1)$, we gauge a 1-form $2\pi\mathbb{Z}$ subgroup out of this 1-form $\mathbb{R}$ global symmetry. The gauging introduces a 2-form $\mathbb{Z}$ gauge field $s_p$, and the flux becomes $F_p=dA_p-2\pi s_p$.
\footnote{Just like gauging an ordinary 0-form symmetry introduces a 1-form gauge field, and the derivative becomes covariant derivative.}
The 1-form $2\pi\mathbb{Z}$ thus becomes a gauge invariance, $A_l\mapsto A_l+2\pi n_l$, $s_p\mapsto s_p+dn_p$ for any $n_l\in\mathbb{Z}$. The size of this gauge group is infinite on each link, so to properly define the path integral measure it is necessary to perform gauge fixing (in contrast to finite size gauge groups, for which gauge fixing is not needed on lattice \cite{Wilson:1974sk}, as we emphasized above), and $A_l\in (-\pi, \pi]$ is a gauge fixing condition---this condition is local and fixes the gauge completely without over-fixing, and is therefore useful for defining the path integral measure in an explicitly local manner. In practical calculations later, it will often be convenient to keep $A_l\in\mathbb{R}$---so that we can use Gaussian integral to solve the theory---but gauge fix $s_p$ instead; such gauge fixing condition depends on the spacetime topology and is incomplete (since a lattice derivative is involved in $s_p\mapsto s_p+dn_p$), but in a given spacetime these tasks are manageable. See \cref{ground_state_degeneracy} for more formal discussions on this.

\subsection{Lattice Action}
\label{lattice_action}
We can construct the Chern-Simons-Maxwell action on either a generic simplicial complex or a regular lattice, for $3$d spacetime of generic oriented topology. However, to solve for the dynamics of the theory---we will see that a Maxwell term is necessary and thus there will always be some non-topological dynamics---we would like to work with a regular lattice so that Fourier transformation can be used, and for simplicity we will restrict to cubic lattice in this paper. We will work with both the Lorentzian and the Euclidean signatures. 

For Lorentzian signature in the continuum, the Maxwell term is $F^{\mu\nu}F_{\mu\nu}/2=c^2\mathbf{B}^2-\mathbf{E}^2$ (as we will see, in Lorentzian signature it will be useful to keep $c$ as a tuning parameter). On the lattice, the Maxwell term will be
\begin{equation}
    S_{\text{Maxwell}}=-\frac{1}{2e^2}\sum_p \eta_p F_p^2
\end{equation}
where, with lattice constants $\Delta t, \Delta x, \Delta y$, we have $\eta_p=(1-i\epsilon)c^2 \Delta t/(\Delta x\Delta y)$ for $xy$-plaquette, $\eta_p=-(1+i\epsilon)\Delta x/(\Delta y\Delta t)$ for $yt$-plaquette and $\eta_p=-(1+i\epsilon)\Delta y/(\Delta x \Delta t)$ for $xt$-plaquette; the $i\epsilon$ prescriptions are to ensure the convergence of the path integral. As long as $\Delta x=\Delta y$, we can set $\Delta t=\Delta x=\Delta y =1$ by rescaling $c^2$ and $e^2$, and we will do so for the remaining of the paper. More generally, gauge invariant higher order terms can be included in the definition of the theory, but they are expected to be irrelevant. Here, we keep the simplest quadratic term so that the theory is solvable.

\begin{figure}[htbp]
    \centering
    \begin{subfigure}[t]{0.36\textwidth}
        \centering
        \includegraphics[width=\textwidth]{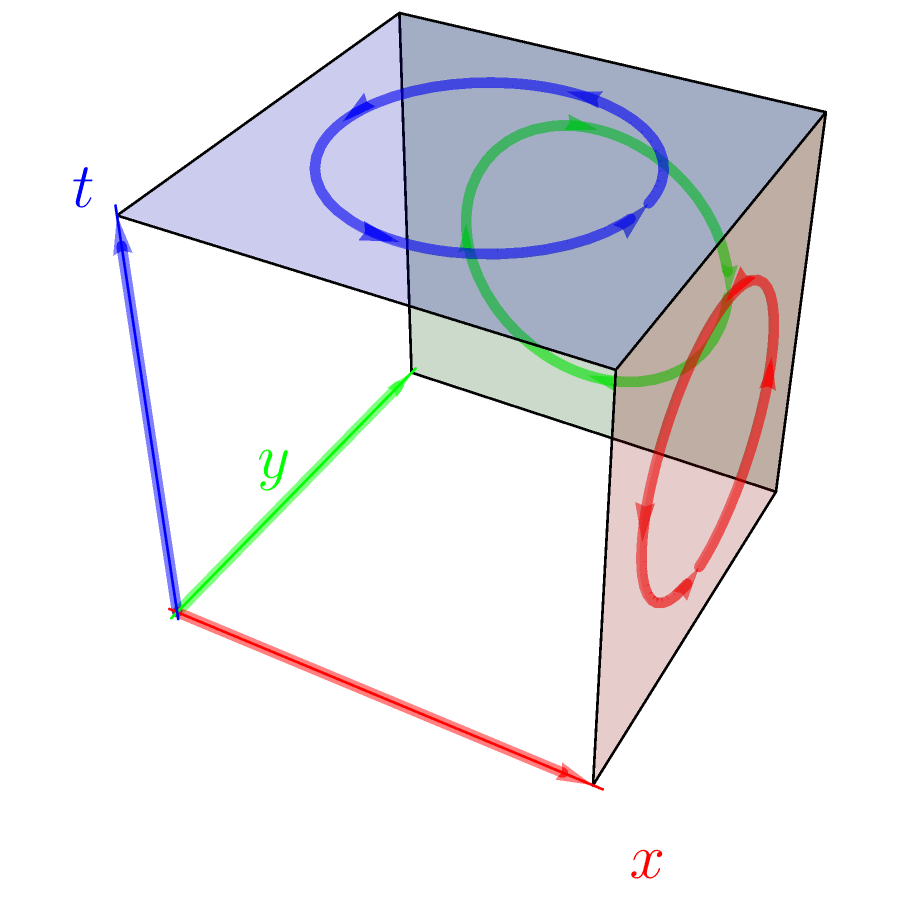}
        \caption{$(A\cup dA)_c$}
    \end{subfigure}
    \begin{subfigure}[t]{0.36\textwidth}
        \centering
        \includegraphics[width=\textwidth]{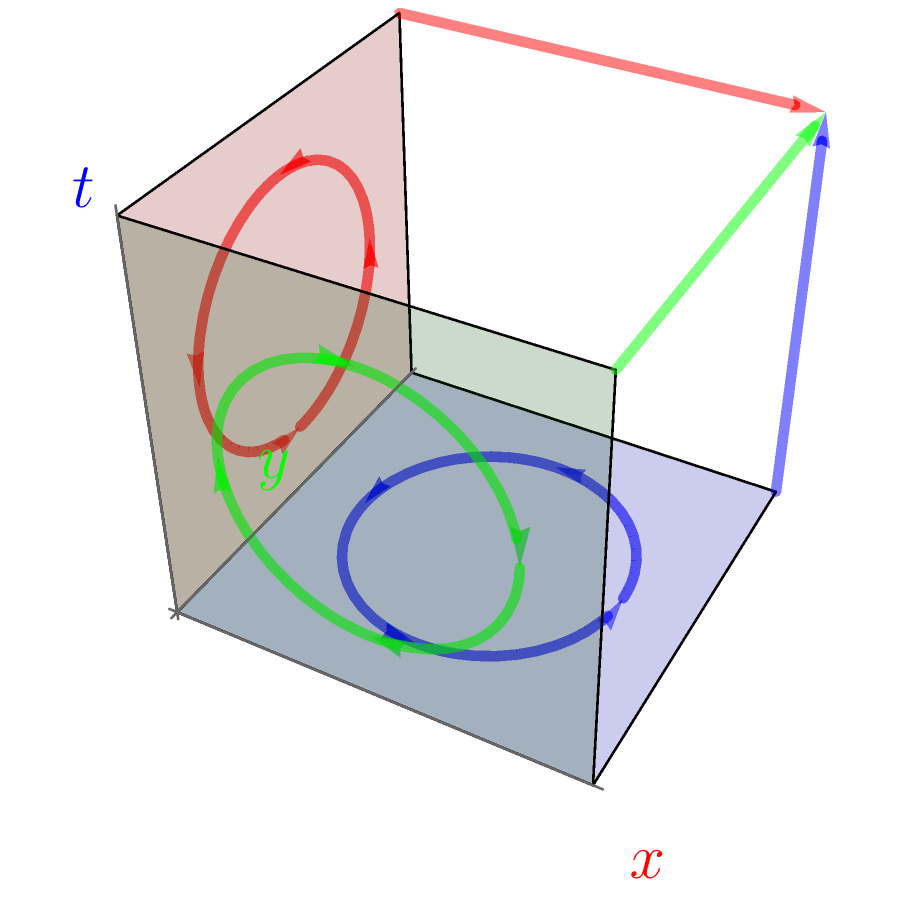}
        \caption{$(dA\cup A)_c$}
    \end{subfigure}
    \caption{Cup Product on Cubic Lattice}
    \label{cup_product_on_cubic_lattice}
\end{figure}

In order to define the Chern-Simons term, the lattice version of wedge product is essential. This product will be the cup product $\cup$. In the continuum, under gauge transformation,
\begin{equation}
    (A+d\phi)\wedge d(A+d\phi)-A\wedge dA=d\phi\wedge dA=d(\phi\wedge dA)
\end{equation}
so the CS term is gauge invariant on closed manifolds. To keep this crucial property on the lattice, the $\cup$ should satisfy the lattice version of Leibniz rule, that for $p$-form $f$ (i.e. $f$ lives on $p$-dimensional lattice cell) and $q$-form $g$, 
\begin{equation}
    d(f\cup g)=df\cup g+(-1)^pf\cup dg.
\end{equation}
On the cubic lattice, one choice of $(A\cup dA)_c$ is shown in \cref{cup_product_on_cubic_lattice}, we will use this choice until otherwise specified in \cref{spin}. For each cube, $(A\cup dA)_c$ is a sum of three terms, each term is $A_l$ on a colored link times $dA_p$ on the plaquette of the same color. When we sum over all the cubes on a spacetime without boundary, gauge invariance of $\sum_c (A\cup dA)_c$ under $\phi_v$ at any vertex $v$ can be checked explicitly. 

It is also important for the theory to respect the 1-form $\mathbb{Z}$ gauge invariance $A_l\mapsto A_l+2\pi n_l$, $s_p\mapsto s_p+dn_p$, which ensures the theory has the topological aspects of a $\text{U}(1)$ gauge theory. This will lead to the quantization of CS level. The lattice CS term reads \cite{Chen:2019mjw,Jacobson:2023cmr}
\begin{equation}
    S_{\text{CS}}=\frac{k}{4\pi}\sum_c \left[(A\cup dA)_c - (A\cup 2\pi s)_c - (2\pi s\cup A)_c\right].
    \label{CS_term}
\end{equation}
(Even if we did not set $\Delta t, \Delta x, \Delta y$ to be $1$, the CS action is independent of them.) Note that unlike the wedge product, in general, for lattice $p$-form $f$ and $q$-form $g$, $f\cup g\neq (-1)^{pq}g\cup f$, therefore the last two terms above cannot be combined; meanwhile, $A\cup dA=dA\cup A - d(A\cup A)$, and the $d(A\cup A)$ term vanishes upon integrating over a closed manifold. 
Under gauge transformation $A_l\mapsto A_l+d\phi_l$, $S_{\text{CS}}$ transforms by
\begin{equation}
    \frac{k}{4\pi}\sum_c\left[
    (d\phi\cup dA)_c-(d\phi\cup2\pi s)_c-(2\pi s\cup d\phi)_c
    \right].
    \label{0formU1transf}
\end{equation}
On a lattice without boundary, using the Leibniz rule this reduces to $(k/2) \sum_c \left[ \phi \cup ds + ds\cup \phi \right]$, which vanishes as $ds=0$ (or equivalently, we can say this transformation can be absorbed by a shift of the Lagrange multiplier $\lambda_c$).
\footnote{If $ds$ is not strictly forbidden by a Lagrange multiplier, then $S_{\text{CS}}$ should also have an extra higher cup product term $A\cup_1 ds$. \cite{Jacobson:2023cmr}.}
\footnote{If the $3$d lattice is embedded in a 4d lattice, then $dS_{\text{CS}}=(k/4\pi)\int F\cup F$, which is $\pi k$ times the abelian instanton density \cite{Sulejmanpasic:2019ytl, Chen:2019mjw,Jacobson:2023cmr}.}
The treatment on lattice with open boundary will be discussed in \cref{boundary_condition}. On the other hand, under 1-form $\mathbb{Z}$ gauge transformation $A_l\mapsto A_l+2\pi n_l$, $s_p\mapsto s_p+dn_p$, $S_{\text{CS}}$ transforms by
\begin{equation}  
    \frac{k}{4\pi}\sum_c \left[
    (2\pi n\cup dA)_c-(2\pi n\cup 2\pi s)_c-(2\pi s\cup 2\pi n)_c-(2\pi dn\cup A)_c-(2\pi dn\cup 2\pi n)_c
    \right].
    \label{1formZtransf}
\end{equation}
Using  $d(n\cup A)=dn\cup A-n\cup dA$, we have
\begin{equation}  
    \exp iS_{\text{CS}}\mapsto \exp \left\{iS_{\text{CS}}-i\pi k\sum_c \bigl[
    (n\cup s)_c+(s\cup n)_c+(dn\cup n)_c
    \bigr]\right\}
\end{equation}
on a lattice without boundary. For $e^{iS_{\text{CS}}}$ to be invariant under the 1-form $\mathbb{Z}$ gauge transformation, we arrive at the requirement of level quantization $k\in 2\mathbb{Z}$ for bosonic CS theory.

Fermionic CS theory with odd level $k$ can also be defined, if the spacetime has a specified spin structure (which is essentially specifying periodic or anti-periodic boundary condition for fermions).
\footnote{An oriented closed $3$d spacetime $\mathcal{M}$ always admits spin structure, but there may be different choices, and the difference is classified by $H^1(\mathcal{M}; \mathbb{Z}_2)$.}
In this case, $e^{iS_{\text{CS}}}$ needs to be modified to $e^{iS_{\text{CS}}} z_\chi[s]$, where $z_\chi[s]$ is a partition function for a Majorana fermion $\chi$ whose worldline is given by the Dirac string $s_p\!\mod 2$ \cite{Gaiotto:2015zta} ($s_p$ satisfying $ds=0$ forms closed loops on the dual lattice). In particular the path integral $z_\chi[s]$ always yields $\pm 1$, where the value depends on both $s\!\!\mod 2$ and the spin structure. Its role is that, under the 1-form $\mathbb{Z}$ gauge transformation, the possible $e^{i\pi}$ transformation in $e^{iS_{\text{CS}}}$ and that in $z_\chi[s]$ always cancel out. The details of the construction of $z_\chi[s]$ on cubic lattice can be found in \cite{Chen:2019mjw}; roughly speaking, 
each cube (i.e. vertex on the dual lattice) that the Majorana worldline moves through contributes a Berry phase of $\pm 1$ depending on which plaquettes on the cube the worldline goes through.

To sum up, in Lorentzian signature, the Chern-Simons-Maxwell partition function on cubic lattice reads
\begin{equation}
\begin{split}
    Z=&\left[\prod_{\text{link} \: l} \int_{-\pi}^\pi \frac{dA_l}{2\pi}\right] \left[\prod_{\text{plaq.} \: p} \sum_{s_p\in\mathbb{Z}} \right]
    \left[\prod_{\text{cube} \: c} \int_{-\pi}^\pi \frac{d\lambda_c}{2\pi} \: e^{i\lambda_c ds_c}\right] \ z_\chi[s]^k \\
    &\exp \left\{-\frac{i}{2e^2}\sum_p \eta_p F_p^2+\frac{ik}{4\pi}\sum_c \left[(A\cup dA)_c - (A\cup 2\pi s)_c - (2\pi s\cup A)_c\right]\right\},
\end{split}
\end{equation}
where $k\in \mathbb{Z}$ is even for bosonic CS (in which case $z_\chi[s]^k=1$) and odd for fermionic CS, and recall $\eta_p=(1-i\epsilon)c^2$ for $xy$-plaquette, $\eta_p=-(1+i\epsilon)$ for $xt$- and $yt$-plaquette.

In Euclidean signature, the partition function becomes
\begin{equation}
\begin{split}
    Z=&\left[\prod_{\text{link} \: l} \int_{-\pi}^\pi \frac{dA_l}{2\pi}\right] \left[\prod_{\text{plaq.} \: p} \sum_{s_p\in\mathbb{Z}} \right]
    \left[\prod_{\text{cube} \: c} \int_{-\pi}^\pi \frac{d\lambda_c}{2\pi} \: e^{i\lambda_c ds_c}\right] \ z_\chi[s]^k \\
    &\exp \left\{-\frac{1}{2e^2}\sum_p F_p^2+\frac{ik}{4\pi}\sum_c \left[(A\cup dA)_c - (A\cup 2\pi s)_c - (2\pi s\cup A)_c\right]\right\},
    \label{euclidean_partition_function}
\end{split}
\end{equation}
where we will always use $c^2=1$ in Euclidean signature unless otherwise specified. It obviously satisfies the reflection positivity requirement that when $\mathcal{M}$ reverses orientation, $Z$ becomes its complex conjugation.

\subsection{Necessity of Maxwell Term}
\label{necessity_of_maxwell_term}
We claimed the Maxwell term is necessary for regularization, both in the continuum and on the lattice. Now we explain this crucial point.

In the continuum, one may naively expect the partition function for ``pure $\text{U(1)}$ CS theory'' to be
\begin{equation}
    Z \ ``\!=\!"\: \int DA \exp \left(i\frac{k}{4\pi}\int AdA\right),
\end{equation}
which can be calculated by evaluating $(\det' d)^{-1/2}$ (where $\det'$ means determinant with zero eigenvalues removed) with some suitable Faddeev-Popov treatment. However, since $d$ has infinitely many eigenvalues, the phase of $(\det' d)^{-1/2}$ needs to be regularized using the eta-invariant \cite{Witten:1988hf}, and the infinitesimal regulator used in defining the eta-invariant is equivalent to having a Maxwell term with arbitrarily small but non-zero coefficient \cite{Bar-Natan:1991fix}. Therefore the CS theory is, in the end, not purely topological as one would naively expect. The presence of the gapless chiral edge mode (i.e. chiral boundary CFT), which is well-defined but not purely topological, is a reminiscence of this fact.

\begin{figure}[htbp]
    \centering
    \begin{subfigure}[t]{0.36\textwidth}
        \centering    \includegraphics[width=0.8\textwidth]{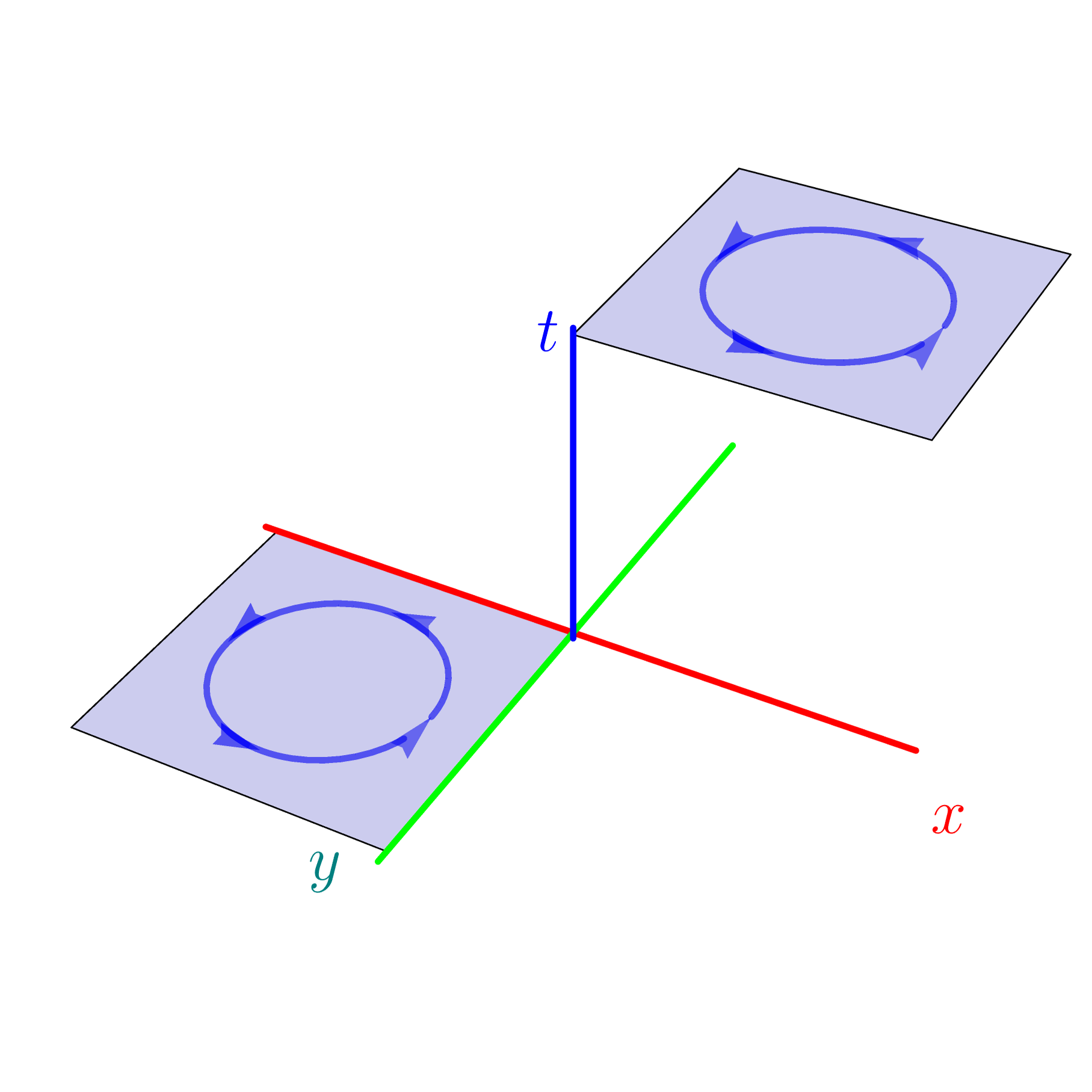}
    \end{subfigure}
    \begin{subfigure}[t]{0.36\textwidth}
        \centering    \includegraphics[width=\textwidth]{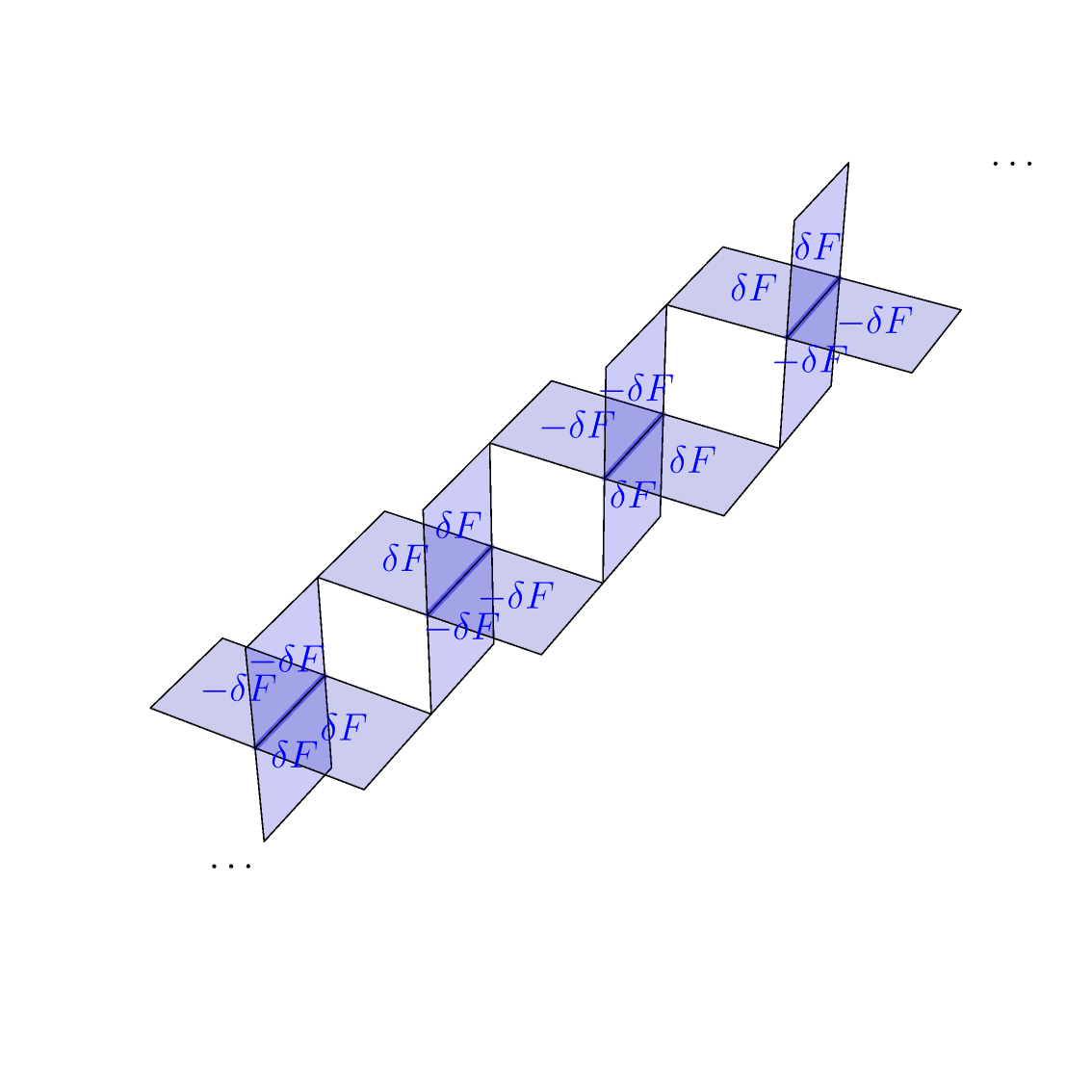}
    \end{subfigure}
    \caption{Left: Real space equation of motion without Maxwell term. We only showed the EoM on the $xy$-plaquettes. Right: Undesired zero mode in the shift of $F$. We only showed the zero mode  with $\delta F$ having $xy$- and $yt$-components (so to ensure $d\delta F=0$); the other two combinations of components also have the same kind of zero mode.}
    \label{eom_with_no_maxwell}
\end{figure}

On the lattice the problem manifests itself more prominently. Let us first consider the case of $\mathbb{R}$ gauge theory and look at the equation of motion, because for a quadratic theory the Gaussian integral is equivalent to taking the equation of motion. Due to the displacement between the link and the plaquette paired up in the cup product, the real space equation of motion (without Maxwell term for now) implies the sum of the gauge flux on two displaced plaquettes, as shown in \cref{eom_with_no_maxwell} (left), equals zero. But this does not mean the flux on each individual plaquette is zero, contrary to our desired equation of motion from the continuum ``pure CS theory''. This leads to extra zero modes, where the gauge flux can shift by interlacing value along the displacing direction without changing the action, as shown in \cref{eom_with_no_maxwell} (right). When we turn the $\mathbb{R}$ gauge theory to Villainized $\text{U(1)}$ gauge theory, such shift of the (still real valued) flux $F$ still leaves the action unchanged. This makes the partition function divergent, regardless of the signature.

Is it possible to remove such undesired zero mode by some constraints in the path integral similar to some gauge fixing condition? Note the shift of $F$ has a non-local profile in the real space, so it cannot be fixed by any local condition. To better demonstrate the non-locality, consider a 3-torus spacetime with $L_x$, $L_y$, $L_\tau$ vertices in each direction, and it turns out that such zero mode will exist unless all of $L_x, L_y, L_\tau$ are odd,
\footnote{This is most easily seen in the momentum space. The extra zero mode occurs at $q_x+q_y+q_\tau=\pi\!\mod 2\pi$, as we will see in the next section (and recall the pure CS term is the same in Lorentzian and Euclidean signature). This can be satisfied as long as at least one of $L_x, L_y, L_\tau$ is even. Then restriction of the zero mode on a plane in the momentum space reflects its non-locality along a line in the real space.}
so the divergence indeed depends on the global details of the lattice. On more general simplical complexes, cup product can also be defined with given branching structure, and whether such zero mode
\footnote{``Such zero mode'' means any non-trivial solution to $M\cap \delta F+ \delta F\cap M =0$ where $M$ is the $3$-chain representing the fundamental class of the manifold}.
exists again depends on all the lattice details. So this problem is totally different from gauge redundancy, which is local because a gauge transformation can be made at each individual vertex (and moreover gauge redundancy is finite rather than divergent for finite dimensional compact Lie group, as we emphasized before).

The problem is not due to our definition of the CS term being a ``bad choice''. In \cite{Berruto:2000dp}, gauge invariant quadratic terms that are odd under reflection have been systematically considered and there is always the same kind of problem. (The gauge theory is $\mathbb{R}$ in \cite{Berruto:2000dp}, but for the same reason as above the problem persists in Villainized $\text{U(1)}$ gauge theory.) So it was argued that some non-topological, reflection even term---with a quadratic Maxwell term being the simplest and most natural choice---must be included.

In summary, a Maxwell term is needed both in the continuum and on the lattice. The advantage of the lattice is that it makes the necessity much more explicit. While the necessity in the continuum implies the necessity is universal. (In comparison, certain doubled CS theories do not require Maxwell term and are realized purely topologically on the lattice \cite{Chen:2019mjw}.) With the Maxwell term, the partition function over a finite spacetime is finite (in the Lorentzian signature, the $i\epsilon$ prescription in $\eta_p$ is important for this to be true) and well-defined in a manifestly local manner.

\subsection{Global Symmetries and Anomalies}
\label{global_symm_anom}

In this subsection we review the manifestation of the global symmetries and anomalies of the CS term on the lattice \cite{Chen:2019mjw, Jacobson:2023cmr}. The Maxwell term respects these symmetries and does not lead to extra anomalies.

The prohibition of monopole means the Dirac quantized flux over space is conserved over time, leading to an ordinary (0-form) $\text{U(1)}$ global symmetry, manifested by \cite{Chen:2019mjw,Sulejmanpasic:2019ytl}
\begin{equation}
\lambda_c \mapsto \lambda_c + \xi_c \mod 2\pi, \ \ \ \ \ \partial \xi_p = 0 \mod 2\pi
\end{equation}
where $\partial$ can be viewed as $d^\star$ on the dual lattice, and this just means $\xi_c$ is a locally constant, i.e. constant over each connected piece of the lattice. (The situation on the boundary will be discussed in the next subsection.) This symmetry has no self-anomaly and can be consistently coupled to a background $\text{U}(1)$ gauge field, in the same way as in doubled CS theory \cite{Chen:2019mjw,Han:2021wsx,Han:2022cnr}; in the present paper we will not pursue this further.

The CS term \cref{CS_term} also has a 1-form $\mathbb{Z}_k$ global symmetry \cite{Chen:2019mjw, Jacobson:2023cmr}
\begin{equation}
    A_l \mapsto A_l + \frac{2\pi}{k} \ell_l \mod 2\pi, \ \ \ \ \ \ell_l \in \mathbb{Z} \mod k, \ \ \ \ \ d\ell_p=0 \mod k
\end{equation}
where the mod $k$ is because that part can be absorbed into the gauged 1-form $\mathbb{Z}$. Here we used the fact that when both $\ell$ and $s$ are closed, $\ell\cup s$ and $s\cup \ell$ only differs by an exact term, i.e. a total derivative.
\footnote{The total derivative is given by $d(\ell \cup_1 s)$. Note that $s$ is only closed after integrating out $\lambda$. Before integrating out $\lambda$, we would also need $\lambda$ to shift by a term $M\cap_1 \ell$, where $M$ is, again, the $3$-chain representing the fundamental class of the spacetime lattice.}
(Again, the situation on the boundary will be discussed in the next subsection.) The interpretation is that the $\text{U(1)}$ holonomy around any non-contractible loop in the space is, and EoM, conserved over time and quantized to $(2\pi/k)\mathbb{Z}_k$ values.

The 1-form $\mathbb{Z}_k$ has self-anomaly. To see this \cite{Chen:2019mjw, Jacobson:2023cmr}, we can try to couple the theory to a background $\mathbb{Z}_k$ gauge field $\mathbf{B}_p \in \mathbb{Z}\mod k$, by replacing $s_p$ in \cref{CS_term} with $s_p+\mathbf{B}_p/k$ (the mod $k$ part can be absorbed into $s$). For simplicity we can further demand the background to be $\mathbb{Z}_k$ flat, $d\mathbf{B}_c=0\mod k$. Now we perform a 1-form $\mathbb{Z}_k$ gauge transformation, i.e. without the $d\ell_p=0 \mod k$ constraint, but demanding $\mathbf{B}_p\mapsto \mathbf{B}_p+d\ell_p \mod k$, we will find the partition function changes by $e^{(i\pi/k)\mathbb{Z}}$ which is not invariant, manifesting the self-anomaly. Associated with this, if we attempt to gauge this 1-form symmetry, i.e. making $\mathbf{B}_p$ dynamical, then we can redefine $\mathbf{B}_p+ks_p$ as $s_p$, and rescale $kA_l$ as $A_l$ accordingly, so that action becomes that of a level-$1/k$ CS, which indeed has vanishing partition function.
\footnote{However, if we demand $\ell_l$ and $\mathbf{B}_p$ to be multiples of $m$ such that $m^2/k\in 2\mathbb{Z}$, the theory will remain invariant. This means such $\mathbb{Z}_{k/m}$ subgroups are non-anomalous. If $m^2/k$ is odd, by including extra fermions in the same way as we discussed before, the theory will also be invariant.}

The global 1-form $\mathbb{Z}_k$ also has a mixed anomaly with the global 0-form $\text{U(1)}$. Since the mod $k$ part of $\mathbf{B}_p$ is absorbed into $s_p$, the Lagrange multiplier factor $e^{i\lambda_c ds_c}$ is no longer well-defined. One choice to make it well-defined is to replace $ds_c$ by $ds_c+d\mathbf{B}_c/k$, but then the global 0-form $\text{U(1)}$ condition becomes $\partial \xi_p = 0 \mod 2\pi k$ instead of $2\pi$, i.e. the periodicity is extended. Another choice is to introduce a Villainzing background $\mathbf{C}_c\in \mathbb{Z}$ for $\mathbf{B}_p$ such that $d\mathbf{B}_c/k-\mathbf{C}_c$ is well-defined in $\mathbb{Z}$, and replace $ds_c$ by $ds_c+\mathbf{C}_c$. But it is impossible to require $\mathbf{C}_c$ to be exact by any local condition, so a generic non-exact $\mathbf{C}_c$ (regardless of whether it is closed) will manifestly break the global 0-form $\text{U(1)}$.
\footnote{This is similar to the lattice manifestation of a familiar mixed anomaly in $S^2$ non-linear sigma model, between the  rotational $\text{SO(3)}$ symmetry and the hedgehog-forbiddening $\text{U(1)}$ symmetry, in relation to the spin-c obstruction by the 3rd integral Stiefel-Whitney class \cite{Chen:2024ddr}.} Alternatively, we can instead introduce a background $\text{U(1)}$ background gauge field \cite{Chen:2019mjw} and find it manifestly breaks the 1-form $\mathbb{Z}_k$.

\subsection{Boundary Condition}
\label{boundary_condition}
Previously we focused on closed spacetime manifolds, now we discuss the cases with boundary. As is familiar from $(0+1)$d quantum mechanics, on a spacetime with boundary, the degrees of freedom on the boundary should be specified, serving as the boundary condition of the path integral. The question is, what are ``the degrees of freedom on the boundary" that are to be specified.

\begin{figure}[htbp]
    \centering
    \begin{subfigure}[t]{0.36\textwidth}
        \centering
        \tikzset{every picture/.style={line width=0.75pt}} 
    
        \begin{tikzpicture}[x=0.75pt,y=0.75pt,yscale=-1,xscale=1]

        \draw  [draw opacity=0][line width=0.75]  (50,50) -- (210,50) -- (210,170) -- (50,170) -- cycle ; \draw  [color={rgb, 255:red, 155; green, 155; blue, 155 }  ,draw opacity=1 ][line width=0.75]  (70,50) -- (70,170)(90,50) -- (90,170)(110,50) -- (110,170)(130,50) -- (130,170)(150,50) -- (150,170)(170,50) -- (170,170)(190,50) -- (190,170) ; \draw  [color={rgb, 255:red, 155; green, 155; blue, 155 }  ,draw opacity=1 ][line width=0.75]  (50,70) -- (210,70)(50,90) -- (210,90)(50,110) -- (210,110)(50,130) -- (210,130)(50,150) -- (210,150) ; \draw  [color={rgb, 255:red, 155; green, 155; blue, 155 }  ,draw opacity=1 ][line width=0.75]  (50,50) -- (210,50) -- (210,170) -- (50,170) -- cycle ;
        \draw [color={rgb, 255:red, 208; green, 2; blue, 27 }  ,draw opacity=1 ]   (130,50) -- (130,90) -- (110,90) -- (110,130) -- (150,130) -- (150,170) ;
        
        \end{tikzpicture}
        \caption{rough lattice boundary}
    \end{subfigure}
    \begin{subfigure}[t]{0.36\textwidth}
        \centering
        \tikzset{every picture/.style={line width=0.75pt}} 
        \begin{tikzpicture}[x=0.75pt,y=0.75pt,yscale=-1,xscale=1]
        
        \draw  [draw opacity=0][line width=0.75]  (50,50) -- (210,50) -- (210,170) -- (50,170) -- cycle ; \draw  [color={rgb, 255:red, 155; green, 155; blue, 155 }  ,draw opacity=1 ][line width=0.75]  (70,50) -- (70,170)(90,50) -- (90,170)(110,50) -- (110,170)(130,50) -- (130,170)(150,50) -- (150,170)(170,50) -- (170,170)(190,50) -- (190,170) ; \draw  [color={rgb, 255:red, 155; green, 155; blue, 155 }  ,draw opacity=1 ][line width=0.75]  (50,70) -- (210,70)(50,90) -- (210,90)(50,110) -- (210,110)(50,130) -- (210,130)(50,150) -- (210,150) ; \draw  [color={rgb, 255:red, 155; green, 155; blue, 155 }  ,draw opacity=1 ][line width=0.75]  (50,50) -- (210,50) -- (210,170) -- (50,170) -- cycle ;
        \draw [color={rgb, 255:red, 208; green, 2; blue, 27 }  ,draw opacity=1 ]   (130,50) -- (150,50) ;
        \draw [color={rgb, 255:red, 208; green, 2; blue, 27 }  ,draw opacity=1 ]   (130,70) -- (150,70) ;
        \draw [color={rgb, 255:red, 208; green, 2; blue, 27 }  ,draw opacity=1 ]   (130,70) -- (130,90) ;
        \draw [color={rgb, 255:red, 208; green, 2; blue, 27 }  ,draw opacity=1 ]   (90,90) -- (110,90) ;
        \draw [color={rgb, 255:red, 208; green, 2; blue, 27 }  ,draw opacity=1 ]   (90,110) -- (110,110) ;
        \draw [color={rgb, 255:red, 208; green, 2; blue, 27 }  ,draw opacity=1 ]   (110,70) -- (110,90) ;
        \draw [color={rgb, 255:red, 208; green, 2; blue, 27 }  ,draw opacity=1 ]   (90,130) -- (110,130) ;
        \draw [color={rgb, 255:red, 208; green, 2; blue, 27 }  ,draw opacity=1 ]   (110,130) -- (110,150) ;
        \draw [color={rgb, 255:red, 208; green, 2; blue, 27 }  ,draw opacity=1 ]   (110,150) -- (130,150) ;
        \draw [color={rgb, 255:red, 208; green, 2; blue, 27 }  ,draw opacity=1 ]   (110,170) -- (130,170) ;
        \draw [color={rgb, 255:red, 208; green, 2; blue, 27 }  ,draw opacity=1 ] [dash pattern={on 4.5pt off 4.5pt}]  (140,50) -- (140,80) -- (100,80) -- (100,140) -- (120,140) -- (120,170) ;

        \end{tikzpicture}

        \caption{smooth lattice boundary}
    \end{subfigure}
    \caption{Cuttings that create rough versus smooth lattice boundaries on a $2$d lattice. Here for the ``rough'' lattice boundary, the red boundary itself looks smooth, but each grey region has an outmost layer of dangling links sticking perpendicularly onto the boundary, which looks rough; for the ``smooth'' lattice boundary, the red boundary itself looks rough, but the outmost layer of each grey region looks smooth. The generalization to $3$d is obvious: A $3$d rough lattice boundary has vertices, links and plaquettes lying on the boundary, and so the outmost layer of each region are made of cubes adjacent to the boundary and plaquettes and links sticking perpendicularly onto the boundary; the opposite for a $3$d smooth lattice boundary, which is equivalent to the rough boundary for the dual lattice.}
    \label{lattice_boundary}
\end{figure}

To answer this, note the path integral for a theory with spacetime locality has the ``glueing property" that, we can cut the total spacetime manifold into many patches, and the path integral evaluated over the total spacetime is equal to taking the product of the path integral evaluated over each patch, and then integrating out the degrees of freedom on the patch boundaries (perhaps with some boundary weight that depends only on these degrees of freedom on the patch boundaries). In our case, let us first consider ``rough lattice boundary'' (see \cref{lattice_boundary}), then
\begin{equation}
    \begin{split}
    Z=&\left[\prod_{\text{link $l$ on boundaries}} \int_{-\pi}^\pi \frac{dA_l}{2\pi}\right] \left[\prod_{\text{plaq. $p$ in boundaries}} \sum_{s_p\in\mathbb{Z}} \right]
    \\[.1cm]
    & \ \left[\prod_{\text{plaq. $p$ in boundaries}} \left( \text{Maxwell weight on $p$}\right) \right]
    \left[ \prod_{\text{patch}} Z_{\text{patch}}\left[A_{l\in\text{boundary}},s_{p\in\text{boundary}}\right]\right] \ .
\end{split}
\end{equation}
Thus $Z_{\text{patch}}[A_{l\in\text{boundary}},s_{p\in\text{boundary}}]$ can be interpreted as the partition function for a spacetime (i.e. the patch) with boundary, and those boundary degrees of freedom to be specified are \begin{equation}    
\left\{\begin{aligned}
    A_{l\in\text{boundary}}&\in (-\pi,\pi]\\
    s_{p\in\text{boundary}}&\in \mathbb{Z} \ .
\end{aligned}\right.
\end{equation}
This is the lattice version of Dirichlet boundary condition. 

Let us make a few remarks:
\begin{itemize}
    \item
    If we want, we can take the square root of the Maxwell weight on the boundary plaquettes, and absorb one such factor into each of the $Z_{\text{patch}}$ of the two neighbouring patches. This changes $Z_{\text{patch}}$ by an overall factor that only depends on the specified boundary conditions, without changing the dynamics inside the bulk of the patch. On the other hand, there is no CS weight on the boundary because there is no cube on the boundary.
    
    \item
    For fermionic CS the above stays the same. This is because the $z_\chi[s]$ is constructed \cite{Gaiotto:2015zta, Chen:2019mjw} by having Majorana worldlines moving along $s_p \!\mod 2$, and the $\pm 1$ Berry phase is contributed by each cube, which always lies inside some patch. Hence it suffices to specify $s_p$ on the boundary.

    \item
    The 0-form $\text{U(1)}$ global symmetry introduced in the previous subsection is respected on the boundary only when $F_{p\in\text{boundary}}=0$, since otherwise there will be flux (the conserved charge) flowing out of the system. And the 1-form $\mathbb{Z}_k$ global symmetry is obviously not respected by any Dirichlet boundary condition that specifies $A_{l\in\text{boundary}}$.
    
    \item
    We can also consider ``smooth lattice boundary" (see \cref{lattice_boundary}), but to handle this we need to first perform a Hubbard-Stratonovich transformation with a field living on the plaquettes; in particular, on the $tx$- and $ty$-plaquettes, the Hubbard-Stratonovich field is interpreted as the canonical momentum of $A$, see \cref{hamiltonian_on_lattice}. This will give the lattice version of Neumann boundary condition. In this paper we will focus on the Dirichlet boundary condition.
    \footnote{Roughly speaking, in the $e^2\to\infty$ limit, the equations of motion says $\Pi^y$ (conjugate momentum of $A_y$, which arises as Hubbard-Stratonovich field)} approaches $-(k/4\pi)A_x$ and $\Pi^x$ (conjugate momentum of $A_x$) approaches $(k/4\pi)A_y$. If we specify the boundary condition by $A_x$ and $\Pi^y$ (in the Hamiltonian formalism they commute, and on the spacetime lattice we use a suitable mix of smooth and rough cuts), then in the $e^2\to\infty$ limit, it is like specifying $A_x$ only; likewise if we exchange $x$ and $y$. This is why, in the so-called ``pure CS theory'' (which really means the $e^2\to\infty$ limit) in continuum, it is usually said ``$A_x$ and $A_y$ are canonical variables, and we only specify one of them in the boundary condition".
\end{itemize}

The specification of boundary condition is therefore not intrinsically tied to gauge invariance, contrary to what is sometimes being said in the continuum context. Indeed, as we emphasized at the beginning, it becomes manifest on the lattice that gauge redundancy is merely a local constant factor that does not require any special treatment \cite{Wilson:1974sk}. However, we can still ask how gauge invariance works in the presence of boundary. First, for any vertex $v$ inside the bulk, performing the $e^{i\phi_v}$ gauge transformation obviously leaves the partition function invariant, likewise for the $2\pi n_l$ shift of $A_l$ for any link $l$ inside the bulk. On the other hand, we can ask how about a boundary condition that is related to the original boundary condition by ``gauge transformation on the boundary'', $e^{i\phi_{v\in \text{boundary}}}$ and $2\pi n_{l\in \text{boundary}}$:
\begin{equation}
    \left\{
    \begin{aligned}
        A'_{l\in \text{boundary}}&=A_{l\in \text{boundary}}+2\pi n_{l\in \text{boundary}}+d\phi_{l\in \text{boundary}}\\[.1cm]
        s'_{p\in \text{boundary}}&=s_{p\in \text{boundary}}+d n_{p\in \text{boundary}}
    \end{aligned}\right. \ .
\end{equation}
According to \cref{0formU1transf} and \cref{1formZtransf}, along with the Leibniz rule and $ds_c=0$ (which is enforced by Lagrangian multiplier in each cube, and there is no cube on the boundary), for bosonic CS we have
\begin{equation}
\begin{split}
&Z[A'_{l\in\text{boundary}},s'_{p\in\text{boundary}}]=\\[.1cm]
&\exp\left[i\frac{k}{4\pi}\sum_{\text{$p\in$ boundary}} (\phi\cup dA-\phi\cup 2\pi s-2\pi s\cup \phi-2\pi n\cup A)_p\right]
Z[A_{l\in\text{boundary}},s_{p\in\text{boundary}}] \ .
\end{split}
\label{guage_transformation_on_boundary}
\end{equation}
i.e. for two boundary conditions differing only by a ``boundary gauge transformation'', the associated partition functions only differ by a constant phase given by the boundary conditions in a local manner, while the dynamics is unaffected. For fermionic CS there is an extra $\pm 1$ phase that also depends only on the boundary conditions.

\

This completes the construction of the lattice model. In the remaining of the paper we solve for the interesting properties of the chiral $\text{U(1)}$ CS theory.

\section{Bulk Spectrum}
\label{bulk_spectrum}
We begin with the bulk spectrum of the theory in Lorentzian signature on an infinite cubic lattice. The concept of ``bulk spectrum'' is well-defined for free theory, so we shall first write the theory in a form that is manifestly free. Since the topology of the spacetime is trivial, any Dirac string field $s_p$ satisfying $ds_c=0$ can be written as $s_p=dn_p$ for some $n_l\in\mathbb{Z}$. Then we can absorb the $n_l$ into $A_l$, so that $A_l-2\pi n_l\in \mathbb{R}$ is now redefined as $A_l\in\mathbb{R}$, such that $F_p=dA_p$ after the redefinition. Thus the theory 
\begin{equation}
    S=-\frac{1}{2e^2}\sum_p \eta_p (dA_p)^2+\frac{k}{4\pi}\sum_c (A\cup dA)_c,
    \label{action}
\end{equation}
is now quadratic in the real valued $A_l$, hence manifestly free. (More systematic treatments, including what happens when the spacetime topology is non-trivial, as well as how the ambiguity in $n_l$ is handled by the real-valued Faddeev-Popov measure of $A_l$, will be discussed in \cref{ground_state_degeneracy}.)

We denote a link $l$ as $(r, \mu)$ where $r\in\mathbb{Z}^3$ is the coordinate of the point on the link with the smallest coordinates, viewed as the starting point of the link, and $\mu$ is the direction it is pointing. Likewise for plaquette $p=(r, \mu\nu)$ and cube $c=(r, txy)$. Then we Fourier transform
\begin{equation}
    A_\mu(q)=\sum_{l=(r, \mu)}A_\mu(r)e^{iq\cdot r}
\end{equation}
where $q_\mu=(-\omega, q_x, q_y)$. After the Fourier transform, the action written in matrix form reads
\begin{equation}
    iS= -\int \frac{dq}{(2\pi)^3}\frac{1}{2} A(q)^\dagger M(q) A(q)
\end{equation}
where
\begin{equation}
    M(q)=i\left[\frac{1}{e^2}d_1(q)^\dagger\eta d_1(q) -\frac{k}{4\pi}(\cup(q)d_1(q)+d_1(q)^\dagger\operatorname{\cup}(q)^\dagger)\right].
    \label{def_M}
\end{equation}
The detailed matrices for $d_1(q)$, $\eta$ and $\cup(q)$ are given in \cref{explicit_form_of_operators_and_structures}. Here $d_1$ has a subscript ``$1$" in order to emphasize that here the lattice derivative $d$ is acting on $1$-form (link variable); later we will also have $d_0$ and $d_2$.

The classical EoM is given by $M(q) A(q)=0$ when we neglect the $i\epsilon$ prescriptions in $\eta$, and the same will be understood in the calculations of spectrum below, unless otherwise emphasized. The $i\epsilon$ prescription ensures that the modes that satisfy the classical EoM when neglecting $i\epsilon$ will not lead to divergence of the partition function when including the $i\epsilon$, rendering a convergent path integral.

\begin{figure}[htbp]
    \centering
    \begin{subfigure}[t]{0.32\textwidth}
        \centering
        \includegraphics[width=\textwidth]{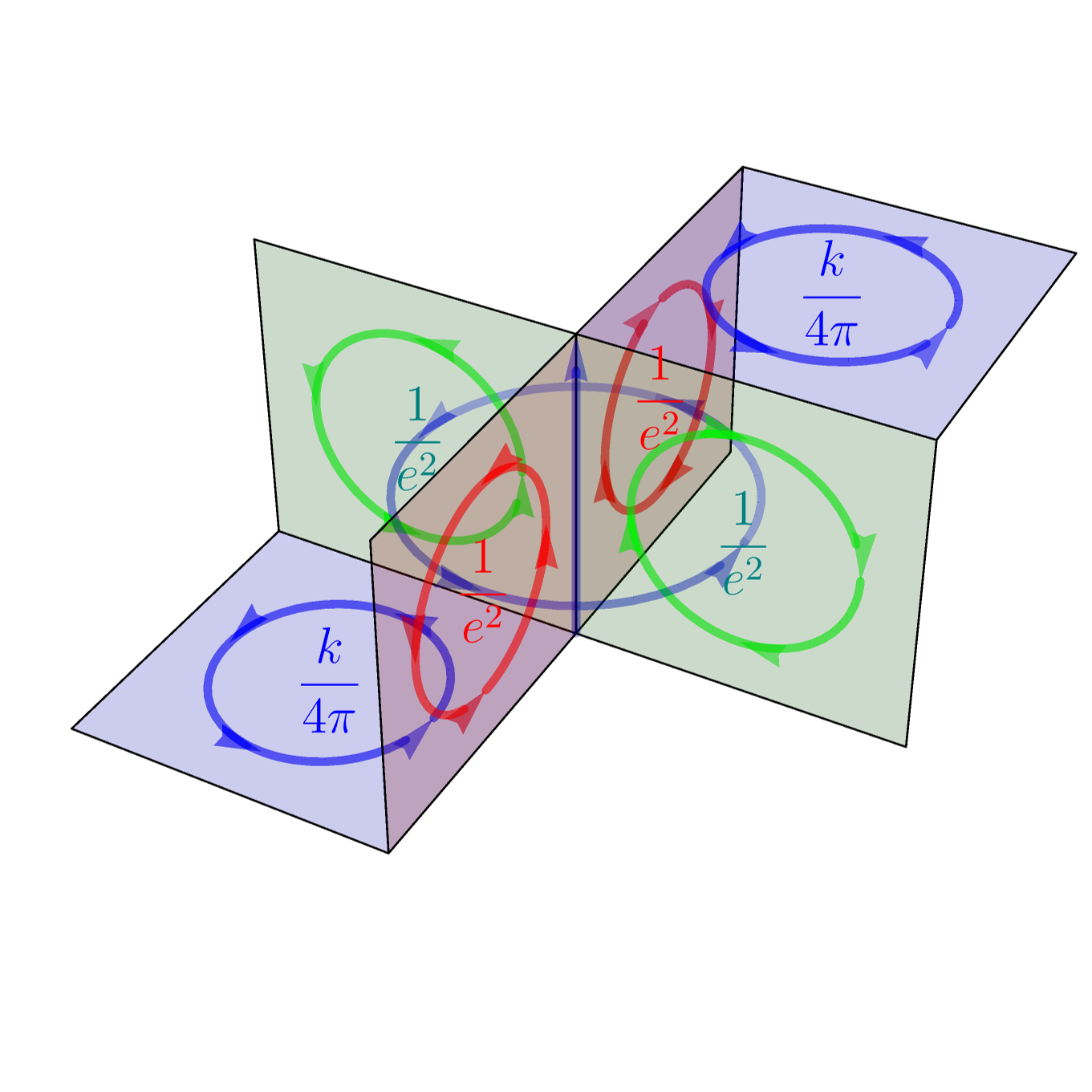}
        \caption{$\frac{\delta S}{\delta A_t(r)}=(iKd_1A)^t(r)$}
    \end{subfigure}
    \begin{subfigure}[t]{0.32\textwidth}
        \centering
        \includegraphics[width=\textwidth]{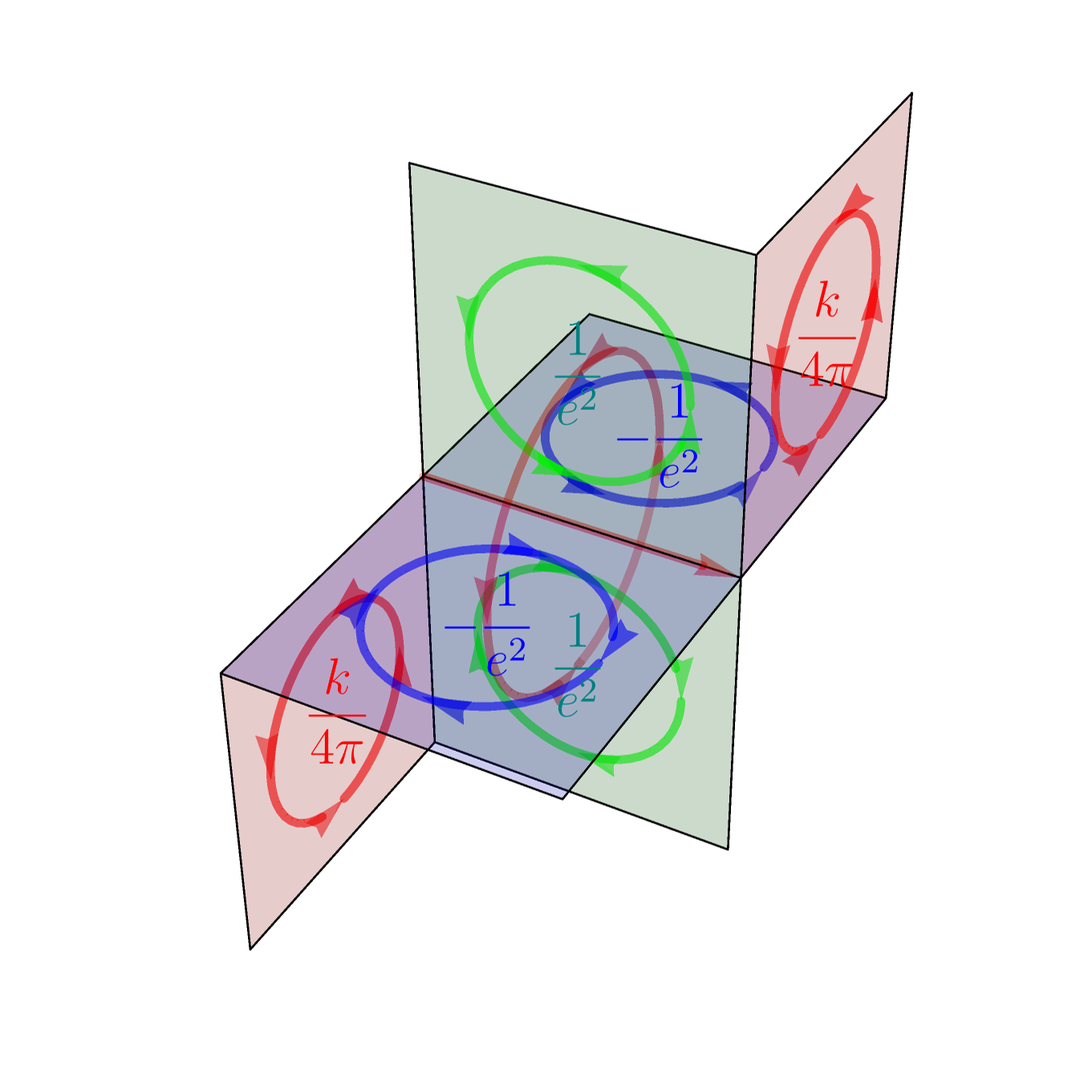}
        \caption{$\frac{\delta S}{\delta A_x(r)}=(iKd_1A)^x(r)$}
    \end{subfigure}
    \begin{subfigure}[t]{0.32\textwidth}
        \centering
        \includegraphics[width=\textwidth]{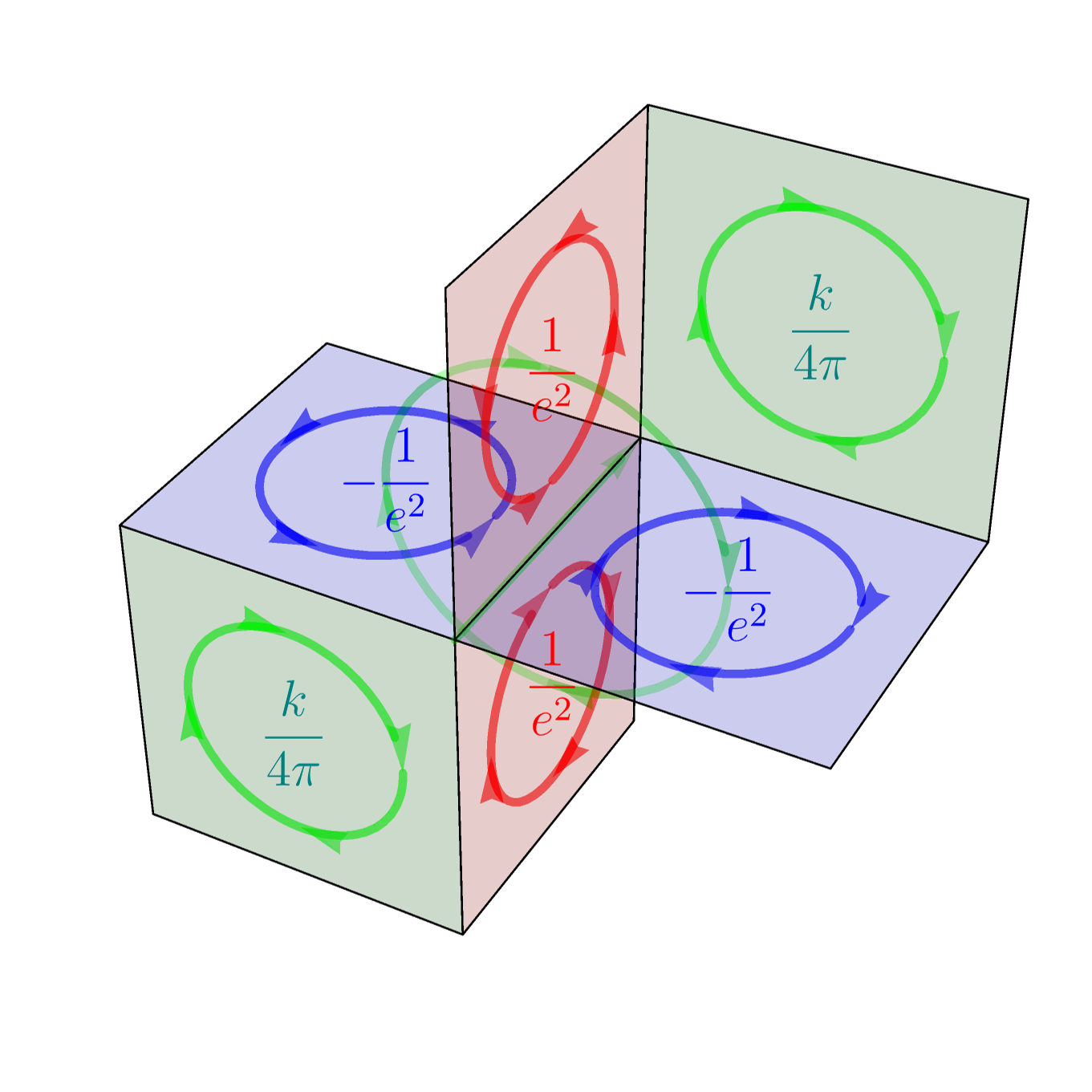}
        \caption{$\frac{\delta S}{\delta A_y(r)}=(iKd_1A)^y(r)$}
    \end{subfigure}
    \caption{Real space EoM: the straight arrows represent the $\delta A(r)$ being varied to derive the EoM, the circular arrows on plaquettes represent the orientation of the $dA$ in the EoM, and the coefficients of the $dA$'s are also labeled on the plaquettes.}
    \label{diagrammatic_representation_of_EoMs}
\end{figure}

The operator $M$ can be further factorized into $M=Kd_1$. 
In \cref{def_M}, all terms are already factorized, except for $d_1^\dagger\cup^\dagger$, and based on \cref{explicit_form_of_operators_and_structures}, we find $d_1^\dagger\cup^\dagger=e^{i(\omega-q_x-q_y)}\cup d_1$, so this term is also factorized. We get
\begin{equation}
    K(q)=i\left[\frac{1}{e^2}d_1(q)^\dagger\eta -\frac{k}{4\pi}\left(1+e^{i(\omega-q_x-q_y)}\right)\operatorname{\cup}(q)\right].
    \label{def_K}
\end{equation}
Note that $K$ is not uniquely defined because we can add $K\mapsto K+Ld_2$ while $Kd_1=M$ remains unchanged. In the below we just fix the choice of $K$ above. This factorization has the advantage that the EoM $Kd_1 A=0$ is now expressed in terms of the flux $dA\equiv d_1 A$ with is gauge invariant. In \cref{diagrammatic_representation_of_EoMs} we present the EoM in real space, from which we can also easily read-off the factorization $MA=Kd_1A$, with $K$ given by the form \cref{def_K}.

Any non-trivial solution to the EoM should satisfy $\operatorname{det} M(q)=0$, which will give the dispersion relation for $q$. However, due to gauge invariance $A\mapsto A+d_0\phi$, for $q\neq0$, among three eigenvalues of $M(q)$ there will always be an zero eigenvalue associated with the eigenvector $\propto d_0(q)$, which corresponds to pure gauge and should be removed from the bulk spectrum. 
\footnote{For $q=0$, $A_x$ is a linear function in $x$, and likewise for $A_y,A_t$, which are unbounded.}
This is equivalent to considering the kernel of $K$ within the image of $d_1$.
The product of the remaining eigenvalues of $M(q)$ for $q\neq 0$ is
\begin{multline}
    \operatorname{det}' M(q)=2(\cos{\omega}+\cos{q_x}+\cos{q_y}-3)\operatorname{det}K(q)|_{\text{image of } d_1}\\
    =\frac{2i}{e^4}(\cos{\omega}+\cos{q_x}+\cos{q_y}-3)G_0^{-1}(q),
    \label{det'M}
\end{multline}
where for later convenience, we have defined
\begin{equation}
    G_0=\frac{i}{(2-2\cos \omega)-c^2(2-2\cos q_x)-c^2(2-2\cos q_y)-m^2 c^4\left(1+\cos(q_x+q_y+\omega)\right)/2},
    \label{def_G_0}
\end{equation}
which is similar to the Green function for a massive scalar field on the lattice. There is a momentum dependent factor $\left(1+\cos(q_x+q_y-\omega)\right)/2$ multiplied to the ``mass term'', in which
\begin{equation}
    mc^2=\frac{ke^2}{2\pi}.
\end{equation}
The bulk spectrum is $\{q|G_0^{-1}(q)=0,q\neq0\}$, which boils down to solving a quadratic equation in $e^{i\omega}$. 

\begin{figure}[htbp]
    \centering
    \begin{subfigure}[t]{0.24\textwidth}
        \centering
        \includegraphics[width=\textwidth]{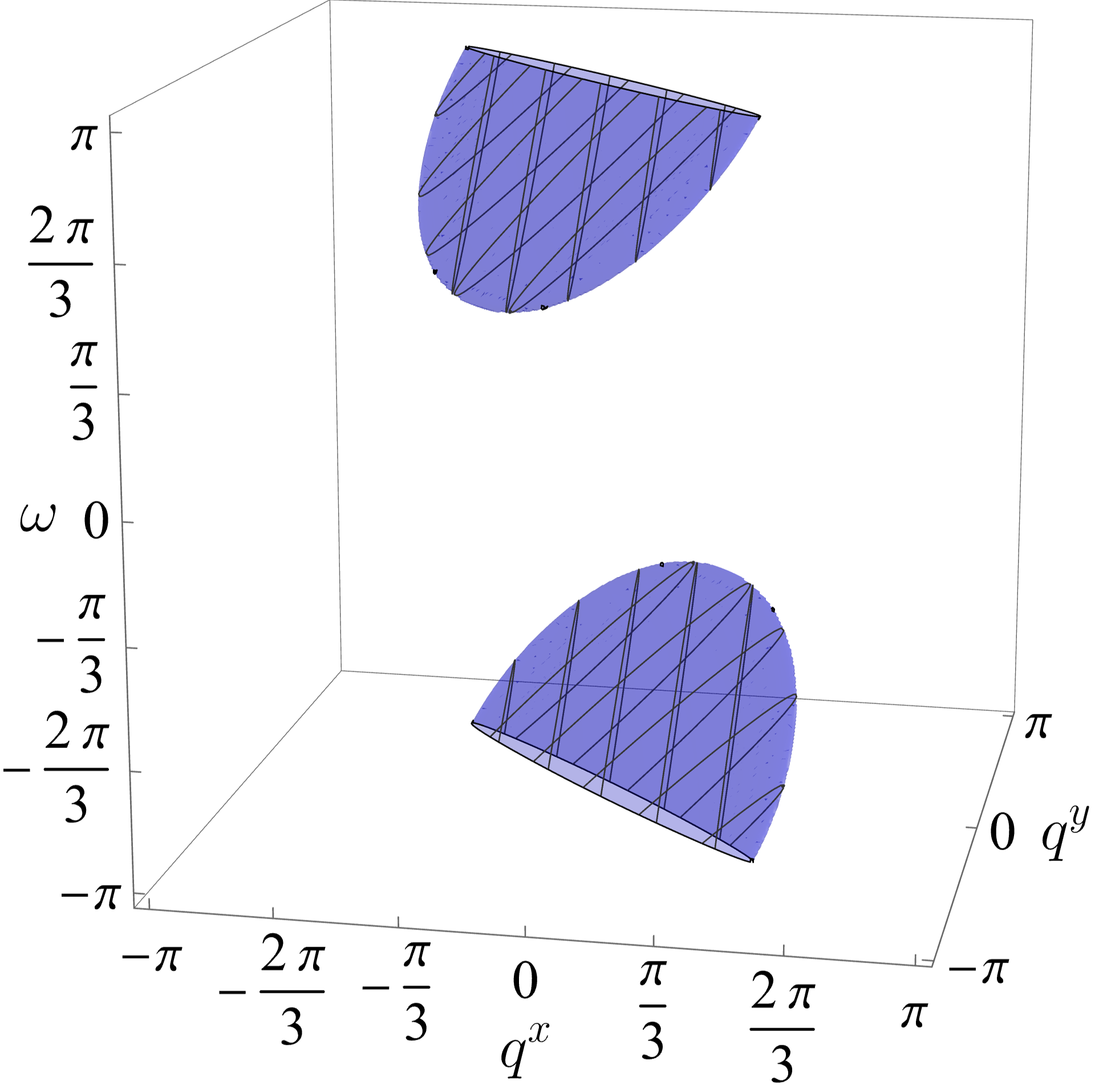}
        \caption{$|m|=20$}
    \end{subfigure}
    \begin{subfigure}[t]{0.24\textwidth}
        \centering
        \includegraphics[width=\textwidth]{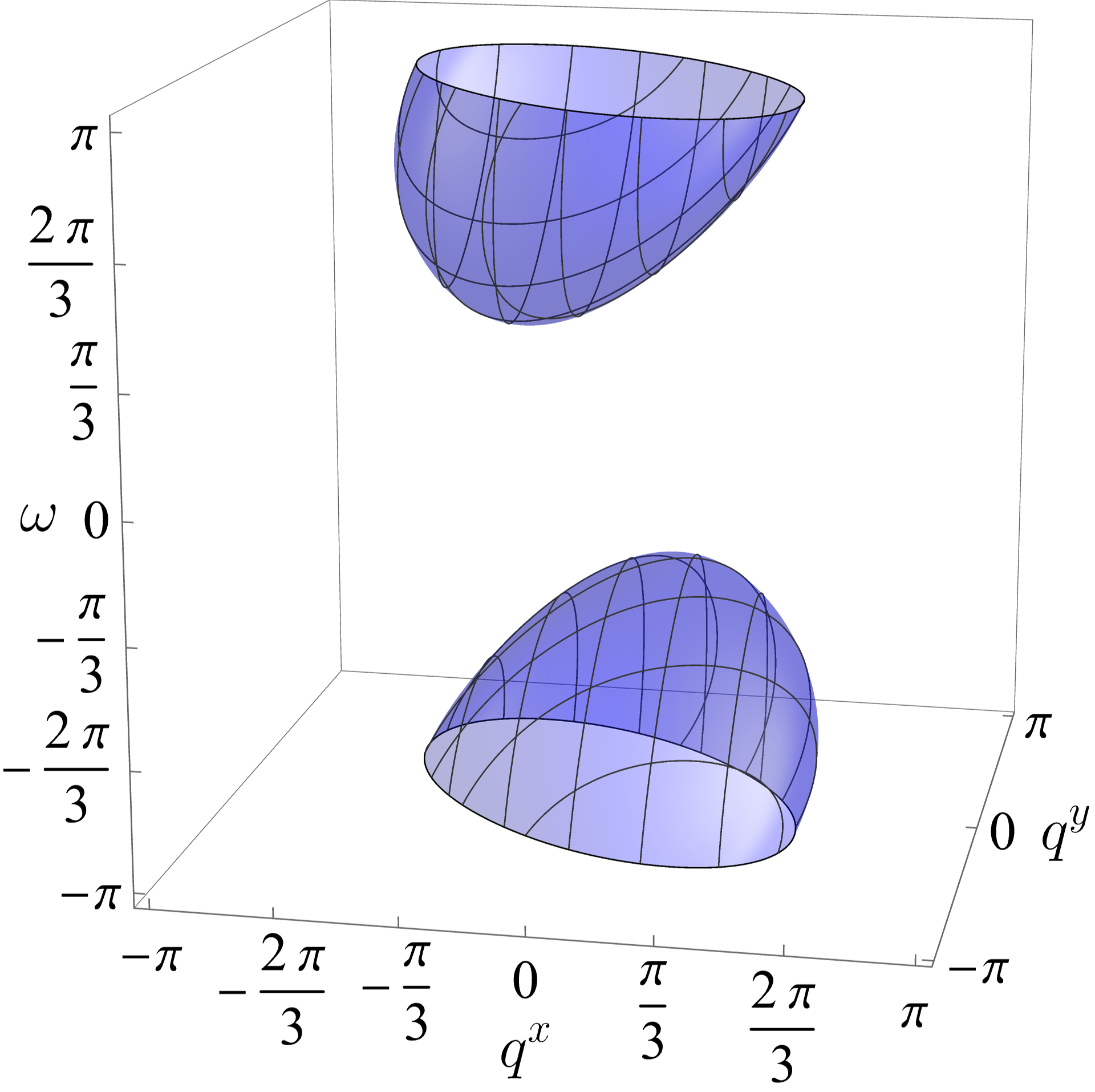}
        \caption{$|m|=2$}
    \end{subfigure}
    \begin{subfigure}[t]{0.24\textwidth}
        \centering
        \includegraphics[width=\textwidth]{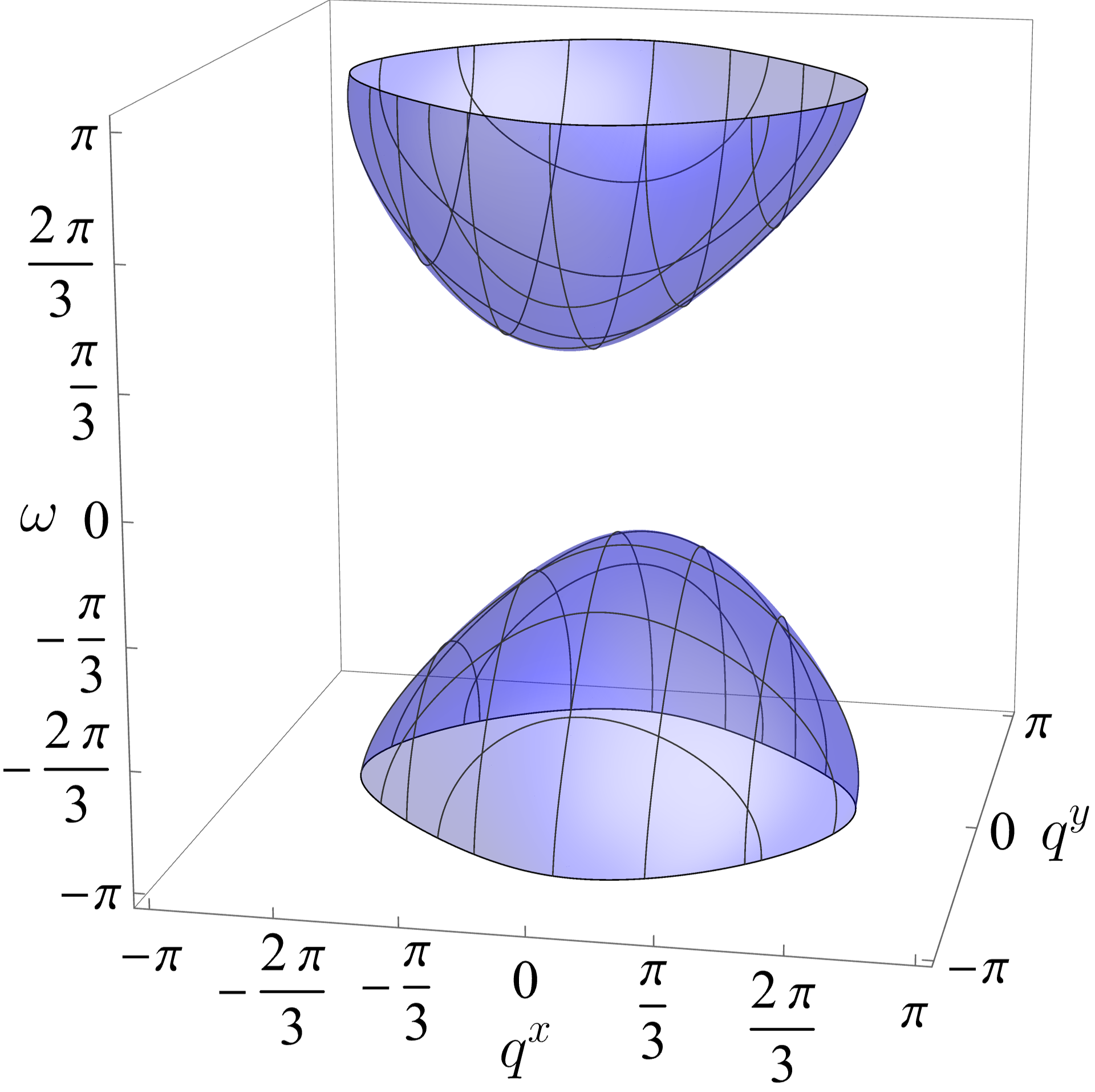}
        \caption{$|m|=1$}
    \end{subfigure}
    \begin{subfigure}[t]{0.24\textwidth}
        \centering
        \includegraphics[width=\textwidth]{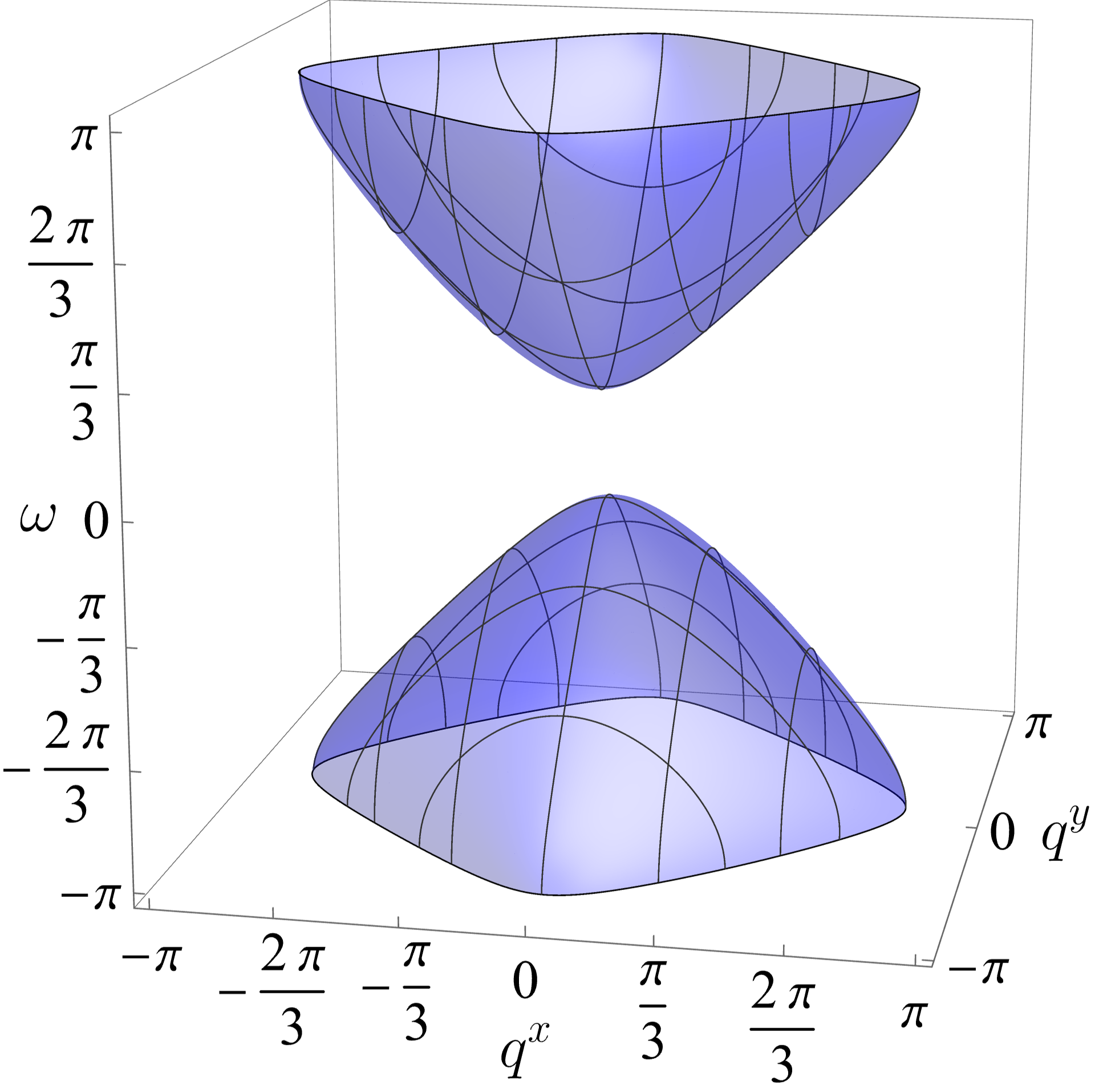}
        \caption{$|m|=0.5$}
    \end{subfigure}
    \caption{Bulk spectrum for $c=1$, tuning $1/e^2$, where $mc^2=ke^2/2\pi$.}
    \label{bulk_spectrum_c_eq_1}
\end{figure}

\begin{figure}[htbp]
    \centering
    \begin{subfigure}[t]{0.24\textwidth}
        \centering
        \includegraphics[width=\textwidth]{plot/5d.png}
        \caption{$c=1$}
    \end{subfigure}
    \begin{subfigure}[t]{0.24\textwidth}
        \centering
        \includegraphics[width=\textwidth]{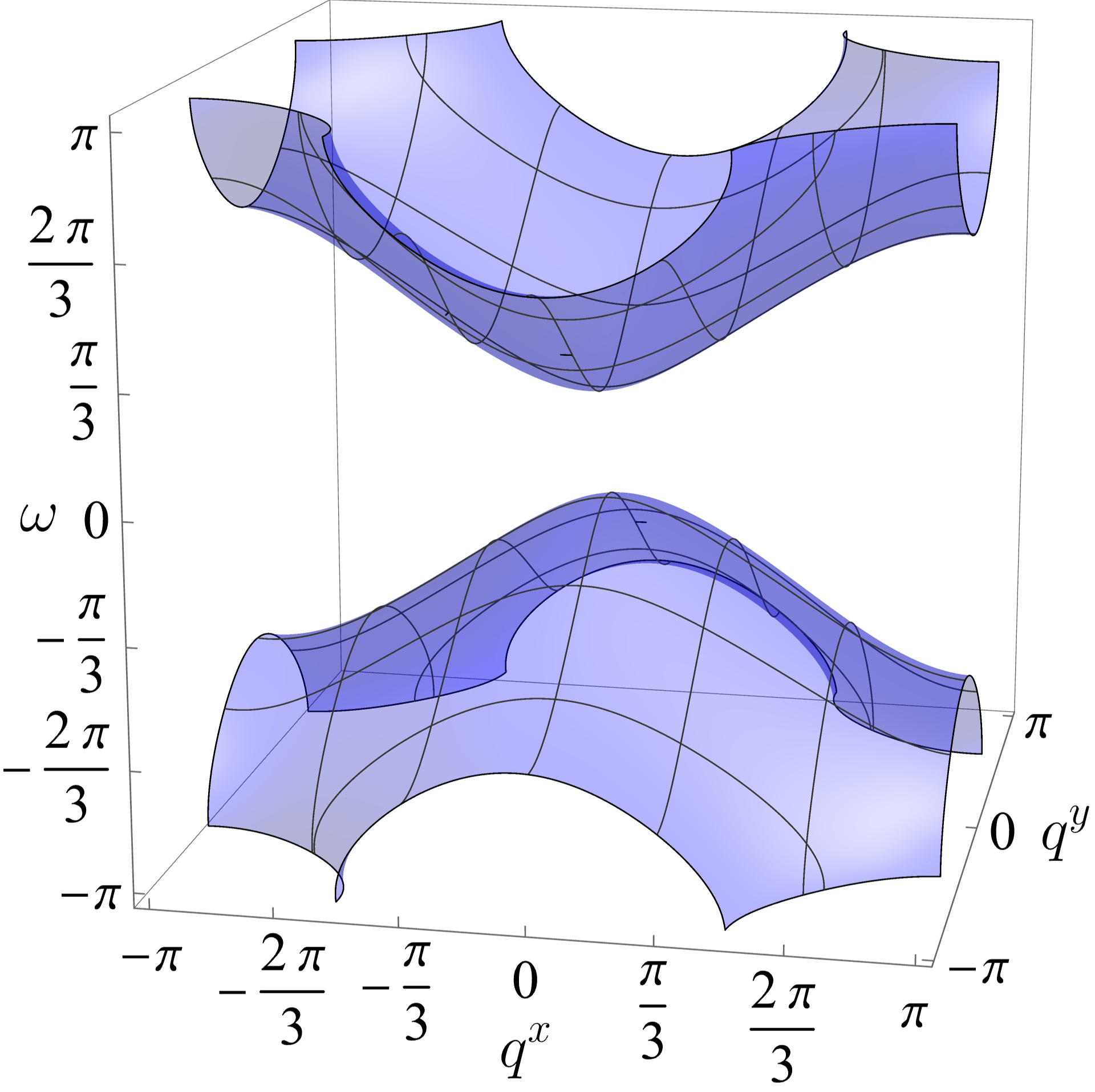}
        \caption{$c=0.8$}
    \end{subfigure}
    \begin{subfigure}[t]{0.24\textwidth}
        \centering
        \includegraphics[width=\textwidth]{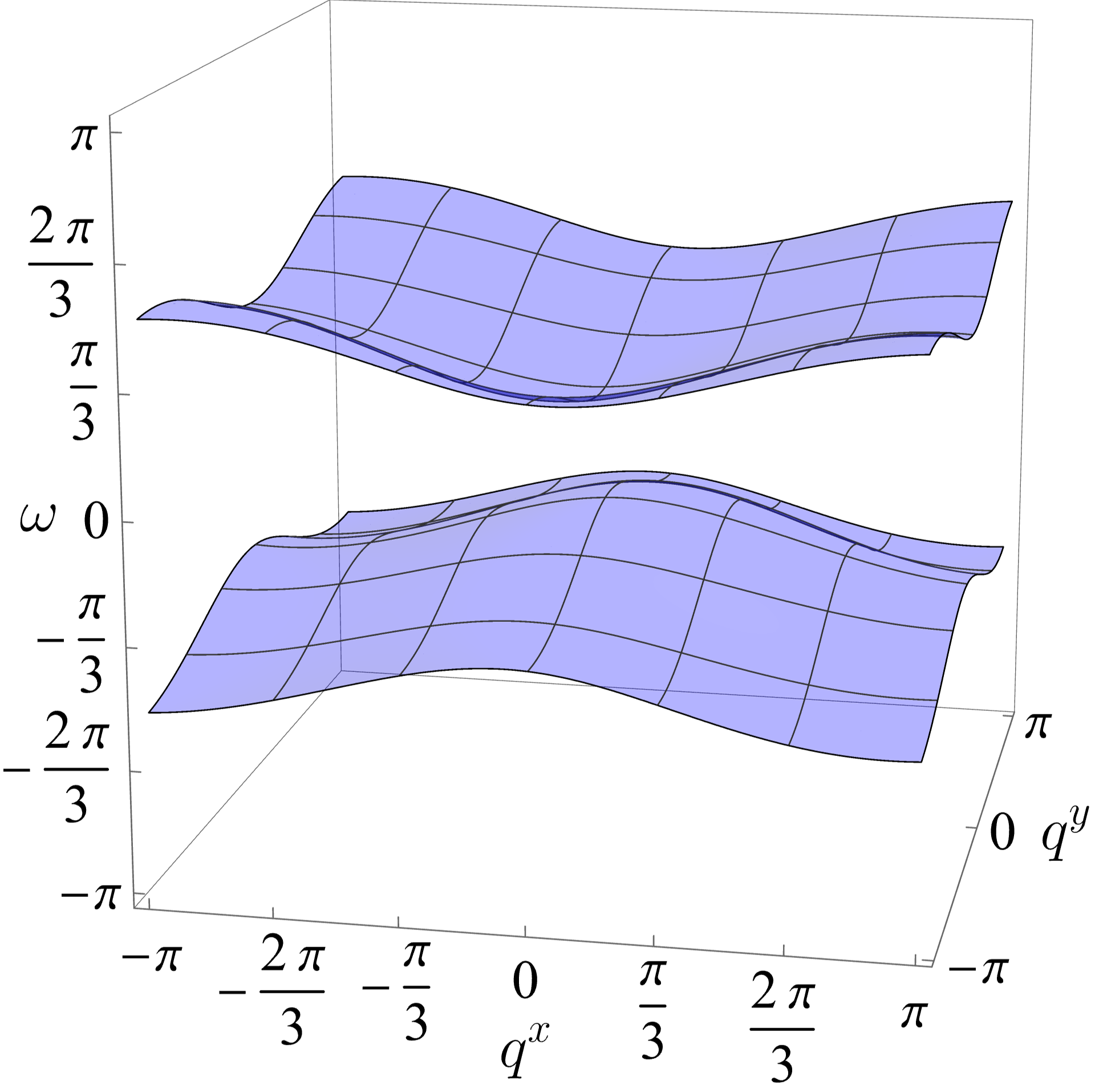}
        \caption{$c=0.5$}
    \end{subfigure}
    \caption{Bulk spectrum for $|m|c^2=|k|e^2/2\pi=0.5$, tuning $c^2$.}
    \label{bulk_spectrum_c_neq_1}
\end{figure}

\begin{figure}[htbp]
    \centering
    \begin{subfigure}[t]{0.24\textwidth}
        \centering
        \includegraphics[width=\linewidth]{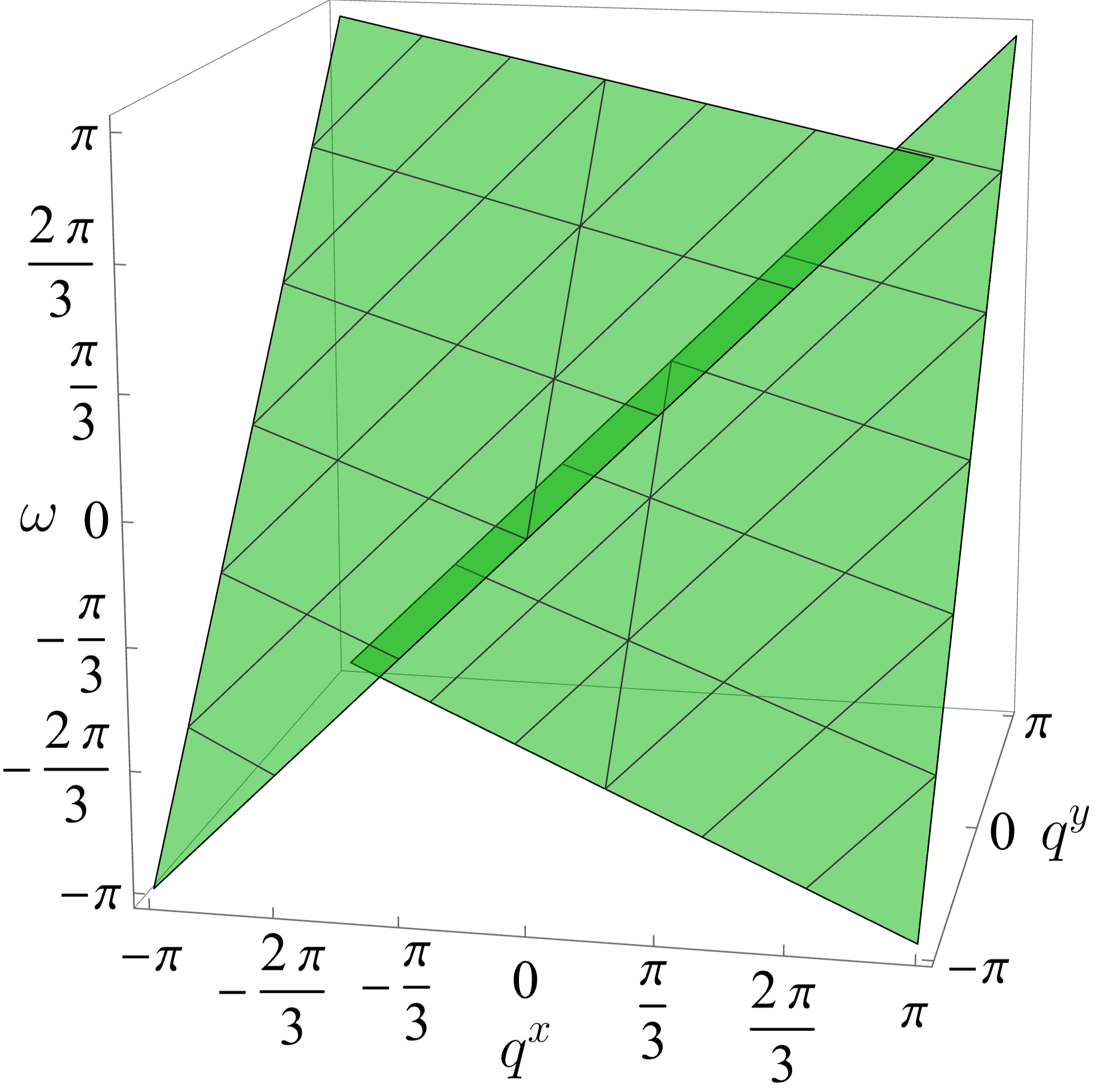}
    \end{subfigure}
    \begin{subfigure}[t]{0.24\textwidth}
        \centering
        \includegraphics[width=\linewidth]{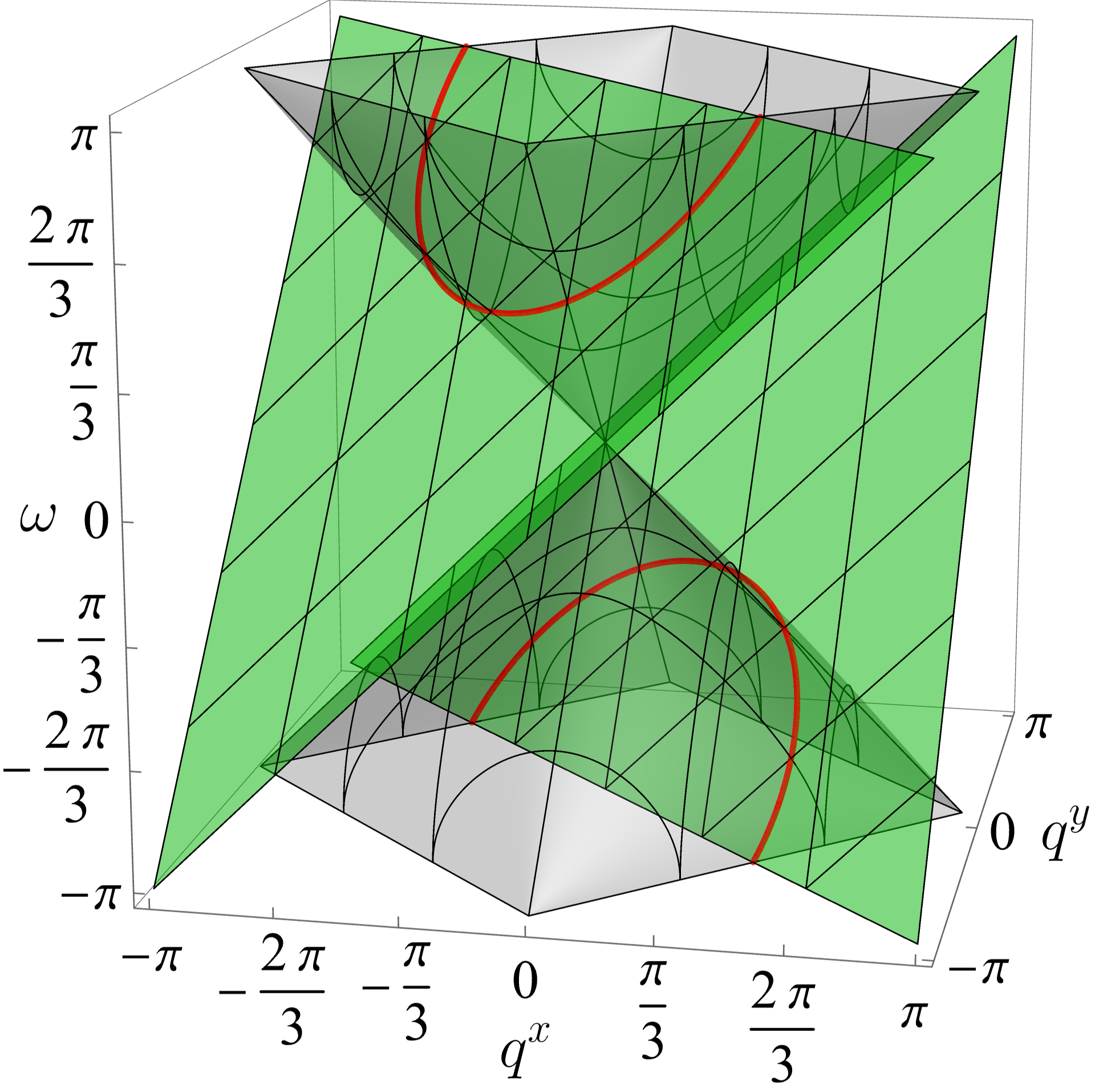}
    \end{subfigure}
    \begin{subfigure}[t]{0.24\textwidth}
        \centering
        \includegraphics[width=\linewidth]{plot/20.png}
    \end{subfigure}

    \caption{Left: Problematic zero mode when $1/e^2=0$. Middle: Conditions for bulk spectrum when $1/e^2\rightarrow 0^+$. Right: Bulk spectrum for $|m|c^2=|k|e^2/2\pi=20$. We have fixed $c=1$.}
    \label{bulk_spectrum_inf}
\end{figure}

In \cref{bulk_spectrum_c_eq_1} we plotted the bulk spectrum for various values of $1/e^2$ and $c^2$. We can note that at fixed $c^2$, as the Maxwell coefficient $1/e^2$ becomes large, the spectrum becomes more like that for a relativistic particle of mass $m$. On the other hand, at fixed $1/e^2$, as $c^2$ becomes small---which is equivalent to $\Delta t/\Delta x$ becoming small---the spectrum becomes more like that of a band theory with spatial lattice and continuous time.

Let us discuss what happens when the Maxwell coefficient $1/e^2$ is small. We know that when the Maxwell term vanishes, the theory has problematic zero modes. More precisely, in this case $\operatorname{det}' M(q)\propto \left(1+\cos(q_x+q_y-\omega)\right)$, so the problematic zero modes occur at 
$q_x+q_y-\omega=\pi\!\mod 2\pi$. This looks like a gapless spectrum, but as we explained it is not a legitimate spectrum, because a legitimate spectrum is only like a ``zero mode'' when we neglect the $i\epsilon$ prescription in the Lorentzian signature, but once the $i\epsilon$ is restored it no longer leads to a divergence of the Lorentzian partition function; by contrast, the problematic zero modes lead to divergence of the partition function regardless of the signature, as we have explained in \cref{necessity_of_maxwell_term}. As long as we have any $1/e^2$, however small, we have legitimate spectrum with order $1$ gap (in the lattice scale). To see this, note that $G_0^{-1}=0$ would always require $(2-2\cos \omega)-c^2(2-2\cos q_x)-c^2(2-2\cos q_y)\geq 0$ (because $m^2c^4(1+\cos(q_x+q_y-\omega))/2$ is always non-negative). When $m\propto e^2\rightarrow\infty$, it further requires $(1+\cos(q_x+q_y-\omega))\rightarrow 0^+$. The overlap of these two conditions dictate the gap to develop at order $1$. See \cref{bulk_spectrum_inf}.

\section{Chiral Edge Spectrum}
\label{chiral_edge_mode}
It is well known that CS theory has non-trivial chiral edge mode. Now let us solve for it in our model.

We should first clarify the meaning of ``edge mode''. Let us compare with a simple wave equation $(\partial_t^2-c^2\partial_x^2-c^2\partial_y^2)\phi=0$ on the manifold $\mathbb{R}^2\times\mathbb{R}_{\geq 0}=\{(t,x,y)|y\geq 0\}$ with boundary $\{(t,x,y)|y=0\}$. Decompose the Dirichlet boundary condition into monochromatic modes of the form $\phi(t,x,0)=\phi(0)e^{-i\omega t+iq_x x}$, then the solution takes the form $\phi(t,x,y)=\phi(0)e^{-i\omega t+iq_x x+iq_y y}$. If $\omega^2-c^2q_x^2<0$, to satisfy the EoM $(-\omega^2+c^2q_x^2+c^2q_y^2)\phi=0$, $q_y$ must take complex value $\pm i\sqrt{q_x^2-c^{-2}\omega^2}$ and we further choose $q_y=i\sqrt{q_x^2-c^{-2}\omega^2}$ to let the solution decay when $y\to\infty$. This decaying solution can be referred to as an edge mode. But this is not the kind of edge mode we have in mind for CS theory, because the $\omega$ and $q_x$ here are fixed by the boundary condition, so it is not a propagating mode with variable $q_x$ and a dispersion $\omega(q_x)$. While the CS theory also has this kind of non-propagating edge mode, it has an additional propagating edge mode. To understand why, note that given a Dirichlet boundary condition, we can add a mode whose solution satisfies $\phi(t,x,0)=0$ (the homogeneous Dirichlet boundary condition), and the original Dirichlet boundary condition is still satisfied. For the simple wave equation, the only solutions satisfying $\phi(t,x,0)=0$ are the sine waves $\propto \sin(q_y y)$ obtained from bulk modes. But for CS theory, there are extra solutions to the homogeneous Dirichlet boundary condition that exponentially decay into the bulk, and this is what we are looking for. Therefore, it suffices to look at the homogeneous Dirichlet boundary condition $F_{p\in\text{boundary}}=0$.

Consider a cubic lattice with vertices at $t, x, y\in\mathbb{Z}$ but $y\geq 0$ only. Just like the treatment in the bulk spectrum case, we first turn the theory into a manifestly free theory form. Since the topologies of both the bulk and the boundary are trivial, for any $s_p$ satisfying $ds_c=0$ in the bulk, we can again write $s_p=dn_p$ for $n_l\in\mathbb{Z}$, and then absorb $n_l$ into $A_l$, so that $A_l-2\pi n_l\in\mathbb{R}$ is now redefined as $A_l\in\mathbb{R}$, and $F_p=dA_p$ after the redefinition. Note that this can be done even for the specified $s_{p\in\text{boundary}}$ and $A_{l\in\text{boundary}}$, because although a 1-form $\mathbb{Z}$ ``boundary gauge transformation'' $n_{l\in\text{boundary}}$ changes the Dirichlet boundary condition, as we explained in \cref{guage_transformation_on_boundary}, under such change of Dirichlet boundary condition the partition function only changes by an overall constant phase that depends on the boundary condition only, while the dynamics, in particular the chiral edge mode, is unaffected. Therefore, now it suffices to specify the homogeneous Dirichlet boundary condition as $dA_{p\in\text{boundary}}=0 \in \mathbb{R}$, with $A_l\in\mathbb{R}$ in \cref{action}.

Since $A_{l\in\text{boundary}}=A_{t,x}(t,x,y=0)$ is fixed rather than dynamical, $\delta S/\delta A_{t,x}(t,x,y=0)$ is not required to be zero under EoM. So if we only Fourier transform $t,x$ but keep $y$ in the real coordinates, we have
\begin{equation}
    \left\{
    \begin{aligned}
        (iKd_1A)^{t,x}(\omega,q_x,y)&=0\\
        (iKd_1A)^{y}(\omega,q_x,y-1)&=0
    \end{aligned}\right.\quad \forall y>0.
    \label{eom_on_lattice_with_edge}
\end{equation}
These are the precise EoM to be solved, in the presence of boundary.

It is a customary practice in boundary value problems solutions to assume $d_1 A$ can be decomposed into the ansatz
\begin{equation}
    (d_1A)(\omega,q_x,y)=\sum_{q_y}e^{iq_y y}(d_1A)(\omega,q_x, q_y)
    \label{d_1A}
\end{equation}
with $|e^{iq_y}|\leq 1$. That is, we will solve the same EoM in \cref{bulk_spectrum},
\begin{equation}
    iK(q)(d_1A)(q)=0,
    \label{iKd_1A}
\end{equation}
except now we also include those $e^{iq_y}$ whose magnitude is smaller than $1$, and moreover we need to impose the boundary condition later. In \cref{proof_of_ansatz} we will carefully address that this customary practice is indeed valid in our present problem.

As in \cref{bulk_spectrum}, the requirement that $K(q)$ has a kernel boils down to
\begin{equation}
    G_0^{-1}(q)=0.
    \label{solving_q}
\end{equation}
According to \cref{def_G_0}, this is a quadratic equation in $z_y=e^{iq_y}$, which we can explicitly solve. Using Vieta's formula, the two solutions $(z_y)_1, (z_y)_2$ satisfy $|(z_y)_1 (z_y)_2|=1$, so either $|(z_y)_1|=|(z_y)_2|=1$, which happens when $\omega, q_x$ admit those bulk modes found in \cref{bulk_spectrum}, or $|(z_y)_1|<1$ and $|(z_y)_2|>1$, which happens when $\omega, q_x$ do no admit those bulk modes, and the $|(z_y)_2|>1$ mode is to be discarded, while the $|(z_y)_1|<1$ mode is the desired edge mode.

Now we shall make use the Dirichlet boundary condition. As we explained at the beginning of this section, it suffices to consider the homogeneous Dirichlet boundary condition, $(dA)_{p\in\text{boundary}}=0$, which hosts the propagating edge modes. Since we are left with the $|e^{iq_y}|<1$ mode, \cref{d_1A} has only one $q_y$ involved in the summation, i.e. 
\begin{equation}
    (d_1A)(\omega,q_x,y)=e^{iq_y y}(d_1A)(\omega,q_x, q_y),
\end{equation}
so the homogeneous Dirichlet boundary condition simply says the $tx$-component vanishes, $(d_1A)_{tx}(\omega,q_x, q_y)=(d_1A)_{tx}(\omega,q_x, y=0)=0$. 

Summarizing the above, the goal is to look for those $\omega, q_x$ in \cref{iKd_1A} that satisfy two conditions:
\begin{itemize}
\item $K(q)$ has a kernel for some $|e^{iq_y}|<1$, and this is given by solving $G_0^{-1}(q)=0$, a quardratic equation in $z=e^{iq_y}$.
\item The zero mode that spans the kernel satisfies $(d_1A)(q)$ such that $(d_1 A)_{tx}(q)=(d_1 A)^y(q)=0$.
\end{itemize}
After some calculations, we find the conditions become
\begin{equation}
    \left\{
    \begin{aligned}
        \sin{\frac{\omega}{2}}&=\pm c\sin{\frac{q_x}{2}}\\
        \tan{\frac{q_y}{2}}&=\frac{\cos\left(\frac{1}{2}(q_x-\omega)\right)}{\pm i \frac{2}{mc}+\sin\left(\frac{1}{2}(q_x-\omega)\right)}\\
        \operatorname{Im} q_y&\geq0
    \end{aligned}\right.\ ,
\end{equation}
where we have $\omega, q_x\in(-\pi,\pi]$, and the ``$\pm$'' in the first and second line should simultaneously take either ``$+$'' or ``$-$''. The solution for $d_1A(\omega, q_x,q_y)_{xy,yt}$ is
\begin{equation}
    (d_1A)_{yt}(\omega, q_x,q_y)=\pm c (d_1A)_{xy}(\omega, q_x,q_y).
\end{equation}
Note that the sign of the CS level $k$ determines the sign of $m$ in the conditions above.

\begin{figure}[htbp]
    \centering
    \begin{subfigure}[t]{0.24\textwidth}
        \centering
        \includegraphics[width=\textwidth]{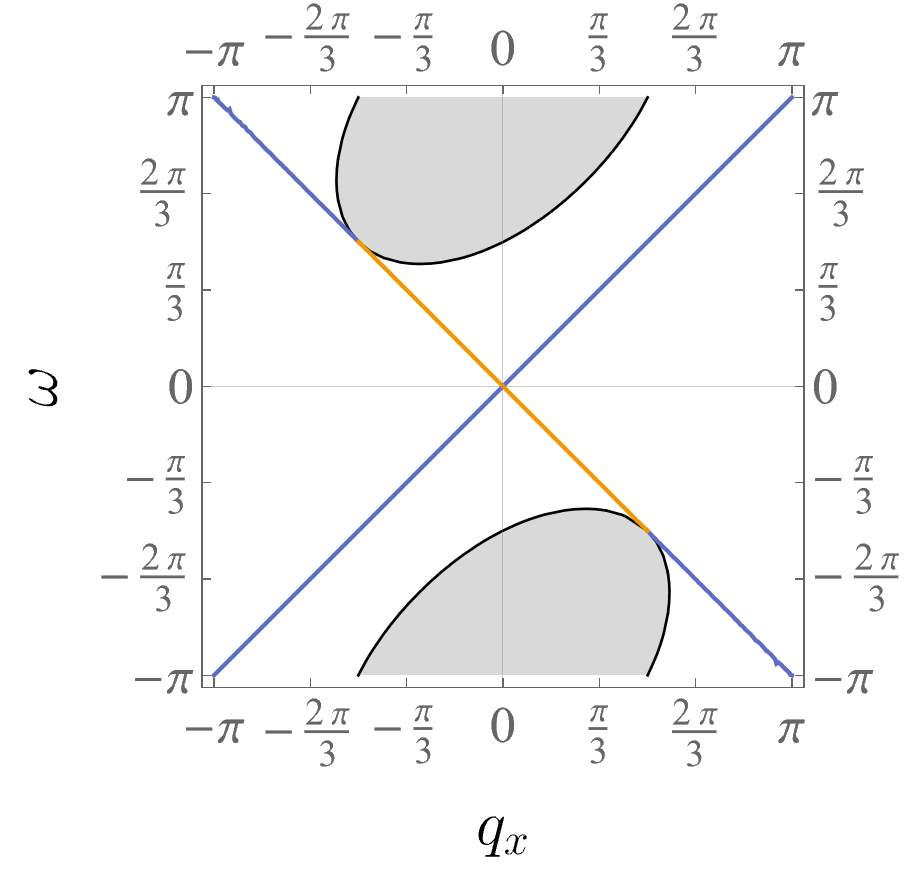}
        \caption{$|m|=20$}
    \end{subfigure}
    \begin{subfigure}[t]{0.24\textwidth}
        \centering
        \includegraphics[width=\textwidth]{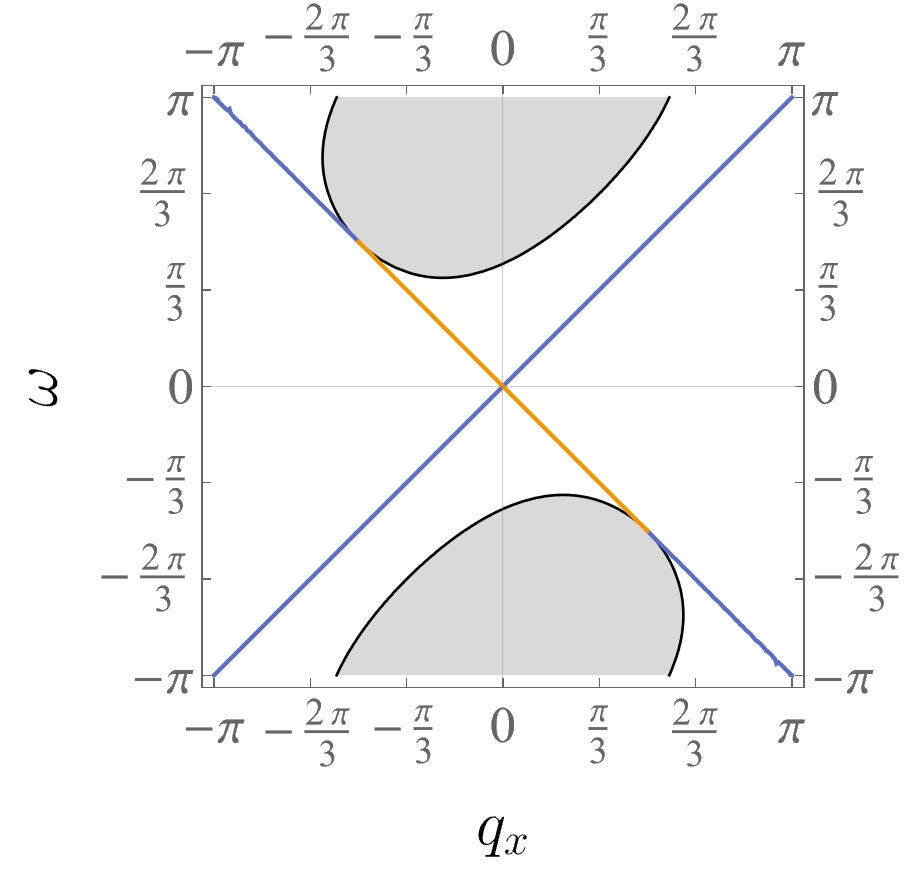}
        \caption{$|m|=2$}
    \end{subfigure}
    \begin{subfigure}[t]{0.24\textwidth}
        \centering
        \includegraphics[width=\textwidth]{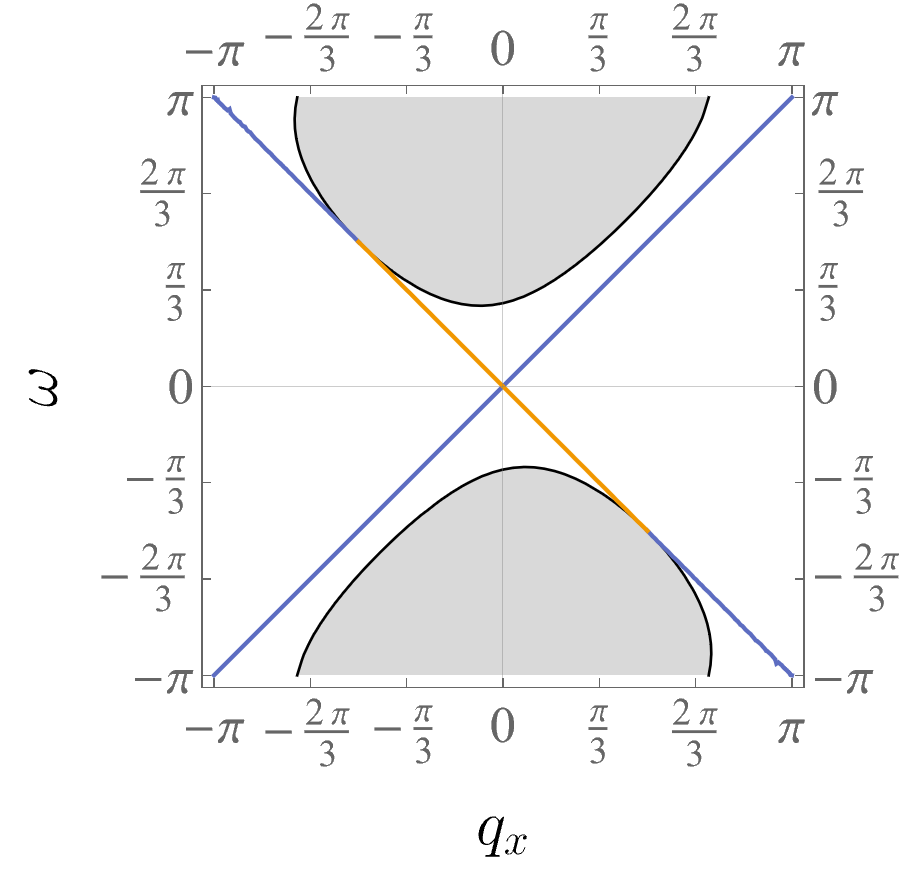}
        \caption{$|m|=1$}
    \end{subfigure}
    \begin{subfigure}[t]{0.24\textwidth}
        \centering
        \includegraphics[width=\textwidth]{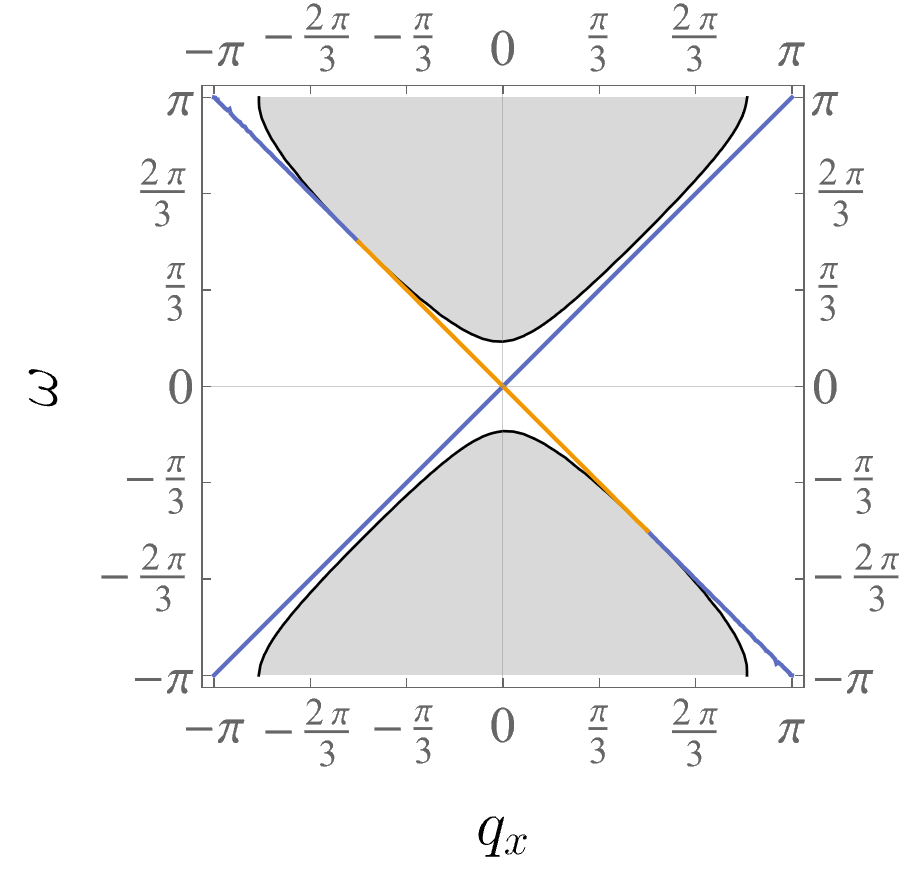}
        \caption{$|m|=0.5$}
    \end{subfigure}
    \caption{Grey is the bulk spectrum, yellow is the chiral edge spectrum if $k>0$, and blue is the chiral edge spectrum if $k<0$ (when bulk is at $y\geq 0$). Here we fix $c=1$, and tune $1/e^2$, with $mc^2=ke^2/2\pi$. (Note it is important that the bulk spectrum is never perfectly symmetric due to the cup product structure, so that the edge spectrum with $\omega=\pm q_x$ can possibly merge into the bulk spectrum.)}
    \label{edge_bulk_spectrum_c_eq_1}
\end{figure}

\begin{figure}[htbp]
    \centering
    \begin{subfigure}[t]{0.24\textwidth}
        \centering
        \includegraphics[width=\textwidth]{plot/edgebulk5d.png}
        \caption{$c=1$}
    \end{subfigure}
    \begin{subfigure}[t]{0.24\textwidth}
        \centering
        \includegraphics[width=\textwidth]{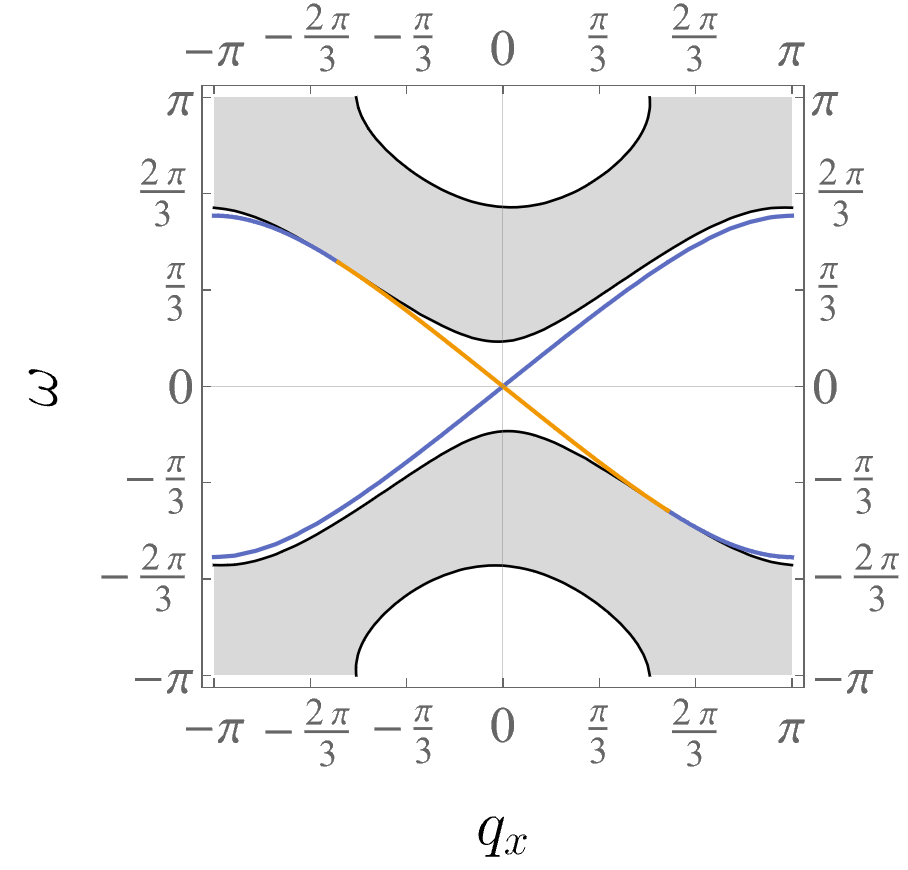}
        \caption{$c=0.8$}
    \end{subfigure}
    \begin{subfigure}[t]{0.24\textwidth}
        \centering
        \includegraphics[width=\textwidth]{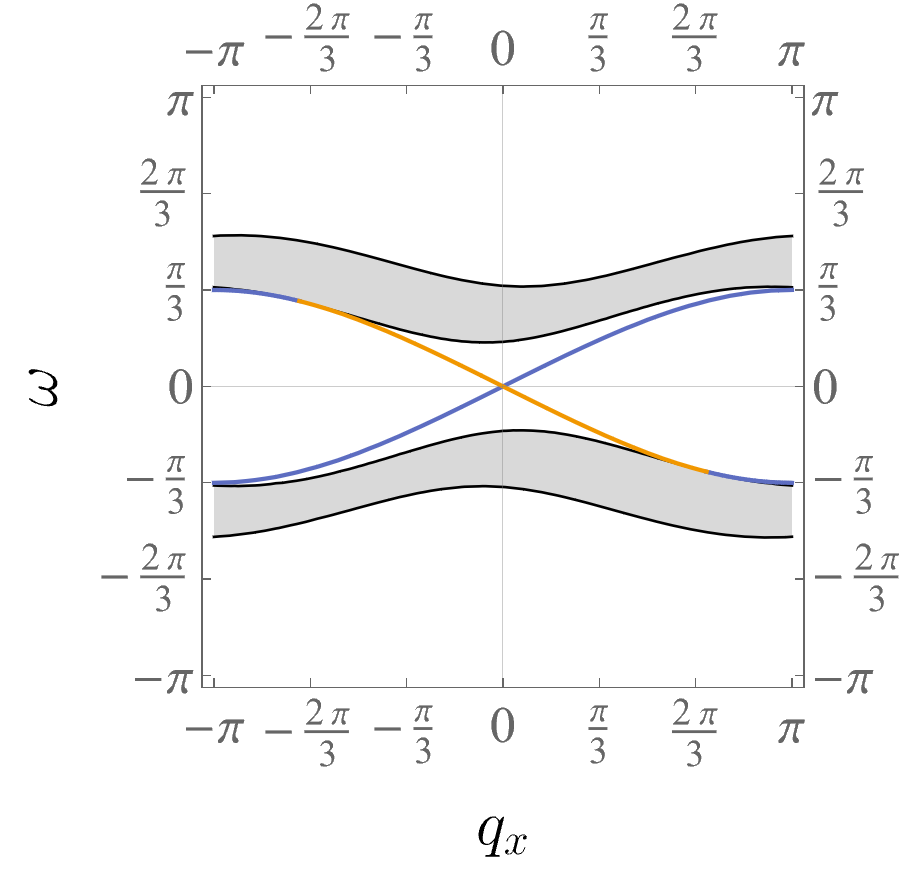}
        \caption{$c=0.5$}
    \end{subfigure}
    \caption{Fix $|m|c^2=|k|e^2/2\pi=0.5$, and tune $c^2$.}
    \label{edge_bulk_spectrum_c_neq_1}
\end{figure}

We plot the bulk spectrum and the chiral edge spectrum together in \cref{edge_bulk_spectrum_c_eq_1} and \cref{edge_bulk_spectrum_c_neq_1}. We can see the edge modes are indeed chiral as desired, with the chirality depending on the sign of $k$. The velocity is approximately $\pm c$ for small momentum $q_x$. (It turns out the chiral edge spectrum always merge into the bulk modes at the intersection of the bulk spectrum and $q_x-\omega=\pm\pi$.)

Interestingly, when $c=1, mc=-2$, i.e. the blue dispersion line in \cref{edge_bulk_spectrum_c_eq_1}(b), for the $\omega=q_x$ branch we have $\tan{q_y/2}=i$, which means the chiral edge mode $(d_1A)(\omega,q_x,y)=(d_1A)(\omega,q_x,q_y)\delta_{y,0}$ is exactly localized at the boundary at $y=0$. We are currently not aware of any special interpretation of this. Note however the chiral edge mode is not entirely localized at the boundary, because there are also the small segments with $\omega=-q_x$ that are not localized.

Later we will see the gravitation anomaly indeed arises from the chiral edge modes. 

\section{Ground State Degeneracy}
\label{ground_state_degeneracy}
The ground state degeneracy is an important characterization of topological order. In particular, for a CS theory on a 2d oriented spatial manifold of genus $g$, the ground state degeneracy should be $|k|^g$. Since we are working with cubic lattice in this paper, we will consider a spatial torus and reproduce the $|k|$ ground state degeneracy.

The ground state degeneracy $D$ on a spatial manifold $\Sigma$ can be extracted from the partition function for a Euclidean spacetime $S^1\times\Sigma$. In this case, the partition function is the trace of the thermal density matrix, $Z=\operatorname{tr}(e^{-\beta H})$, with $\beta$ the length of the $S^1$ Euclidean time direction. If the ground state energy is $0$, this gives the ground state degeneracy $D$ in the $\beta\to\infty$ limit. More generally, if the ground state energy is non-zero, but extensive in the spatial volume $V$ as $V\to\infty$, then as $\beta\rightarrow \infty$, the partition function will approach $D e^{-\varepsilon_0 \mathcal{V}}$, where $\varepsilon_0$ is the ground state energy density, and $\mathcal{V}=\beta V$ is the spacetime volume. The ground state free energy $F = -\ln Z/\beta = \varepsilon_0 V - \ln D/\beta$, so $\ln D$ is indeed the ground state entropy.
\footnote{There is a subtlety in the discussion of this paragraph. The ``degenerate ground states'' usually have tiny energy splits exponentially small in (some positive power of) the spatial linear size $L\sim V^{1/d}$. So when we take $\beta\rightarrow \infty$, we need to simultaneously take $L\rightarrow \infty$ not too slowly, in order to ensure $\beta$ times the splits to still be vanishing.

This subtlety does not appear in our CS-Maxwell theory of interest in this paper, because the exact 1-form symmetry and its self-anomaly introduced in \cref{global_symm_anom} ensure the ground state degeneracy to be exact. This follows from the projective representation of the anomalous symmetry in the Hamiltonian formalism. Here we will not elaborate on this since in this paper we mainly work with the Lagrangian formalism.

Furthermore, while our CS-Maxwell theory of interest here is gapped, if a theory of interest is ``gapless'', it means the ``gapless excitation energy'' is polynomially small in $L$, so we need to take $L\rightarrow \infty$ not too fast, in order to ensure $\beta$ times the ``gapless excitation energy'' still diverges, hence extracting the ground states only.}

In the present case, we calculate the partition function on a cubic lattice $\mathbb{Z}_{\beta}\times\mathbb{Z}_{L_x}\times\mathbb{Z}_{L_y}$ for the three-torus spacetime manifold $\mathbb{T}^3$, and we expect the result to take the form $Z=|k|e^{-\varepsilon_0 \mathcal{V}}$ as $\mathcal{V}=\beta L_x L_y\rightarrow \infty$, where $\varepsilon_0$ approaches a constant with corrections vanishing faster than $1/\mathcal{V}$.

In the previous two sections, since the topologies of the spacetime were trivial, and moreover we were only focusing on the EoM, it was easy to turn $A_l$ into real-valued in order to manifest the free theory nature of our theory. Now, the spacetime topology is non-trivial, and moreover we need to carefully compute the entire partition function. We still want to transform the path integral into a form that appears as a free theory, but we must ensure the path integral measure remains unchanged---in particular we have to keep track of the topological properties as well as any Jacobians arising from the transformations.

For a general Villainized $\text{U}(1)$ gauge theory in arbitrary dimensions, the path integral is of the form
\begin{equation}
    Z=\left[\prod_l \int_{-\pi}^\pi \frac{dA_l}{2\pi}\right] \left[\prod_p\sum_{s_p\in\mathbb{Z}}\right]e^{-S[A_l, s_p]} \ .
\end{equation}
The path integral weight is invariant if we replace $A_l$ and $s_p$ by
\begin{equation}
    \left\{
    \begin{aligned}
        A'_{l}&=A_{l}+2\pi n_{l}+d\phi_{l}\\[.1cm]
        s'_{p}&=s_{p}+d n_{p}
    \end{aligned}\right. \ ,
\end{equation}
for any $\phi_v\in(-\pi,\pi], n_l\in\mathbb{Z}$ as claimed in \cref{lattice_action}. Therefore we can rewrite the partition function as the average over all these transformations
\begin{equation}
    Z=\frac{\left[\prod_v \int_{-\pi}^\pi \frac{d\phi_v}{2\pi}\right]\left[\prod_l \sum_{n_l\in\mathbb{Z}}\right]\left[\prod_l \int_{-\pi}^\pi \frac{dA_l}{2\pi}\right]\left[\prod_p\sum_{s_p\in\mathbb{Z}} \right]}{\left[\prod_v \int_{-\pi}^\pi \frac{d\phi_v}{2\pi}\right]\left[\prod_l \sum_{n_l\in\mathbb{Z}}\right]}
    e^{-S[A_l',s_p']}.
\end{equation}
\footnote{Here we involved a cancellation of infinite factors $\sum_{n_l}$ between the numerator and denominator. To make sense of this, consider $\int dx \sum_n e^{i2\pi nx} f(x) = f(x=0)$ when $f(x)$ is some function supported between $[-1, 1]$, which means $\sum_n e^{i2\pi nx}=\delta(x)$ for $-1\leq x \leq 1$. When we combine $n$ and $A$ into $A'$, we have $\int_{-\infty}^\infty (dA'/2\pi) e^{iAx}=\delta(x)$.}
In the numerator, the sum over $n_l\in\mathbb{Z}$, the integral over $\phi_v\in(-\pi,\pi]$ (which manifestly yields $1$), and the integral over $A_l\in(-\pi,\pi]$ can be combined into the integral of $A_l'\in\mathbb{R}$ with trivial Jacobian. Moreover, $\sum_{s_p\in\mathbb{Z}}=\sum_{s'_p\in\mathbb{Z}}$. So the measure becomes
\begin{equation}
    \frac{\left[\prod_{l} \int_{-\infty}^{\infty} \frac{dA_l'}{2\pi}\right] \left[\prod_p\sum_{s'_p\in\mathbb{Z}} \right]}{\left[\prod_{v} \int_{-\pi}^\pi \frac{d\phi_v}{2\pi}\right]\left[\prod_{l} \sum_{n_l\in\mathbb{Z}}\right]} .
\end{equation}
By a similar treatment as that turning $A$ and $s$ into $A'$ and $s'$, we can consider $\phi'_v=\phi_v+2\pi\kappa_v\in \mathbb{R}$ for $\kappa_v\in\mathbb{Z}$, and $n'_l=n_l-d\kappa_l$, which gives the same gauge transformation. The measure becomes
\begin{equation}
     \frac{\left[\prod_{l} \int_{-\infty}^{\infty} \frac{dA_l'}{2\pi}\right]}{\left[\prod_{v} \int_{-\infty}^{\infty}  \frac{d\phi_v'}{2\pi}\right]}\frac{ \left[\prod_p\sum_{s'_p\in\mathbb{Z}}\right]\left[\prod_v\sum_{\kappa_v\in\mathbb{Z}} \right]}{\left[\prod_{l} \sum_{n'_l\in\mathbb{Z}}\right]}.
\end{equation}
We can readily note here that the first factor is the Faddeev-Popov measure for a real-valued (rather than $\text{U(1)}$-valued) gauge field.
\footnote{On a generic spacetime, the Faddeev-Popov measure for a real gauge field may diverge, because the flat holonomy in $A'$ takes real rather than $\text{U(1)}$ values; or it might vanish, because the constant (in each connected component) $\phi'$ that does not transform $A$ also takes real rather than $\text{U(1)}$ values. But as we will see below, these diverging and vanishing factors will be cancelled out (in the sense of the previous footnote) by the second factor of the integer summations, so to restore the finiteness of the Villainized $\text{U(1)}$ path integral that we started with.}

To proceed, we can note that the 1-form $\mathbb{Z}$ gauge transformation part of the $s'$ summation should be cancelled by part of the $n'$ summation, while the 0-form $\mathbb{Z}$ gauge transformation part of the $n'$ summation should be cancelled by part of the $\kappa$ summation. More exactly,
\begin{equation}    
\frac{\left[\prod_p\sum_{s'_p\in\mathbb{Z}} \right]\left[\prod_v\sum_{\kappa_v\in\mathbb{Z}} \right]}{\left[\prod_{l} \sum_{n'_l}\in\mathbb{Z}\right]}=\frac{\sum_{ds'}\sum_{[s']}\sum_{dn'}\phantom{\sum_{[n']}}\sum_{d\kappa}\sum_{[\kappa]}}{\phantom{\sum_{ds'}\sum_{[s']}} \sum_{dn'}\sum_{[n']}\sum_{d\kappa}\phantom{\sum_{[\kappa]}}}
\label{summation_of_discrete_dofs}
\end{equation}
where we used the abbreviation 
\begin{equation}
    \sum_{ds'}=\left[\left(\prod_c \sum_{m_c\in\mathbb{Z}}\right)\:\text{s.t. $\exists s'$ where $m=ds'$}\right]=\frac{\left[\left(\prod_c\sum_{m_c\in\mathbb{Z}}\right)\text{s.t. $dm=0$}\right]}{\sum_{[m]}}    
\end{equation}
and likewise for $\sum_{dn'}$ and $\sum_{d\kappa}$. Here $[m]$ is a class in the cohomology $H^3(M; \mathbb{Z})$, $[s']$ a class in $H^2(M; \mathbb{Z})$, $[n']$ a class in $H^1(M; \mathbb{Z})$, and $[\kappa]$ in $H^0(M; \mathbb{Z})$. For a three-torus, $H^3(M; \mathbb{Z})\cong\mathbb{Z}\ni[m]$ means the spacetime has only one connected component, and having such a factor in the denominator means the total monopole charge in each connected component is constrained (to zero); $H^2(M; \mathbb{Z})\cong\mathbb{Z}^3\ni[s']$ classifies the non-contractible loops of Dirac strings on the dual lattice; $H^1(M; \mathbb{Z})\cong\mathbb{Z}^3 \ni[n']$ classifies non-contractible surfaces on the dual lattice, and having such a factor in the denominator means removing the volume of large gauge transformations; $H^0(M; \mathbb{Z})\cong\mathbb{Z}\ni[\kappa]$ again means the spacetime has only one connected component, and such a factor in the numerator can be thought of as ``large global transformation''.

We emphasize that in \cref{summation_of_discrete_dofs} there is no Jacobian when we cancel factors in the denominator and the numerator, because the $d$'s are acting on integer-valued fields. This might seem puzzling, because if these were real-valued integrals, then there should be Jacobians of the $d$'s acting on real-valued fields. In fact there is no contradiction, because the overall Jacobian actually corresponds to the torsion parts in the cohomology---which for three-torus will turn out to simply give a factor of $1$, see \cref{torsion_in_homology_cohomology_theory}.

Applying the above to our Euclidean CS-Maxwell theory \cref{euclidean_partition_function}, we can see the monopole-forbidding $\lambda_c$ integral just removes the $\sum_{ds'}$ in \cref{summation_of_discrete_dofs}, so the partition function becomes
\begin{equation}
    \begin{split}
     Z&=\frac{\left[\prod_{l} \int_{-\infty}^{\infty} \frac{dA_l'}{2\pi}\right]}{\left[\prod_{v} \int_{-\infty}^{\infty}  \frac{d\phi_v'}{2\pi}\right]}\frac{\sum_{[\kappa]}}{\sum_{[n']}} \ \sum_{[s']} \ (z_\chi[[s']^{\text{rep}}])^k \\
     &\exp \left\{-\frac{1}{2e^2}\sum_p F_p^2+\frac{ik}{4\pi}\sum_c \left[(A'\cup dA')_c - (A'\cup 2\pi [s']^{\text{rep}})_c - (2\pi [s']^{\text{rep}}\cup A')_c\right]\right\}.
    \end{split}
    \label{partition_function_gsd}
\end{equation}
where $[s']^{\text{rep}}_p$ is a representative element of the class $[s']\in H^2(M; \mathbb{Z})\cong \mathbb{Z}^3$, i.e. a representative non-contractible Dirac string running on the dual lattice. Everything discussed so far applies to arbitrary lattice for spacetime manifolds with arbitrary oriented topology. This is the general procedure to make the Villainized $\text{U(1)}$ CS-Maxwell theory appear manifestly free. (This procedure requires knowledge of the global topology of the spacetime, and therefore the result is not expressed in terms of local fields, but involve topological classes.)

\begin{figure}[htbp]
    \centering
    \includegraphics[width=0.5\linewidth]{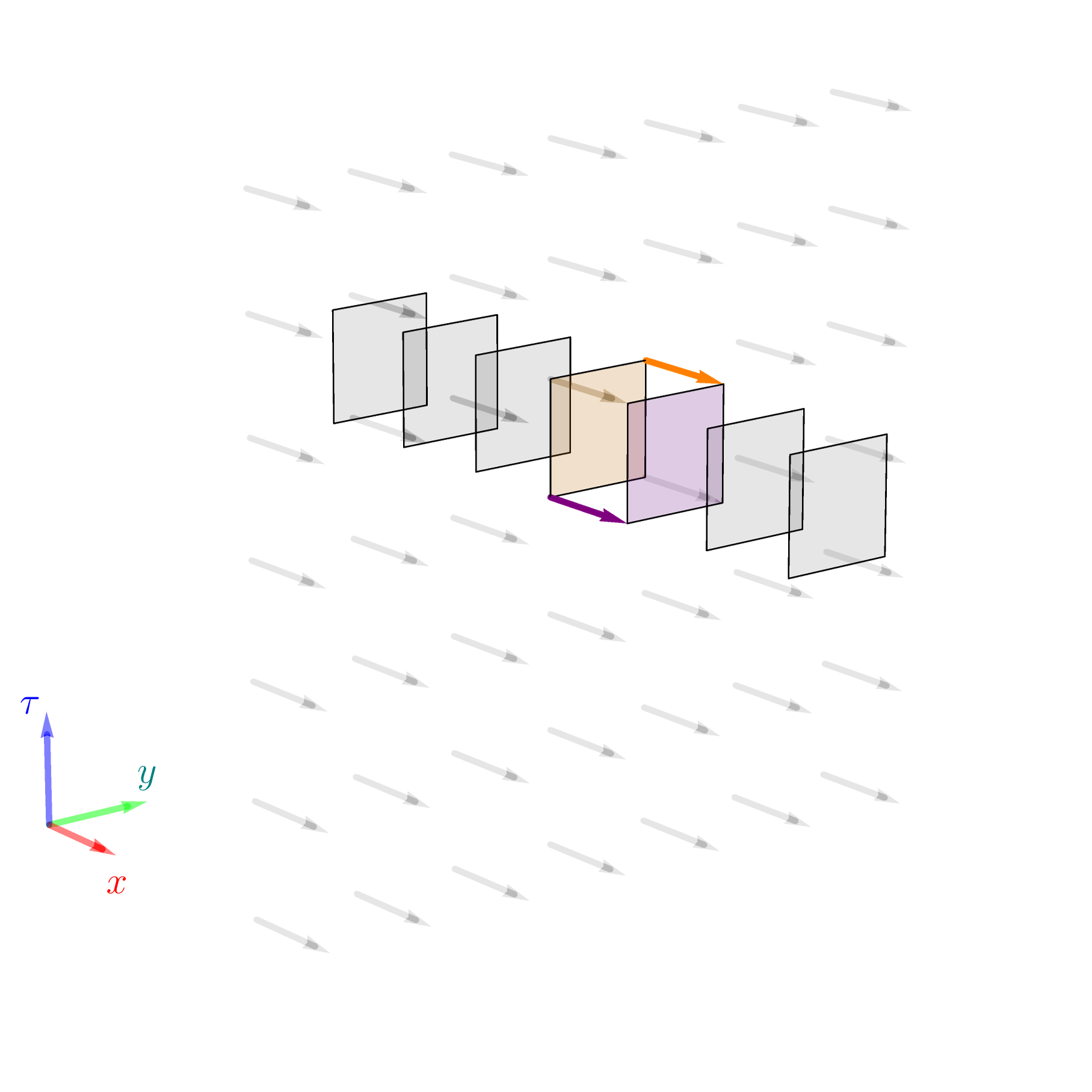}
    \caption{The indicated plaquettes form the representative Dirac string, on which $[s]^{\text{rep}}_p$ equals some common integer $\sigma$. The indicated links are where the flat gauge field fluctuation $\delta A'_l$ equals a common value $\theta\!\mod 2\pi$. The two pairs of links and plaquettes in orange and purple are where the $A\cup 2\pi s+2\pi s\cup A$ in the CS term contributes a phase of $e^{ik\theta \sigma}$. Integrating over $\theta$ leads to $\sigma=0$. Similarly for the other two directions on the 3-torus. Therefore, only the trivial class $[s]=0$ contributes to the partition function.}
    \label{s_a_cup}
\end{figure}

We can further simplify the above by considering those flat fluctuations $\delta A'_l$ of $A'_l$ such that $d\delta A'=0$. Such $\delta A'_l$ do not contribute to the $A'\cup dA'$ term, but they couple to the non-contractible $[s']^{\text{rep}}$ such that, on a three-torus, these $\delta A'_l$ fluctuations make the path integral vanish unless $[s]$ is the trivial class, for which we can choose $[s]^{\text{rep}}_p=0$; see \cref{s_a_cup} (For more general spacetime manifolds, only those $[s]$ that belong to the torsion part of $H^2(M; \mathbb{Z})$ will contribute, so it suffices to let summation of $[s]$ run over the torsion part; see Appendix \cref{torsion_in_homology_cohomology_theory}.) Thus, for a three-torus, we finally have
\begin{equation}
     Z=\frac{\left[\prod_{l} \int_{-\infty}^{\infty} \frac{dA_l'}{2\pi}\right]}{\left[\prod_{v} \int_{-\infty}^{\infty}  \frac{d\phi_v'}{2\pi}\right]}\frac{\sum_{[\kappa]}}{\sum_{[n']}} \ \exp \left\{-\frac{1}{2e^2}\sum_p (dA')_p^2+\frac{ik}{4\pi}\sum_c (A'\cup dA')_c\right\} \ ,
\end{equation}
a real gauge field Faddeev-Popov Gaussian integral. The flat holonomy fluctuations of the real-valued gauge field $d\delta A'=0$ are not constrained by the action once they have picked out $[s]=0$, and thus they contribute an infinite factor which, by construction, cancels the $\sum_{[n']}$; similarly, the uniform fluctuations $d\delta\phi'=0$ that do not transform $A'$ contribute an infinite factor which, by construction, cancels the $\sum_{[\kappa']}$.

Performing the Faddeev-Popov Gaussian integral, we find (see \cref{a_more_rigorous_calculation_of_the_partition_function} for a rigorous treatment)
\begin{equation}
    Z=\frac{\sqrt{\det'(d_0^T d_0)}}{\sqrt{\det'(2\pi M_\text{E})}} \ ,
    \label{gsd}
\end{equation}
where, as usual, $\det'$ stands for the product of all non-zero eigenvalues. $M_E$ is the Euclidean analogue of $M$, the matrix coupling two $A'$s. On a cubic lattice 3-torus,
\begin{equation}
    M_\text{E}(q)=\left[\frac{1}{e^2}d_1(q)^\dagger\eta_{\text{E}} d_1(q) -\frac{ik}{4\pi}(\cup(q)d_1(q)+d_1(q)^\dagger\operatorname{\cup}(q)^\dagger)\right]
    \label{M_E_def}
\end{equation}
after Fourier transformation. As in the previous section, we can factor $M_\text{E}(q)=K_\text{E}(q) d_1(q)$, and we will see the ground state degeneracy is independent of the details of $K_E(q)$ as long as it has some basic properties.

To proceed, we use the key fact that
\begin{equation}
    R=\frac{\det'(d_2) \det'(d_0)}{\det'(d_1)}
\end{equation}
is a topological invariant of the manifold---a special case of the Reidemeister torsion \cite{Adams:1996yf},
\footnote{More general Reidemeister torsion can involve covariant derivative with a flat background gauge field, generalizing the ordinary derivative here.

A continuum counterpart is the Ray-Singer torsion, which is well-known in the continuum CS context \cite{Schwarz:1978cn,Witten:1988hf,Guadagnini:2014mja}. It is proven that the Reidemeister torsion and the Ray-Singer torsion are equal \cite{muller1978analytic,cheeger1979analytic}.}
and is given by the size of the torsion parts of the cohomology classes---indeed because those ``would-have-been-Jacobians'' in \cref{summation_of_discrete_dofs} becomes the torsion parts of the cohomology classes; see \cref{torsion_in_homology_cohomology_theory}.
For three-torus, the torsion part is trivial, so $R=1$. We can also check this  explicitly by evaluating the determinants, following \cref{a_more_rigorous_calculation_of_the_partition_function}.

Since our lattice and dual lattice are the same, we have $\det'(d_2)=\det'(-d_0^T)$.
\footnote{If the dual lattice and the lattice do not appear the same, we expect $\det'(d_2)$ and $\det'(-d_0^T)$ to only differ by an extensive factor $e^{-\text{const.}\mathcal{V}}$ which does not affect the ground state degeneracy.}
Combining this with $R=1$, we end up with 
\begin{equation}
    Z=\frac{1}{\sqrt{\prod_{q\neq 0}[-\det'(2\pi K_\text{E}(q)|_{\text{image of $d_1(q)$}})]}} \ ,
    \label{Z_K}
\end{equation}
where we have projected $K_\text{E}(q)$ to the image of $d_1(q)$. For any $q\neq 0$, $d_1(q)$ projects a 3-dimensional vector space to a 2-dimensional image. At $q=0$, which corresponds to the fluctuation of flat holonomies, we have $K_\text{E}(q=0)=(ik/2\pi)\mathbf{1}_{3\times 3}$ but the image of $d_1(q)$ is trivial. We can include $q=0$ by writing
\begin{equation}
    \prod_{q\neq 0}[-\operatorname{det}'(2\pi K_\text{E}(q)|_{\text{image of $d_1(q)$}})]=\frac{\prod_{q} [-\operatorname{det}'(2\pi K_\text{E}(q)|_{2d-\text{image of $d_1(q)$}})]}{k^2} \ .
\end{equation}
where in the numerator we pretend we also projected $K_\text{E}(q=0)$ to a 2-dimensional subspace, so we need to remove the extra factor of $k^2$. Thus we have
\begin{align}
Z= |k| e^{-\sum_q h(q) }
\end{align}
where $h(q)=(1/2)\ln[-\det'(2\pi K_{\text{E}}(q)|_{2d-\text{image of $d_1(q)$}})]$.
All we need to know is that $h(q)$ is a smooth periodic function in $q$, so that when the system size is large, we can replace $\sum_q h(q)$ by $\mathcal{V} \int_{-\pi}^\pi \frac{d^3q}{(2\pi)^3} \, h(q)$, with error vanishing faster than any polynomial in $1/\mathcal{V}$ by the Euler-Maclaurin formula.
\footnote{We can see this directly without referring to the general formula. For simplicity suppose $q\in (-\pi, \pi]$ has only one component, as the generalization to three components is obvious. $h(q)$ being a smooth periodic function in $q$ means $h(q)=\sum_{\rho\in\mathbb{Z}} h(\rho) e^{-iq\rho}$ has its Fourier components $h(\rho)$ vanishing superpolynomially in $1/|\rho|$ as $|\rho|$ becomes large. Now, $\sum_{q\in (2\pi/L)\mathbb{Z}_L} h(q)=L\sum_{\rho\in L\mathbb{Z}} h(\rho)$; on the other hand, $L\int^\pi_{-\pi} (dq/2\pi) h(q)=Lh(\rho=0)$. So the error is $L\sum_{\rho\in L\mathbb{Z}, \rho\neq 0} h(\rho)$, which vanishes superpolynomially in $1/L$.}
Hence the ground state degeneracy is indeed $|k|$.
(Moreover, $\sum_q h(q)/\mathcal{V}$ is indeed real, as is desired for ground state energy.
\footnote{This is to say $Z$ is positive. Naively, it seems the reflection positivity requirement of a Euclidean theory (which is equivalent to the unitarity requirement of a Lorentzian theory) is enough to ensure this. Reflection positivity of the Euclidean theory says, when the spacetime, along with the extra structures on it needed for defining the theory, can be viewed as the gluing of a manifold (with the said structures) with its reflection, then $Z$ is positive if non-zero. While the 3-torus itself is indeed such a manifold, the cup product structure on it involves a certain shifted direction, which transforms under lattice reflection, so unfortunately we cannot directly use the reflection positivity argument.

It is not hard to see $Z$ is real. First suppose all of $L_x, L_y, \beta$ are even. Given an $A'$ configuration, consider another configuration $\tilde{A}'$ such that $\tilde{A}'_l=-A'_{\tilde{l}}$, where $\tilde{l}=(-r-\hat{\mu}, \mu)$ for $l=(r,\mu)$. Using the fact that $A\cup dA=dA\cup A$, we can show the contribution of $\tilde{A'}$ to $Z$ is equal to the complex conjugate of that of $A'$. Hence $Z$ is real. When, say, $L_x$ becomes odd instead of even, $\tilde{l}$ becomes $(-r+\hat{x}-\hat{\mu}, \mu)$. Likewise when $L_y, \beta$ become odd.

To see $Z$ is positive, we can directly check the expression of the determinant of $K_E$. We are currently unaware of (though we expect there to be) a more general argument that ensures the positivity of $Z$ on a three-torus.}
)

\section{Gravitational Anomaly}
\label{gravitational_anomaly}
\subsection{Review of a Continuum Calculation}
An important consequence of the chirality of the continuum CS theory is the \emph{gravitational anomaly} \cite{Witten:1988hf}. That is, naively we expect the continuum CS theory to be independent of the metric, but any sensible regularization will essentially involve a tiny Maxwell term \cite{Bar-Natan:1991fix} as we discussed in \cref{necessity_of_maxwell_term}; as a result, the partition function will have a universal metric dependence---known as the \emph{gravitational anomaly}---in the limit as the Maxwell term becomes infinitesimal. It turns out that this universal metric dependence can be cancelled by adding a suitable ``gravitational CS" counter-term \cite{Witten:1988hf}, but such counter-term itself depends on the choice of trivialization of the tangent bundle of the manifold, and this dependence is called \emph{framing anomaly}. Therefore, there is a trade-off between whether the theory has the gravitational or the framing anomaly; but either way the gravitational/framing anonaly is a manifestation of the chirality of the theory. 

\begin{figure}[h!]
    \centering
    \begin{subfigure}{0.3\textwidth}
        \centering   \includegraphics[width=\textwidth]{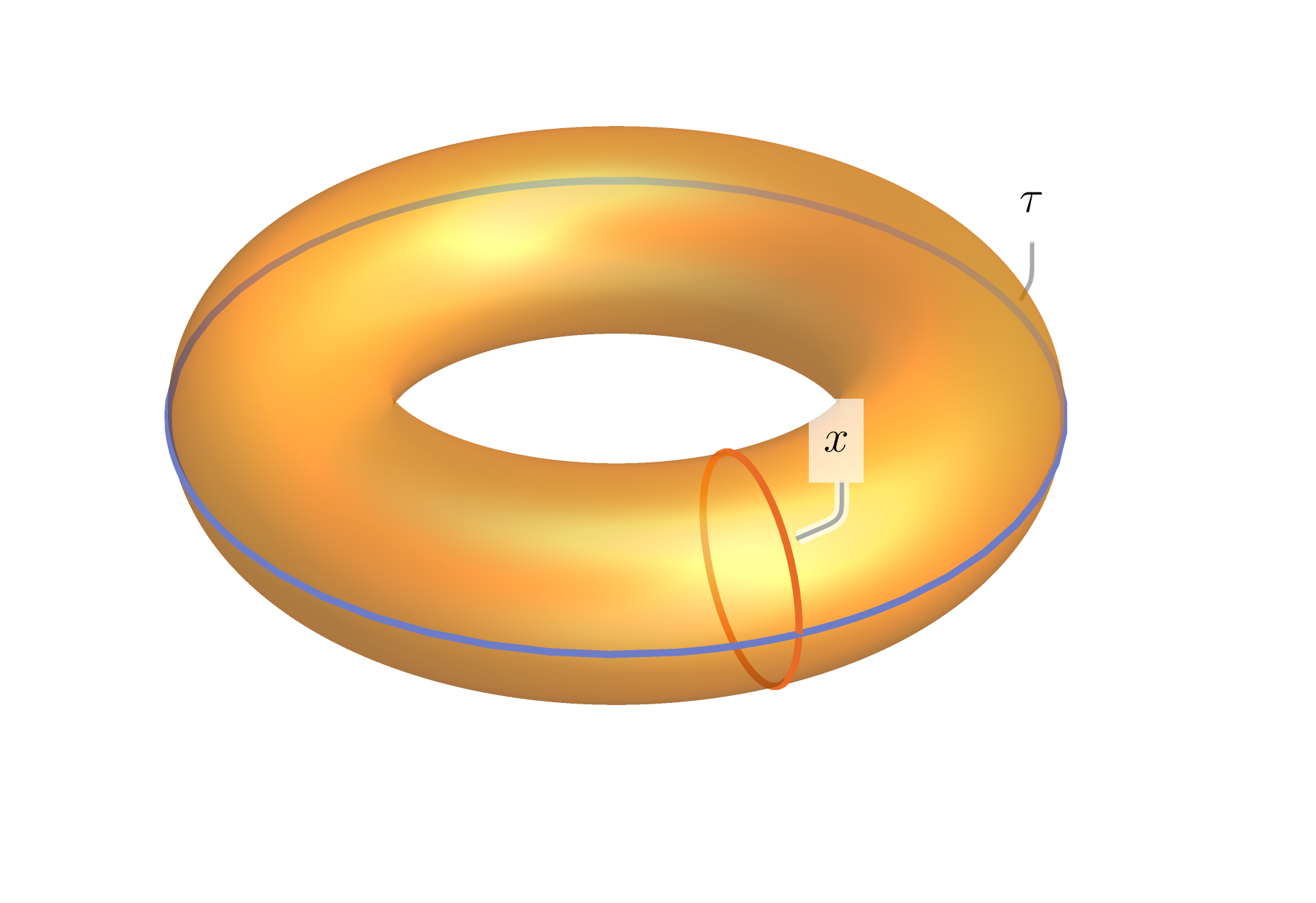}
        \caption{$\delta x=0$}
    \end{subfigure}
    \begin{subfigure}{0.3\textwidth}
        \centering        \includegraphics[width=\textwidth]{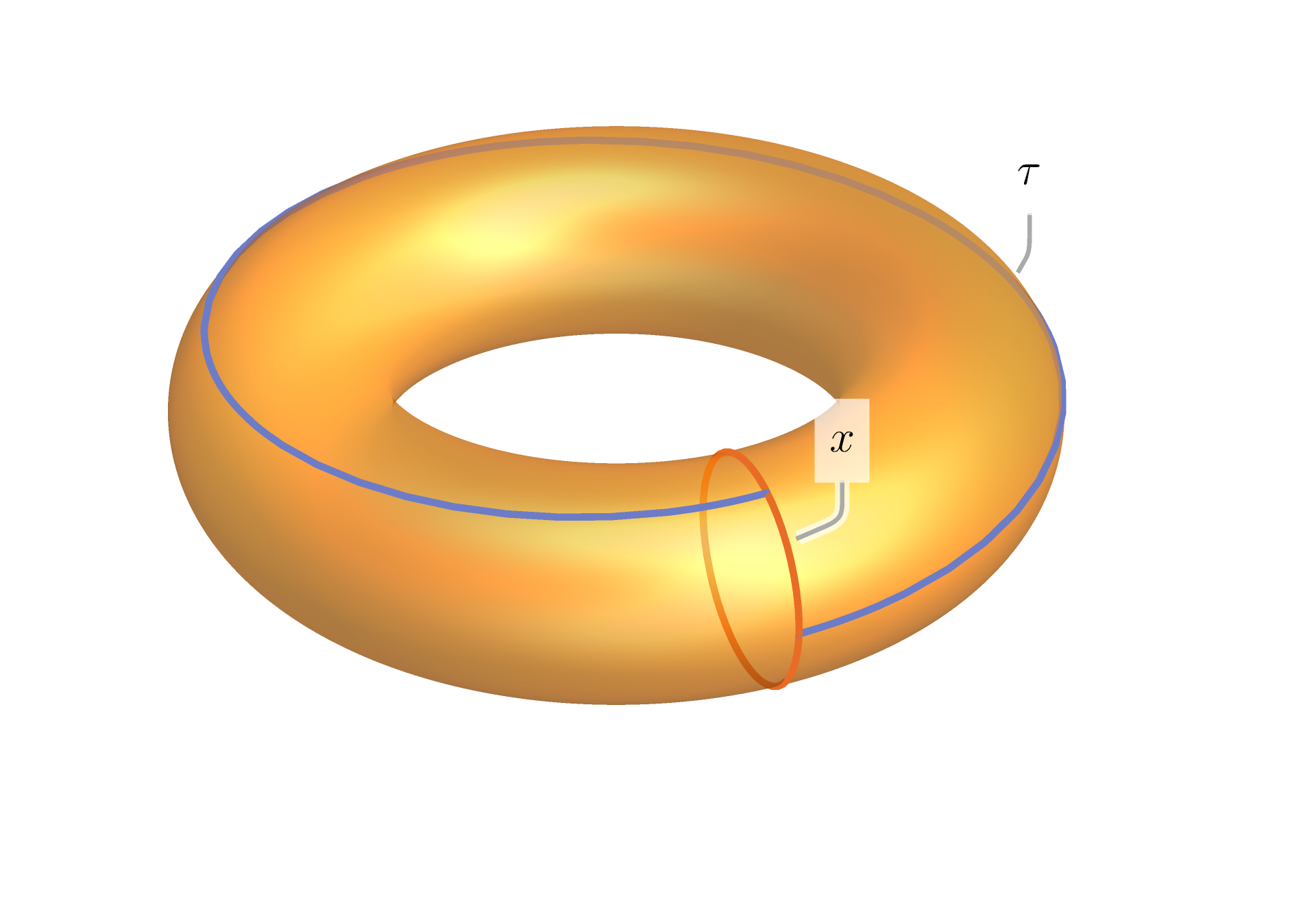}
        \caption{$0<\delta x<L_x$}
    \end{subfigure}
    \begin{subfigure}{0.3\textwidth}
        \centering        \includegraphics[width=\textwidth]{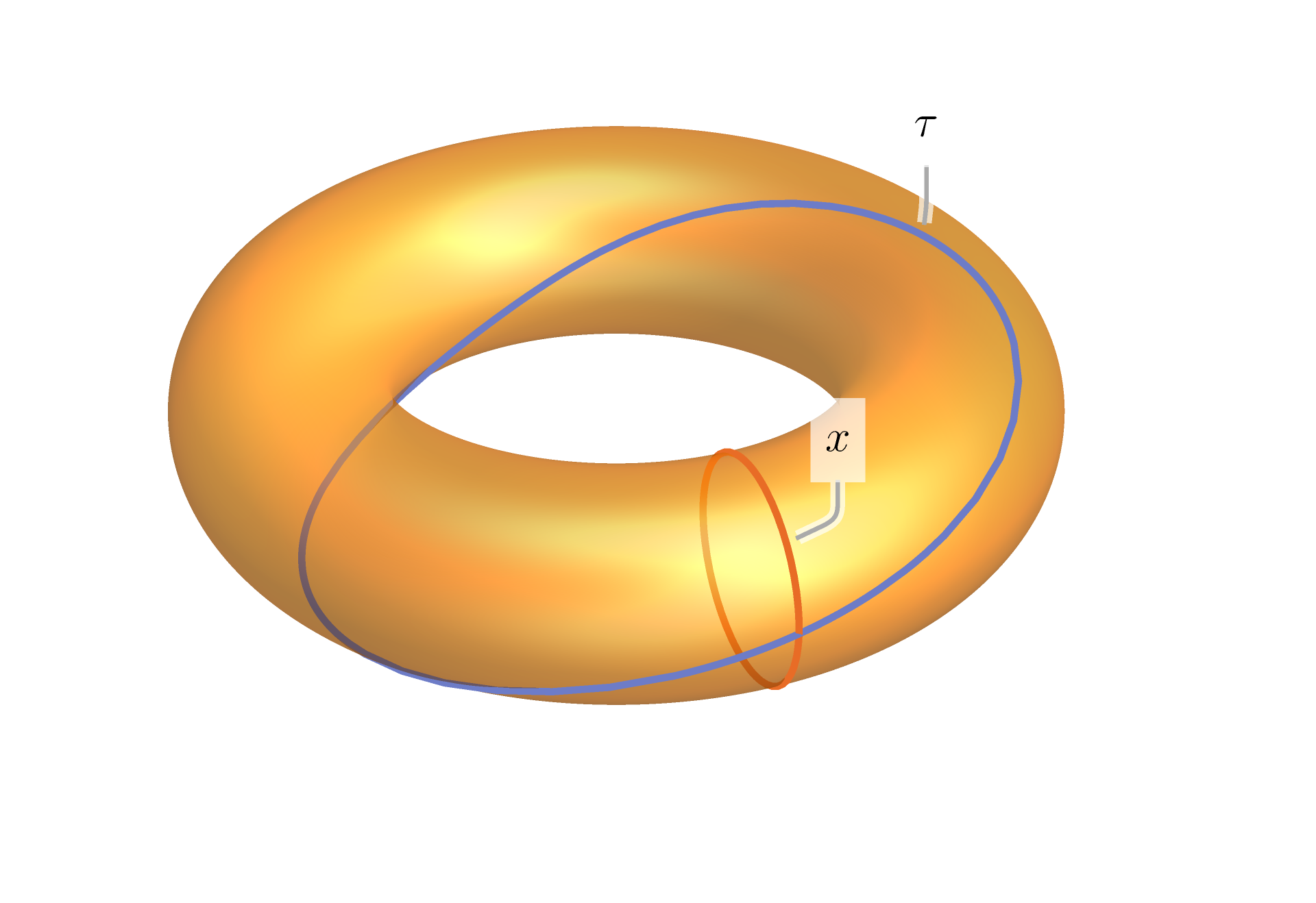}
        \caption{$\delta x=L_x$}
    \end{subfigure}
    \caption{Twisting the boundary condition on the solid torus.}
    \label{twist}
\end{figure}

A canonical way to see the gravitational/framing anomaly and its relation to chiral edge mode is to consider a solid torus Euclidean spacetime $D_2\times S^1$, whose boundary is a torus $S^1\times S^1$.
\footnote{A change of framing on a solid torus can induce a change of framing on more general manifolds through Dehn surgery, so it suffices to focus on the case of solid torus \cite{Witten:1988hf}.}
Let us call the two circles on the boundary the $x$ and the $\tau$ direction, with circumferences $L_x$ and $\beta$ respectively. Originally, we identify $(\tau, x)\sim (\tau, x+L_x)$ and $(\tau, x)\sim (\tau+\beta, x)$. But now for the second condition suppose we identify $(\tau, x)\sim (\tau+\beta, x+\delta x)$ instead, see \cref{twist}. We want to probe how the phase of the partition function responds to this change. When $\delta x$ is small, this is a change of the global holonomy, or say boundary condition, of the metric (although the local curvature of the metric is unchanged), so the phase response is seen as ``gravitational''. When $\delta x$ gradually increases to $L_x$, the metric becomes the same as the original, but we have changed a trivialization of the tangent bundle (a large gauge transformation of coordinates), and the accumulated phase is seen as due to ``framing''.
\footnote{This is much like ``treading a $2\pi$ flux'' in the Laughlin argument or Thouless pump for electrical response.}

Let us review the evaluation the phase response in the continuum. Note that evaluating the partition function with such a global change of the metric is equivalent to evaluating the partition function with an insertion of translation operator:
\begin{equation}
    Z=\operatorname{tr}e^{-\beta H+iP_x\delta x} \ .
\end{equation}
With fixed $L_x$ and large $\beta$ (in particular $\beta\gg L_x/c$), the phase response becomes
\begin{equation}
    \langle e^{iP_x\delta x} \rangle = e^{i\langle P_x \rangle \delta x}
\end{equation}
where we are evaluating the zero-point momentum $\langle P_x \rangle$ in the ground state on $D_2$. Just like the familiar zero-point energy, each classical harmonic mode with momentum $q_x$ contributes a $q_x/2$ to the zero-point momentum. In the continuum CS, the only available classical harmonic modes are the chiral edge modes, labeled by momenta satisfying $q_x\in 2\pi\mathbb{Z}/L_x$ and $\operatorname{sgn}(q_x)=-\operatorname{sgn}(k)$ (we only count half of all $q_x$---those with positive energy $\omega=-\operatorname{sgn}(k) q_x>0$ in the classical spectrum---when counting the classical harmonic modes, because the creation/annihilation operators satisfy $a_{-q_x}=a_{q_x}^\dagger$). Therefore we are evaluating
\begin{equation}
\langle P_x \rangle = \sum_{q_x \: (\omega(q_x)>0)} \frac{q_x}{2} = -\frac{\operatorname{sgn}(k)}{2}\sum_{n\geq0} \frac{2\pi n}{L_x}
\label{cont_Px_naive}
\end{equation}
which apparently diverges. To make sense of this, some smooth regulating function $f(q_x)$ with a soft cutoff $\Lambda$, such that $f(q_x)\rightarrow 1$ for $q_x\ll \Lambda$ and $f(q_x)\rightarrow 0$ for $q_x \gg \Lambda$, is introduced, so that we are evaluating
\begin{equation}
\begin{split}
    \langle P_x f(P_x) \rangle = &-\frac{\operatorname{sgn}(k)}{2}\sum_{n\geq0} \frac{2\pi n}{L_x}f\left(\frac{2\pi n}{L_x}\right)\\
    =& -\frac{L_x}{2\pi} \frac{\operatorname{sgn}(k)}{2} \int_0^\infty dq_x q_x f(q_x) - \frac{2\pi \, \operatorname{sgn}(k)}{24 L_x}\left.\left(\frac{d}{dq_x}q_xf(q_x)\right)\right|_{q_x=0}^{\infty} + \frac{1}{L_x} \mathcal{O}\left(\frac{1}{L_x\Lambda}\right)\\
    =& -\frac{L_x}{2\pi} \frac{\operatorname{sgn}(k)}{2} \int_0^\infty dq_x q_x f(q_x) + \frac{2\pi \, \operatorname{sgn}(k)}{24 L_x} + \frac{1}{L_x} \mathcal{O}\left(\frac{1}{L_x\Lambda}\right)
\end{split}
\label{cont_Px}
\end{equation}
where in the second line we used the Euler-Maclaurin formula. In the limit where $L_x\Lambda\gg 1$, the last term drops out. The first term, extensive in $L_x$, can be removed by a constant shift (of order $\mathcal{O}(\Lambda^2)$)---which can be viewed as normal ordering---in the definition of the local boundary momentum density operator. Thus we are left with the universal, non-extensive contribution $2\pi \operatorname{sgn}(k)/24L_x$, which is independent of the details of $f$ and the soft cutoff scale $\Lambda$. As we can see in the Euler-Maclaurin formula, the universal term comes from the kink of $\Theta(q_x) q_x f(q_x)$ at $q_x=0$ (here $\Theta$ is the Heaviside step function). This is the precise physical meaning of the seemingly bizarre formula ``$1+2+3+\cdots=-1/12$".

Thus, the universal gravitational response of the partition function to $\delta x$ is
\begin{equation}
    \langle e^{i :P_x f(P_x): \delta x} \rangle=e^{i\langle :P_x f(P_x): \rangle \delta x}=e^{i\frac{2\pi}{24}\frac{\operatorname{sgn}(k) \delta x}{L_x}}
    \label{continuum_anomaly}
\end{equation}
where $:\!O\!:$ means normal ordering an operator $O$. When $\delta x=L_x$, the response is interpreted as framing dependence. From here we can identify that the edge mode has a chiral central charge of $\operatorname{sgn}(k)$ \cite{Verlinde:1988sn}.

\subsection{Lattice Calculation and Interpretation}
The continuum calculation involves an artificial soft UV regulator, though the universal result does not depend on the details of it. The spirit of a lattice theory is different. Once the lattice theory has been defined, there is no subtlety in the UV and there would be no place to artificially impose any further regulator. All we can do is to calculate suitable observables and extract the physical meaning from the results.

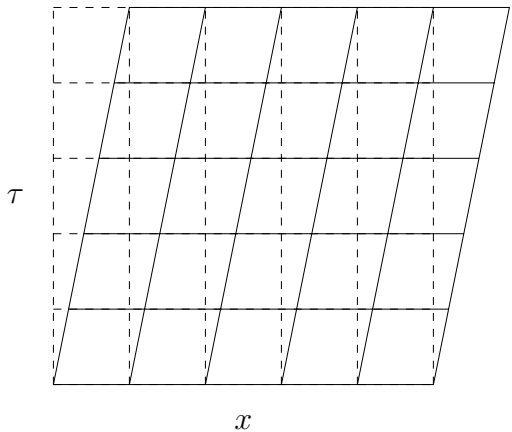
\begin{figure}
    \centering
    \begin{tikzpicture}
    \draw[step=1cm,dashed,very thin] (0,0) grid (5,5);
    \draw[step=1cm,black,very thin]
        \foreach \x in {0, ..., 5} {
          (\x, 0) -- (\x+1, 5)
        }
        \foreach \y in {0, ..., 5} {
          (0+\y/5, \y) -- (5+\y/5, \y)
        };
        \node at (2.5,-0.5) {$x$};
        \node at (-0.5,2.5) {$\tau$};
    \end{tikzpicture}
    \caption{Gridding Twist. Here $\delta x =1$.}
    \label{anomaly_grid}
\end{figure}

We still consider a Euclidean spacetime with a torus boundary, and then twist the periodic identification. Instead of a solid torus which is hard to work with using cubic lattice, we will consider a $S^1\times S^1\times \mathbb{R}_{\geq 0}$ Euclidean spacetime (the continuum result would be the same as a solid torus). The untwisted lattice would be $\mathbb{Z}_{\beta}\times\mathbb{Z}_{L_x}\times\mathbb{Z}_{\geq 0}$ with boundary $\mathbb{Z}_{\beta}\times\mathbb{Z}_{L_x}\times\{0\}$, and the ``twisted holonomy of the metric", say with $\delta x=1$, is realized by twisting the lattice gridding as shown in \cref{anomaly_grid} \cite{Tu:2012nj}. We will only consider the gravitational anomaly, with $\delta x$ of order $1$; the understanding of framing anomaly on the lattice will be deferred to a subsequent work \cite{XuChen_preparation}, as we will mention again at the end of this section and in the last section.

Readily from here, we can see a crucial difference with the continuum calculation. The continuum calculation really is computing the ground state expectation of $e^{iP_x f(P_x) \delta x}$, which only translates low momentum modes but not high momentum modes, i.e. consider an arbitrary function in $x$, the operator $e^{iP_x f(P_x) \delta x}$ will translate the envelope shape of the function by $\delta x$, but not any wiggly details below the length scale $1/\Lambda$. By contrast, in \cref{anomaly_grid} for the lattice, we are literally translating everything, so all $q_x$ momentum modes in the Brillouin zone, or say the entire function in $x$ with details that can be defined down to the lattice scale, will be translated by exactly the same amount $\delta x$. In the already UV complete lattice theory, there is just no natural way to consider a twist that only translates the envelope but not the details of a function. So the physical interpretation of the gravitational anomaly must be slightly different from that in the continuum calculation. We anticipate the lattice calculation will lead us towards a more UV complete, and in some sense more physical understanding of the seemingly mysterious gravitational anomaly.

Let us first summarize what we will find and what the physical interpretation is. We will compute the phase due to the $\delta x$ twist as shown in \cref{anomaly_grid},
\begin{equation}
    \frac{Z_{\text{twisted}}}{Z_{\text{no twist}}} = e^{i\mathcal{A}} \ ,
    \label{grav_anomaly_full}
\end{equation}
and find the phase $\mathcal{A}$ to have non-universal dependence on $L_x$ due to the UV physics---even the non-extensive in $L_x$ terms are non-universal, and roughly speaking, the non-universality in these terms depends on how the chiral edge mode merges into the bulk mode after Wick rotation. However, we can extract the IR physics by comparing the results at two different temperatures, $1\ll L_x/c\ll \beta_1$ and $1\ll \beta_2 \ll L_x/c$ (we also assume $L_x\gg 1$ regardless of $c$):
\begin{equation}
    \left.\frac{Z_{\text{twisted}, \beta_1}}{Z_{\text{no twist}, \beta_1}} \right/ \frac{Z_{\text{ twisted}, \beta_2}}{Z_{\text{no twist}, \beta_2}} = e^{i\mathcal{A}_{\text{IR}}} \ ,
    \label{grav_anomaly_ratio}
\end{equation}
and it turns out $\mathcal{A}_{\text{IR}}$ has the expected universal $2\pi \operatorname{sgn}(k) \delta x/24L_x$ behavior. Essentially the ``high temperature'' $\beta_2$ contribution is playing the role of the regulating function $f(P_x)$ in the continuum that removes the UV physics.

Now we perform the actual calculation.

We are going to use a somewhat unusual method to calculate the partition function. To demonstrate how our method works, we take harmonic oscillator on Euclidean lattice $\mathbb{Z}_\beta$ as an example. We may write its partition function as
\begin{equation}
\begin{split}
    Z_{\text{h.o.}}&= \left[\prod_v \int \frac{dx_v}{\sqrt{2\pi}}\right] e^{-\frac{1}{2}\sum_l (d_0x)_l^2-\frac{1}{2}\sum_v \omega_0^2 x_v^2}=\frac{1}{\sqrt{\det(d_0^Td_0+\omega_0^2)}}\\
    &=\prod_{q_\tau\in 2\pi\mathbb{Z}_\beta/\beta}\frac{1}{\sqrt{2-2\cos q_\tau+\omega_0^2}},
    \end{split}
\end{equation}
where $d_0$ is just the time derivative on the $1d$ $\mathbb{Z}_\beta$ lattice. To actually work out the product, we view $z=e^{iq_\tau}$ as a complex variable, so we have
\begin{equation}
\begin{split}
    Z_{\text{h.o.}}&=\prod_{z\: (z^\beta=1)}\frac{1}{\sqrt{2-z-1/z+\omega_0^2}}\\
    &=\prod_{z\: (z^\beta=1)}\frac{1}{\sqrt{(z-e^{2\sinh^{-1}(\omega_0/2)})(e^{-2\sinh^{-1}(\omega_0/2)}-z)/z}}.
    \end{split}
    \label{SHO_example}
\end{equation}
The denominator vanishes at $z=e^{\pm2\sinh^{-1}(\omega_0/2)}$. While in fact $z$ cannot take these values because $z$ is on the unit circle, it is still useful to think of the entire complex plain of $z$. Then $e^{\pm2\sinh^{-1}(\omega_0/2)}$ are determined by roots of the equation $d_0^T(z) d_0(z)+\omega_0^2=0$ (and in the continuum limit, this will correspond to the equation of motion in the Lorentzian signature after Wick rotation). The key point of this method is: we are finding the roots instead of the eigenvalues of $d_0^T d_0+\omega_0^2$ in order to compute the determinant, and the former is a much easier task in more general problems. Now, to arrive at the determinant, we take the product over $z$, using the fact that for any given complex number $z_0$,
\begin{equation}
    \prod_{z\: (z^\beta=1)}(z_0-z)=z_0^\beta-1.
    \label{z_product}
\end{equation}
Thus
\begin{equation}
\begin{split}
    Z_{\text{h.o.}}
    &=\frac{1}{\sqrt{(1-e^{2\beta \sinh^{-1}(\omega_0/2)})(e^{-2\beta \sinh^{-1}(\omega_0/2)}-1)}}\\
    &=\frac{1}{2\sinh(\beta\sinh^{-1}(\omega_0/2))}.
\end{split}
\end{equation}
Notice that in the $\omega_0\ll 1$ limit, i.e. when the imaginary time becomes continuous, $e^{\pm 2\sinh^{-1}(\omega_0/2)}$ becomes the exponential of energy modes $e^{\pm\omega_0}$. In this limit, we have
\begin{equation}
    Z_{\text{h.o.}}=\frac{1}{2\sinh(\beta\omega_0/2)}
\end{equation}
which indeed agrees with $\sum_{n=0}^{\infty}e^{-\beta(n+1/2)\omega_0}$ that comes from directly diagonalizing the Hamiltonian.

We now apply this method to our CS-Maxwell theory of interest. Recall from the previous section the partition function is evaluated as
\begin{equation}
    Z=\frac{\sqrt{\det'(d_0^T d_0)}}{\sqrt{\det'(2\pi M_\text{E})}} \ .
\end{equation}
On the non-twisted lattice, a Fourier transform of the $\tau, x$-directions can be readily performed; on the other hand, we keep the semi-infinite $y$-direction in the real space coordinates. We have
\begin{equation}
    Z=\frac{\sqrt{\det'(d_0^T d_0)}}{\sqrt{\det'(2\pi M_\text{E})}}=\prod_{q_\tau,q_x}\frac{\sqrt{\det'(d_0^\dagger(q_\tau,q_x) d_0(q_\tau,q_x))}}{\sqrt{\det'(2\pi M_\text{E}(q_\tau,q_x))}}=\prod_{q_\tau,q_x}Z(q_\tau,q_x),
\end{equation}
where the determinant is being taken in the space of $y$-coordinates and link directions; there is no phase ambiguity in taking the square roots, because the square roots come from Gaussian integral, so the branch cut is always placed along the negative real axis. Note that on a twisted lattice, $(\tau,x,y)$ and $(\tau+\beta,x+\delta x,y)$ are identified. Thus we have
\begin{equation}
    e^{iq_\tau\beta+iq_x\delta x}=1,
\end{equation}
in other words, $q_\tau$ takes $2\pi n/\beta-q_x\delta x/\beta, n=0,1,\cdots \beta-1$ values.

Let us focus on $Z(q_\tau, q_x)$ for fixed $q_\tau$ and $q_x$. If we view $z=e^{iq_\tau}$ and $w=e^{iq_x}$ as complex variables, both the numerator and the denominator are series of $z,1/z,w,1/w$. Following the key idea explained below \cref{SHO_example}, to evaluate the determinant, instead of finding the eigenvalues, we can more easily look for the roots: 
\begin{equation}
    Z^2(z,w)=C(w) z^{n_0} \frac{\prod_{z_i\:\text{roots of}\:\det'd_0^\dagger d_0}(z-z_i(w))}{\prod_{z_j\:\text{roots of}\:\det'M_\text{E}}(z-z_j(w))}
\end{equation}
where $C(w)$ is a function that only depends on $w$, and $n_0$ is some integer which is formally divergent---this is due to the $y$-direction being semi-infinite in size, but as we will see this will not pose a problem when we take the ratio \cref{grav_anomaly_ratio} in the end to extract the IR physics.

Before we proceed, let us first have some intuition about what the $z_j(w)$'s mean. Writing $z_j(w)=e^{-\omega_j(q_x)}$, we find $q_\tau$ has poles at $i\omega_j(q_x)$. The intuition can be drawn when $c\ll 1$, where the Euclidean time direction essentially becomes continuous so that we can appeal to the familiar Wick rotation which analytically continues $q_\tau$ to $-iq_t=i\omega$. In this $c\ll 1$ limit, we can recognize that $\omega_j(q_x)$ are just the spectra we found in \cref{bulk_spectrum} and \cref{chiral_edge_mode}, with $j$ being the real and complex values of $q_y$ that give normalizable modes. Away from the $c\ll 1$ limit, while the Wick rotation is no longer exact, the intuition thus built is still helpful for understanding the calculations below.

We compute \cref{grav_anomaly_full} as
\begin{equation} 
\begin{split}
    \frac{Z^2_{\delta x}}{Z^2_{\delta x=0}}&=
    \prod_{w\: (w^{L_x}=1)}\frac{\prod_{z\:(z^\beta w^{\delta x}=1)}Z(z,w)}{\prod_{z'\:(z'^\beta=1)}Z(z',w)}\\
    &=\prod_{w\:(w^{L_x}=1)}\left[\frac{\prod_{z\:(z^\beta w^{\delta x}=1)}z^{n_0}}{\prod_{z'\:(z'^\beta=1)}z'^{n_0}}\prod_{z_i}\frac{\prod_{z\:(z^\beta w^{\delta x}=1)}(z-z_i)}{\prod_{z'\:(z'^\beta=1)}(z'-z_i)}\prod_{z_j}\frac{\prod_{z'\:(z'^\beta=1)}(z'-z_j)}{\prod_{z\:(z^\beta w^{\delta x}=1)}(z-z_j)}\right],
\end{split}
\end{equation}
where the $C(w)$ coefficients cancel out as the products over $w$ are the same in the numerator and denominator. Note the dependence on $\delta x$ is only in the difference in the conditions for $z$ and $z'$. The $z^{n_0}/z'^{n_0}$ factor gives rise to
\begin{equation}
    \prod_{w\:(w^L_x=1)}\frac{\prod_{z\:(z^\beta w^{\delta x}=1)} z^{n_0}}{\prod_{z'\:(z'^\beta=1)}  z'^{n_0}}=\prod_{w\:(w^L_x=1)} w^{-n_0\delta x}= \left[(-1)^{(L_x-1) \delta x} \right]^{n_0},
\end{equation}
Although $n_0$ is formally divergent, we can see this is independent of $\beta$, hence will be cancelled out when taking the ratio \cref{grav_anomaly_ratio} for the IR result.

Now we proceed with the other factors, again using \cref{z_product}, and get
\begin{equation}
    \frac{Z^2_{\delta x}}{Z^2_{\delta x=0}}=(-1)^{n_0 (L_x-1) \delta x}\prod_{w\:(w^{L_x}=1)}\left[\prod_{z_i}\frac{(z_i^\beta(w)-w^{-\delta x})}{(z_i^\beta(w)-1)}\prod_{z_j}\frac{(z_j^\beta(w)-1)}{(z_j^\beta(w)-w^{-\delta x})}\right].
\end{equation}
The problem has now been reduced to determining the roots $z_i(w)$ of $\det'd_0^\dagger(z,w) d_0(z,w)$ (with $w$ fixed) and $z_j(w)$ of $\det'M_\text{E}(z,w)$. That is we need to find those $z_i$ in the complex plain for which there exists $\phi(y)$ such that $d_0^\dagger(z_i,w)d_0(z_i,w)\phi=0$ with boundary condition $\phi(0)=\phi(\infty)=0$, and $z_j$ for which there exists $A(y)$ such that $M_\text{E}(z_j,w)A=0$ with boundary condition $A_{\tau,x}(0)=A_{\tau,x,y}(\infty)=0$. 

It is intuitive that we will mainly care about the roots that correspond to the ``chiral edge modes'' when viewed as ``Wick rotated'' to the  Lorentzian signature (see discussions above); later we will come back and justify that the bulk modes are indeed unimportant. Everything is now quite similar to what we have done in \cref{chiral_edge_mode}, so here we directly give the result. $d_0^\dagger(z_i,w)d_0(z_i,w)$ is just a Laplacian operator, and has no edge mode, while the edge mode roots for $M_\text{E}(z_j,w)$ satisfy
\begin{equation}
    \left\{
    \begin{aligned}
        \sin{\frac{q_\tau}{2}}&=\pm i c\sin{\frac{q_x}{2}}\\
        \tan{\frac{q_y}{2}}&=\frac{\cos\left(\frac{1}{2}(q_x+q_\tau)\right)}{\pm i \frac{2}{mc}+\sin\left(\frac{1}{2}(q_x+q_\tau)\right)}\\
        \operatorname{Im} q_y&\geq0
    \end{aligned}\right.\ ,
    \label{edge_mode_conditions}
\end{equation}
similar to the Lorentzian signature edge mode. For each $w=e^{iq_x}$, there is only one of the roots $z_j(w)$ that corresponds to chiral edge mode, and we shall call this root  $z_{\text{edge}}(q_x)$:
\begin{equation}
    z_{\text{edge}}(q_x)=
    \left\{
    \begin{aligned}
        &e^{-2\sinh^{-1}(c\sin(q_x/2))}& -\pi\leq q_x<q_c\\
        &e^{2\sinh^{-1}(c\sin(q_x/2))}& q_c<q_x\leq \pi
    \end{aligned}
    \right.,
\label{z_edge}
\end{equation}
according to the first equation of \cref{edge_mode_conditions}; which sign in $\pm$ to choose for each $q_x$ is determined non-trivially by the second and third equations of \cref{edge_mode_conditions} and it turns out the choices are separated by the critical point
\begin{equation}
    q_c=-2\cot^{-1}(mc^2/2).
\end{equation}
We can define the contribution by such chiral edge mode to the phase of $Z_{\delta x}^2/Z_{\delta x=0}^2$ as
\begin{align}    
\varphi(q_x)&=\operatorname{arg}\frac{z_{\text{edge}}(q_x)^{\beta}-1}{z_{\text{edge}}(q_x)^{\beta}-e^{-iq_x\delta x}} \nonumber \\[.2cm]
&=\left\{
    \begin{aligned}
        &\tan^{-1}\frac{\sin{q_x\delta x}}{\cos{q_x\delta x}-e^{-2\beta\sinh^{-1}(c\sin(q_x/2))}}& -\pi\leq q_x<q_c\\[.2cm]
        &\tan^{-1}\frac{\sin{q_x\delta x}}{\cos{q_x\delta x}-e^{2\beta\sinh^{-1}(c\sin(q_x/2))}}& q_c<q_x\leq \pi
    \end{aligned}
    \right.\ 
\label{varphi}
\end{align}
so that
\begin{equation}
    2\mathcal{A}= n_0 (L_x-1)\pi\delta x +  \sum_{q_x}\varphi(q_x) + (\mbox{bulk mode contributions}).
    \label{anomaly}
\end{equation}

\begin{figure}[t!]
    \centering
    \begin{subfigure}{0.44\textwidth}
        \centering
        \includegraphics[width=\textwidth]{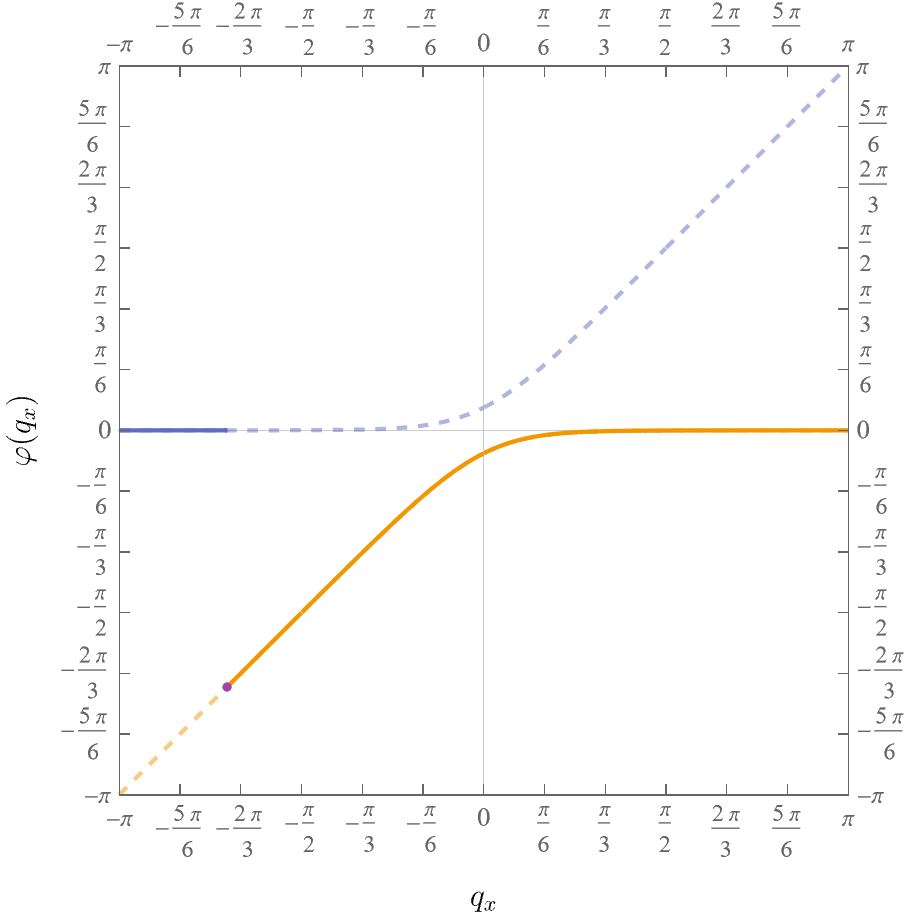}
        \caption{$\varphi(q_x)$}
    \end{subfigure}
    \begin{subfigure}{0.44\textwidth}
        \centering
        \includegraphics[width=\textwidth]{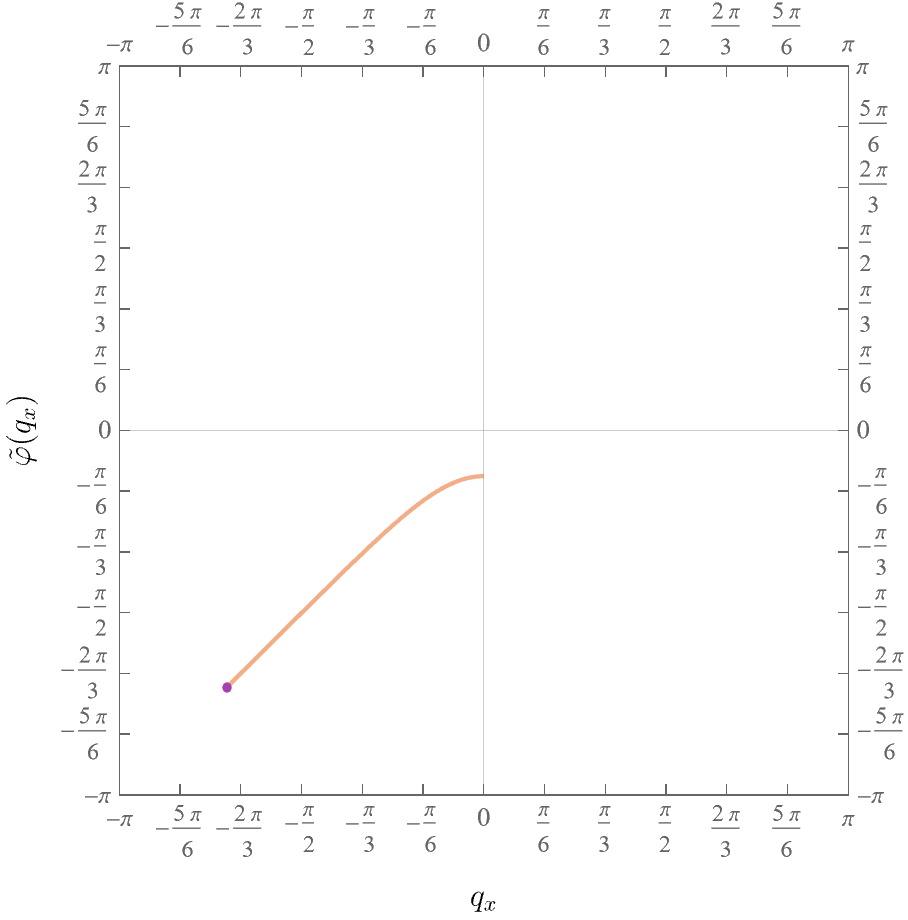}
        \caption{$\tilde\varphi(q_x)$}
    \end{subfigure}
    \caption{The plots are made under the set of parameters $\beta c=5, mc^2=1, c=1, \delta x=1$.  $\varphi(q_x)$ is represented by the solid curves in the left panel, with the dashed curves being the segments that are excluded by the $\operatorname{Im} q_y>0$ condition. The marked out point $q_c$ is given by the $\operatorname{Im} q_y>0$ condition in \cref{edge_mode_conditions}: for $q_x<q_c=-2\cot{(mc^2/2)}$, $\sin (q_\tau/2)=ic\sin(q_x/2)$, while for $q_x>q_c=-2\cot{(mc^2/2)}$, $\sin (q_\tau/2)=-ic\sin(q_x/2)$.}
    \label{anomaly_phase}
\end{figure}

We would like to plot the $\varphi(q_x)$, the edge mode contribution to the phase, and compare it to the summand of \cref{cont_Px} in the continuum Hamiltonian formalism. However, to make sense of the comparison, it turns out here we need to look at $\tilde\varphi(q_x)=\varphi(q_x)+\varphi(-q_x)$ (for half of all $q_x$). This is because in the Lagrangian formalism we employ here, $\pm q_x$ are different Fourier modes, but in the Hamiltonian formalism, they are the creation and annihilation of the same excitation mode.

In \cref{anomaly_phase} we plot $\varphi(q_x)$ by the solid lines in the left panel (the dashed lines are the segments that are excluded by the $\operatorname{Im}(q_y)\geq 0$ condition), and $\tilde\varphi(q_x)$ in the right panel, under one set of parameters. (We will plot under more sets of parameters below.) There is an apparent discontinuity at $q_c$ indicated by the thick solid dot. From the $\tilde\varphi(q_x)$ plot it is not hard to recognize the physical meaning of it---this is where the edge mode merges into the bulk mode.

What are the crucial features that we should focus on in the plot? We should focus on those discontinuities and kinks, for the same reason as how the universal term in \cref{cont_Px} arises from the kink. Terms in the phase that are proportional to system size $L_x$ are non-universal and unimportant; we only care about the leading non-extensive part of anomaly phase, that arise from the discontinuities and kinks, due to the Euler-MacLaurin formula. Before we present the precise, non-trivial calculation that extracts these effects, let us pictorially get an intuition of how tuning the parameters changes the appearance of the discontinuity and kinks.

\begin{figure}[h!]
    \centering
    \begin{subfigure}{0.35\textwidth}
        \centering        \includegraphics[width=\textwidth]{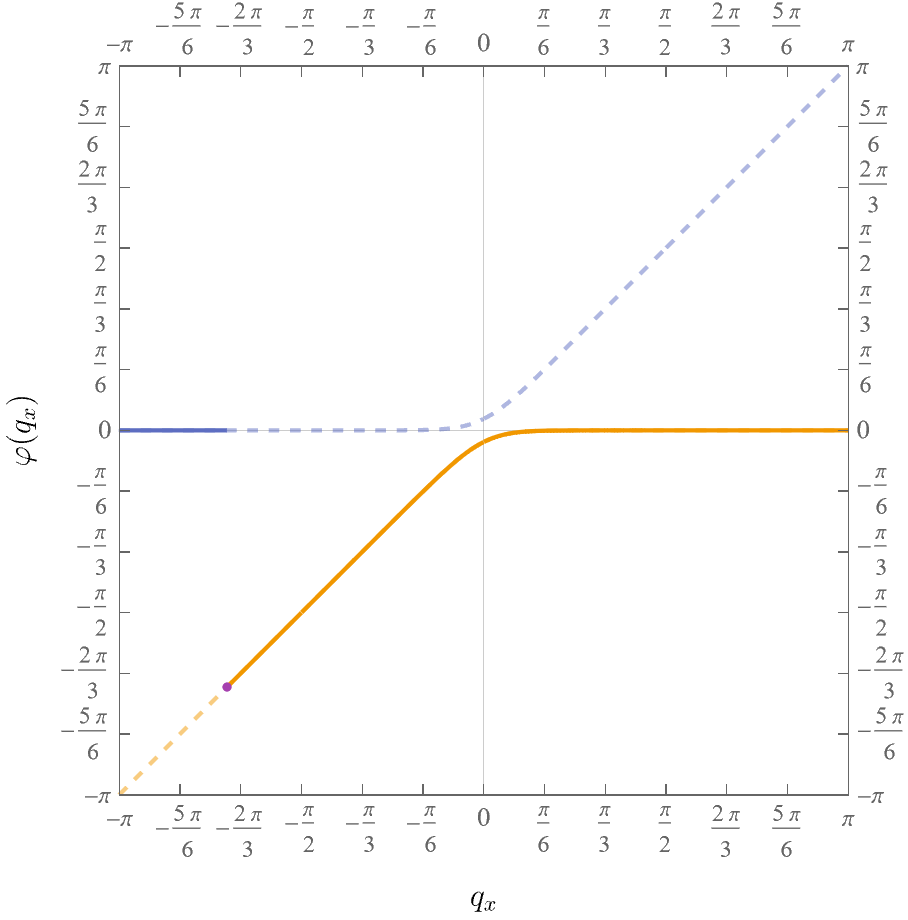}
        \caption{$\beta=10$}
    \end{subfigure}
    \begin{subfigure}{0.35\textwidth}
        \centering      \includegraphics[width=\textwidth]{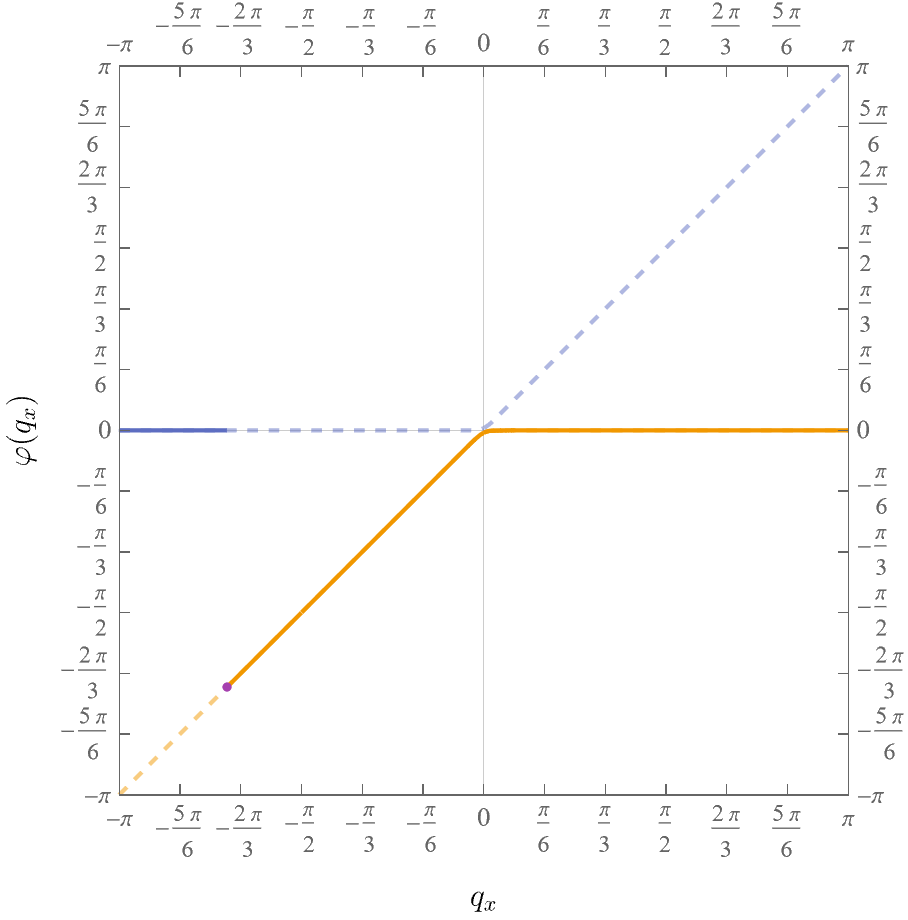}
        \caption{$\beta=50$}
    \end{subfigure}
    \caption{Keep $mc^2=1, c=1, \delta x=1$ and tune $\beta$.}
    \label{anomaly_phase_beta}
\end{figure}
\begin{figure}[h!]
    \centering
    \begin{subfigure}{0.35\textwidth}
        \centering      \includegraphics[width=\textwidth]{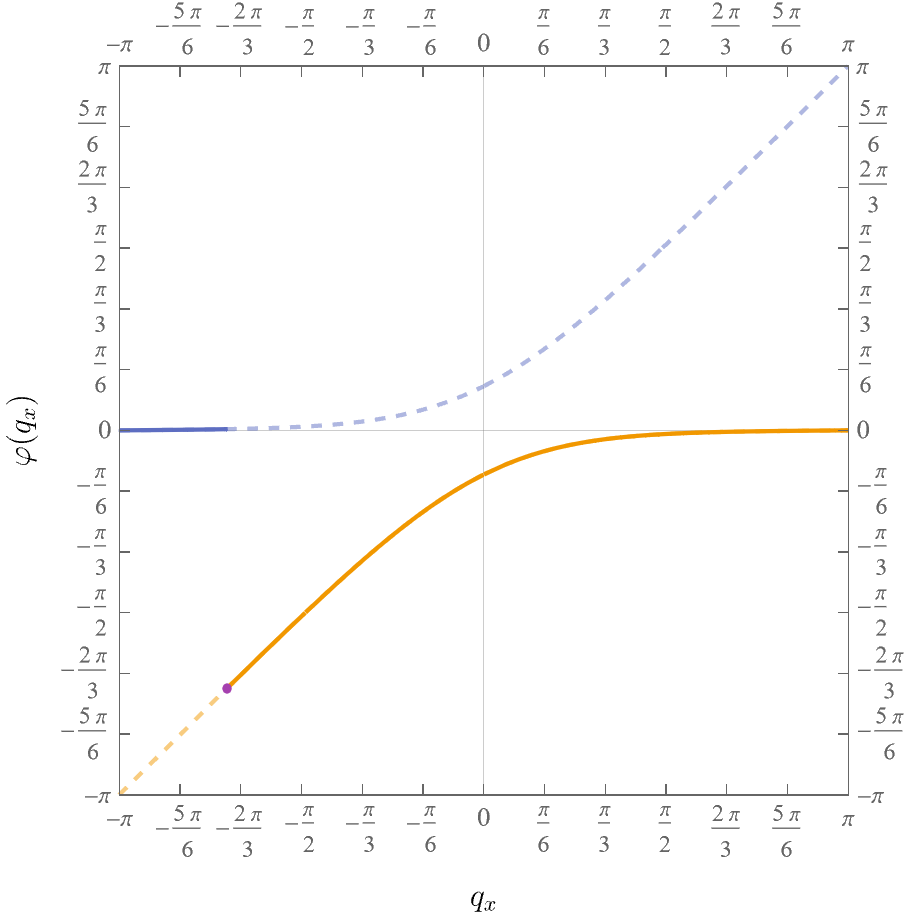}
        \caption{$c=0.5$}
    \end{subfigure}
    \begin{subfigure}{0.35\textwidth}
        \centering
        \includegraphics[width=\textwidth]{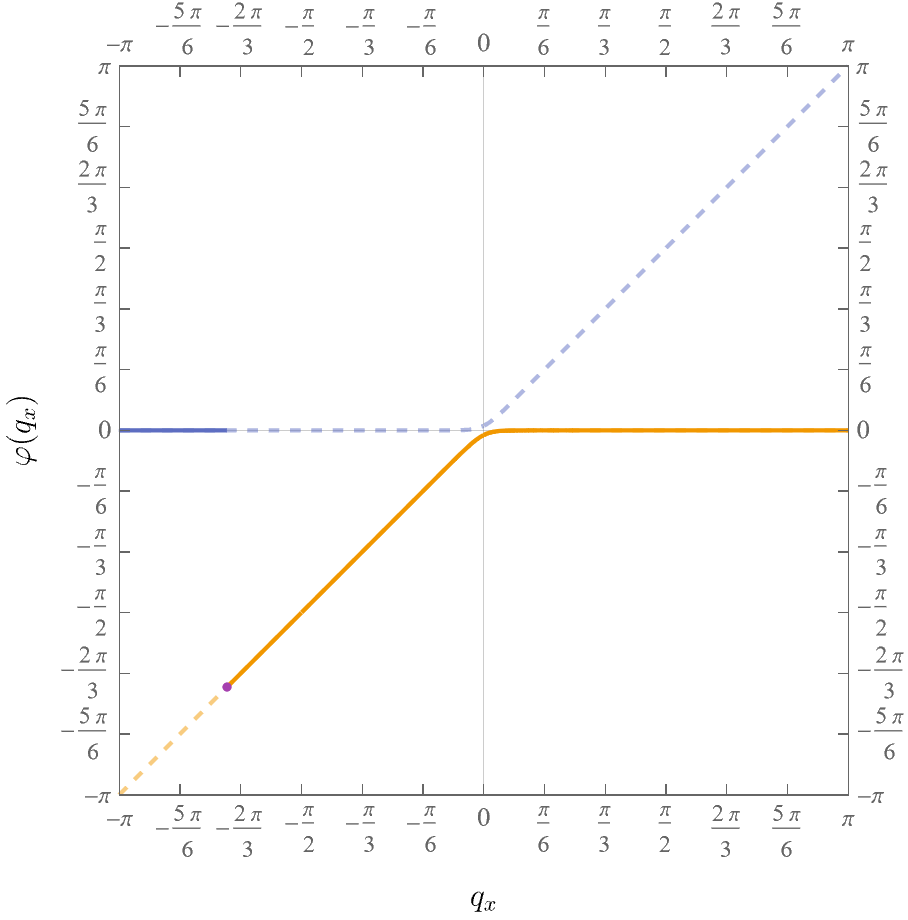}
        \caption{$c=5$}
    \end{subfigure}
    \caption{Keep $\beta=5,m c^2=1, \delta x=1$ and tune $c$.}
    \label{anomaly_phase_c}
\end{figure}

In \cref{anomaly_phase_beta}, we look at the $\beta$ dependence of $\varphi(q_x)$. We note the effect mainly occurs at small $|q_x|$, within a range of $1/\beta c$, as can be read-off from \cref{varphi}. As $\beta$ increases, the behavior near $q_x=0$ becomes more and more like a kink, even though for finite $\beta$ it never becomes a real kink. However, considering that $q_x$ takes values with step size $2\pi/L_x$, we expect that when $1/\beta \ll c/L_x$, the behavior near $q_x=0$ becomes indistinguishable from having a real kink---this is the important point that we are going to rigorously extract. In \cref{anomaly_phase_c}, we can see tuning $c$ has a similar effect as tuning $\beta$.
\footnote{Note when $c\gg 1$, we not only need $L_x\gg 1$, we also need $L_x/c\gg 1$ to extract the correct universal behavior, as can be seen from the argument of $\sinh^{-1}$ in $z_{\text{edge}}$.}

\begin{figure}[h!]
    \centering
    \begin{subfigure}{0.35\textwidth}
        \centering   \includegraphics[width=\textwidth]{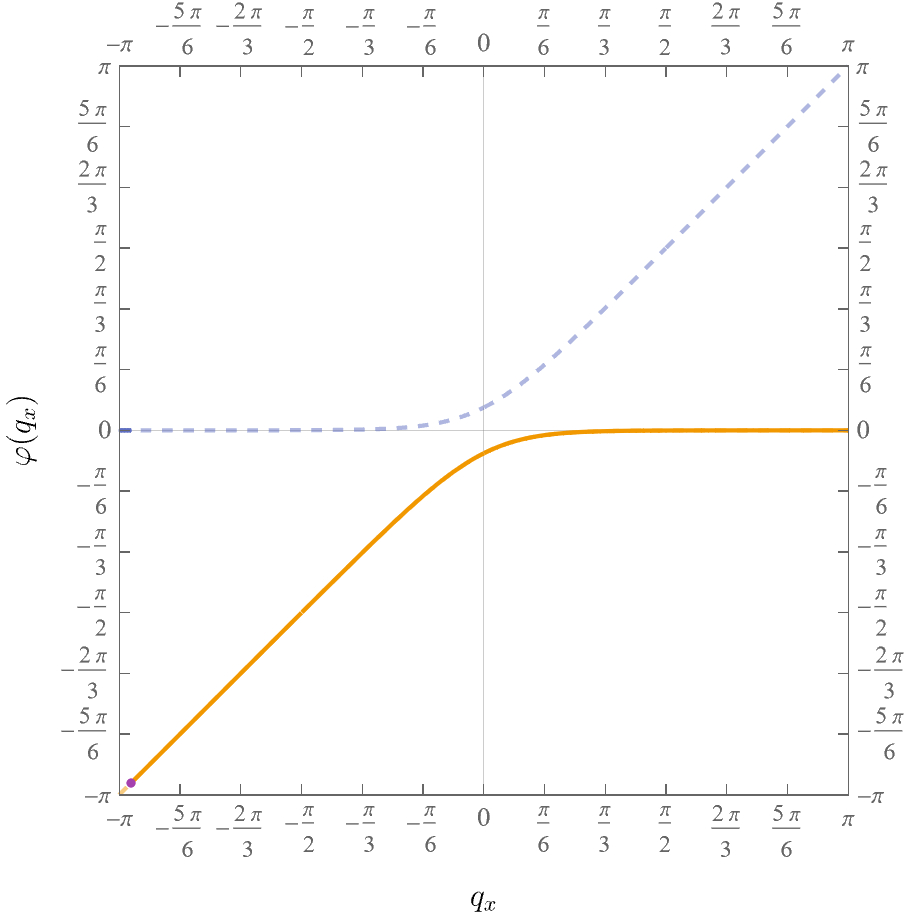}
        \caption{$mc^2=0.1$}
    \end{subfigure}
    \begin{subfigure}{0.35\textwidth}
        \centering        \includegraphics[width=\textwidth]{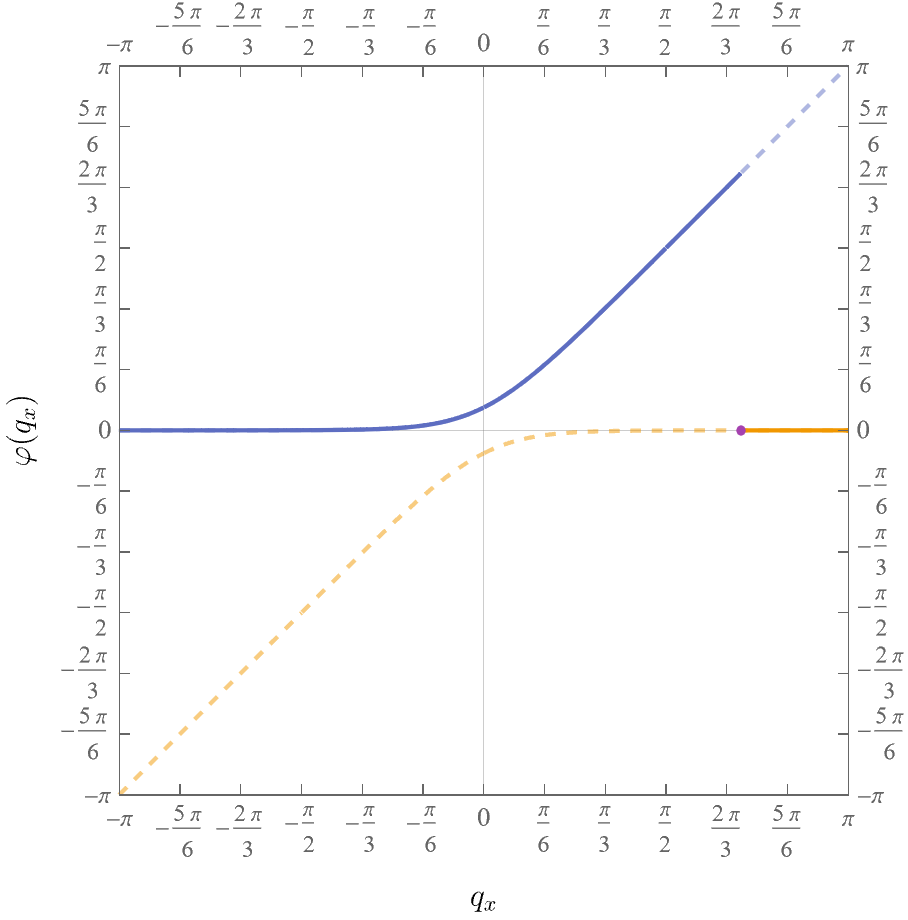}
        \caption{$mc^2=-1$}
    \end{subfigure}
    \caption{Keep $\beta=5, c=1, \delta x=1$ and tune $mc^2$.}
    \label{anomaly_phase_m}
\end{figure}

In \cref{anomaly_phase_m}, changing $mc^2$ will move the discontinuous point $q_c=-2\cot^{-1}(mc^2/2)$ which is interpreted as where the chiral edge mode merges into the bulk mode. Notice the sign of $mc^2$, i.e. the sign of the CS level $k$, determines whether the curve ends in the first quadrant or in the third quadrant, which corresponds to the direction of chiral edge mode. 

Now let us rigorously extract the universal non-extensive contribution to the anomaly phase.

Naively applying the original Euler-Maclaurin formula, we get
\begin{equation}
\begin{split}
\sum_{q_x}\varphi(q_x)=& L_x\int_0^{2\pi}\frac{dq_x}{2\pi}\varphi(q_x)-P_1(1)(\varphi(2\pi)-\varphi(0))\\
&+\frac{1}{2}\frac{2\pi}{L_x}P_2(1)(\varphi'(2\pi)-\varphi'(0))\\
&-\frac{1}{2}\frac{2\pi}{L_x}\int_{0}^{2\pi} dq_x P_2\left(\frac{q_xL_x}{2\pi}\right)\varphi''(q_x)
\end{split}
\label{euler_maclaurin_original}
\end{equation}
where $P_n$ are periodic Bernoulli polynomials, whose magnitude is $\mathcal{O}(1)$, satisfies $\int_0^1 dx P_n(x)=0$, $P_n(x+1)=P_n(x)$, and 
most importantly $P_n'(x)=nP_{n-1}(x)$. Since $\varphi$ is a periodic function, the $P_1, P_2$ terms vanish. It seems that we can always do integral by part
\begin{equation}
    \sum_{q_x}\varphi(q_x) \overset{\text{naive}}{=} \frac{1}{3!}\left(\frac{2\pi}{L_x}\right)^2\int_{0}^{2\pi} dq_x P_3\left(\frac{q_xL_x}{2\pi}\right)\varphi'''(q_x),
\end{equation}
and this process can be iterated. Thus no non-extensive part which is finite order polynomial in $1/L_x$ exists, just like the analysis in the end of \cref{ground_state_degeneracy}. However, notice that $\varphi$ in fact is not a smooth function, due to the discontinuity of $\varphi$ and $\varphi'$ at $q_c$. After taking this into account, the Euler-Maclaurin formula should read
\begin{equation}
\begin{split}
    \sum_{q_x}\varphi(q_x)=& L_x\int_0^{2\pi}\frac{dq_x}{2\pi}\varphi(q_x)\\[.1cm]
    &+P_1\left(\frac{q_cL_x}{2\pi}\right)(\varphi(q_c^+)-\varphi(q_c^-))-\frac{1}{2}\frac{2\pi}{L_x}P_2\left(\frac{q_cL_x}{2\pi}\right)(\varphi'(q_c^+)-\varphi'(q_c^-))\\[.1cm]
    &-\frac{1}{2}\frac{2\pi}{L_x}\int_{0}^{2\pi} dq_x P_2\left(\frac{q_xL_x}{2\pi}\right){\varphi''}_r(q_x)
\end{split}
\label{euler_maclaurin}
\end{equation}
where ${\varphi''}_r(q_x)$ is $\varphi''(q_x)$ but with all delta functions and derivative of delta functions at $q_c$ removed:
\begin{equation}
    {\varphi''}_r(q_x)=\varphi''(q_x)-(\varphi(q_c^+)-\varphi(q_c^-))\delta'(q_x-q_c)-(\varphi'(q_c^+)-\varphi'(q_c^-))\delta(q_x-q_c).
\end{equation}
The first line of \cref{euler_maclaurin} is the extensive in $L_x$ term, the second line are non-extensive terms that arise from the discontinuity at $q_c$, and we will see below the third line will capture the ``effective kink'' at $q_x=0$ that develops in \cref{anomaly_phase_beta}, \cref{anomaly_phase_c}.

From the second line of \cref{euler_maclaurin}, we can see the chiral edge mode contributes to $\mathcal{A}$ a non-universal term of order $\mathcal{O}(1)$ in $L_x$, and a non-universal term of order $\mathcal{O}(1/L_x)$, because $P_1, P_2$ are of order $1$, but their particular values are non-universal due to the detailed dependence on $q_c \mod 2\pi/L_x$. This explains our statement below \cref{grav_anomaly_full}. Fortunately, from \cref{varphi}, we can see $\varphi(q_x)$ is exponentially close to either $q_x\delta x$ or $0$ for most values of $q_x$---those that satisfy $|q_x|\gg 1/\beta c$ (and the correction is exponentially small, $e^{-\beta c q_x}$). Therefore, the terms in the second line of \cref{euler_maclaurin} are largely independent of $\beta$ up to exponentially small error (here we assumed $q_c$ to be $\mathcal{O}(1)$), ensuring that the treatment \cref{grav_anomaly_ratio} can remove these non-universal UV terms.

It remains to extract information from the last term of \cref{euler_maclaurin}. As $\varphi$ is close to either $q_x\delta x$ or $0$ up to error $e^{-\beta c q_x}$, we can see $\varphi''$ is exponentially small unless $q_x$ is of order $1/\beta c$ or below. Moreover, when $q_x=0$, $|\varphi(0)|=\delta x/(\beta c)$. This tells us that when $\beta c\gg 1$, ${\varphi''}_r$ is a bump function with a width of $1/\beta c$, and when $\beta c\to\infty$, it becomes a delta function $-\operatorname{sgn}(k)\delta x\delta(q_x)$. So in the $\beta \gg L_x/c\gg 1$ limit, we may estimate the last term as (all periodic Bernoulli polynomials are of order $1$)
\begin{equation}
    \frac{1}{12}\frac{2\pi}{L_x}\delta x\operatorname{sgn}(k)+\frac{1}{L_x}\mathcal{O}(L_x/\beta c)
\end{equation}
where the first term reflects the ``effective kink'' at $q_x=0$ that we saw developed in \cref{anomaly_phase_beta}, \cref{anomaly_phase_c}.
\footnote{The $\mathcal{O}(L_x/\beta c)$ comes from $(1/L_x)\int dq_x (L_x/(2\pi))P_2'(0) (q_xL_x/(2\pi)) e^{-q_x\beta c}$, where we use Taylor series to estimate $P_2(q_xL_x/(2\pi))-P_2(0)$.}
On the other hand, when $L_x/c\gg \beta \gg 1$ we can integral the last term in \cref{euler_maclaurin} by part and get
\begin{equation}
    \frac{1}{6}\left(\frac{2\pi}{L_x}\right)^2\int_{0}^{2\pi} dq_x P_3\left(\frac{q_xL_x}{2\pi}\right){{\varphi''}_r}'(q_x).
\end{equation}
And we know $|{{\varphi''}_r}'|$ is order $\beta^2 c^2$ with a width of $1/\beta c$. So the result of this integral is of order $\beta c/L_x^2$.

To sum up, given a ``low'' temperature $\beta_1\gg L_x/c\gg 1$ and an  ``intermediate high'' temperature $ L_x/c\gg\beta_2\gg 1$, we find the chiral edge mode contribution to $2\mathcal{A}_\text{IR}$ defined in \cref{grav_anomaly_ratio} to be
\begin{equation}
\begin{split}
&\sum_{q_x}\left[\left.\varphi(q_x)\right|_{\beta_1}-\left.\varphi(q_x)\right|_{\beta_2} \right] \\
=&L_x\int_0^{2\pi}\frac{dq_x}{2\pi}\left[\left.\varphi(q_x)\right|_{\beta_1}-\left.\varphi(q_x)\right|_{\beta_2} \right]\\[.2cm]
    &+\frac{1}{12}\frac{2\pi}{L_x}\delta x\operatorname{sgn}(k)+\frac{1}{L_x}\mathcal{O}(L_x/\beta_1 c)+\frac{1}{L_x}\mathcal{O}(\beta_2 c/L_x)+\mathcal{O}(1/L_x^2)+ \mathcal{O}(e^{-\beta c q_c}) \ .
\end{split}
\end{equation}
Now we can assemble the $\mathcal{A}_{\text{IR}}$. In \cref{anomaly}, we already mentioned that the $n_0$ term is manifestly independent of $\beta$, hence does not contribute to $\mathcal{A}_{\text{IR}}$. The bulk mode contributions are also unimportant, for two reasons: first, it is easy to see the bulk mode contribution to the phase does not develop discontinuity or kinks in $q_x$, hence making no leading non-extensive contribution to $\mathcal{A}$; moreover, the bulk mode always has a gap of order $mc^2\sim \mathcal{O}(1)$ and hence its contribution depends little on $\beta$, hence the contribution to $\mathcal{A}_{\text{IR}}$ is always exponentially suppressed by $\beta$. Therefore we have the universal IR result from the chiral edge mode
\begin{equation}
    \mathcal{A}_{\text{IR, leading non-extensive} }=\frac{2\pi}{24}\frac{\operatorname{sgn}(k)\delta x}{L_x}
\end{equation}
as desired. (Also note the extensive piece scales as $L_x (c\beta_2)^2$, much like the $L_x \Lambda^2$ in the continuum calculation.)

Let us summarize what has happened in physical terms. We look at the chiral edge mode (understood in the Lorentzian signature after Wick rotation) contribution to the phase $\mathcal{A}=\mathrm{arg} (Z_{\text{twisted}}/Z_{\text{no twist}})$. The contribution has: 1) a part that is extensive in $L_x$, 2) some non-universal non-extensive parts coming from $q_x=q_c$ at which the chiral edge mode merges into the bulk mode, and 3), when $\beta_1\gg L_x/c$ (i.e. zero temperature limit), a universal non-extensive part $(2\pi/24)(\operatorname{sgn}(k)\delta x/L_x)$ coming from $q_x=0$; on the other hand, when $L_x/c\gg\beta_2\gg 1$ (i.e. in the high temperature limit, as long as it is still much lower than the temporal lattice scale, or the bulk mode energy gap scale), the last part vanishes. Therefore, when we compare the difference between these two limits of temperatures, the non-universal non-extensive parts from $q_x=q_c$ cancel out, which can be interpreted as the removal of UV contribution. We are left with some residual extensive part, plus the $(2\pi/24)(\operatorname{sgn}(k)\delta x/L_x)$ from $q_x=0$, interpreted as the universal IR contribution. Just as in \cref{continuum_anomaly}, the coefficient $(2\pi/24)(\operatorname{sgn}(k)/L_x)$ in front of $\delta x$ can be interpreted as the universal IR contribution to zero-point momentum (``zero-point" since $\beta_1\gg L_xc/c$ corresponds to zero temperature).

If we switch the roles of the $x$- and the $\tau$-directions, i.e. perform a $\delta\tau$ rather than a $\delta x$ twist, the same calculation will extract the universal IR contribution to the thermal Hall current $(2\pi/24)(\operatorname{sgn}(k)/\beta^2)$ \cite{Kane:1996bje}.
\footnote{The universal IR phase we will get is $(2\pi/24)(\operatorname{sgn}(k) \delta \tau/\beta)$, which is interpreted as $\beta J^E_{\text{Hall}} \delta \tau$.}
This contribution will come from comparing the $L_{x, 1}\gg \beta c$ limit to the $\beta c\gg L_{x, 2}\gg 1$ limit---the former situation means taking the thermodynamical limit so that the chiral edge mode becomes a continuous spectrum, as is necessary in the calculation of the universal thermal Hall current \cite{Kane:1996bje}.

In retrospect, what we summarize in physical terms here is not surprising---all the calculations could have been performed in a much more familiar model in condensed matter physics: Consider a non-interacting integer Chern insulator on a spatial lattice, and perform the same $Z_{\text{twisted}}/Z_{\text{no twist}}$ calculation with continuous Euclidean time (which would further simplify the calculation compared to discrete Euclidean time), we will get the same result and same physical interpretation. The whole point of the calculation in this section, however, is indeed to reveal that the seemingly mysterious gravitation anomaly is manifested in a general level-$k$ CS-Maxwell theory on spacetime lattice in such a physical manner.

\

We have demonstrated the gravitational anomaly in a physical, UV complete manner, it is then natural to ask how to understand the framing anomaly when $\delta x=L_x$. More exactly, when we directly apply $\delta x=L_x$ in \cref{varphi} the phase obviously vanishes---or more intuitively, if we take $\delta x=L_x$ in \cref{anomaly_grid}, nothing has really been done to begin with. So in what sense can we get a framing anomaly phase of $(2\pi/24)\operatorname{sgn}(k)$? The problem here is more complicated than in the continuum, because here the extensive piece $L_x\int_0^{2\pi}\frac{dq_x}{2\pi}\left[\left.\varphi(q_x)\right|_{\beta_1}-\left.\varphi(q_x)\right|_{\beta_2} \right]$ no longer appears linear in $\delta x$ (see \cref{varphi}) as $\delta x$ becomes large. This means even if we extract the change of $\mathcal{A}_{\text{IR}}$ ``bit by bit'' from a $\delta x$ twist to a $\delta x+1$ twist, the change still depends on $\delta x$, making it trickier to analyze what has happened as $\delta x$ increases towards $L_x$.
\footnote{If the extensive piece were linear in $\delta x$ it would not be so hard to understand what has happened. We demonstrate the idea with a simplifying example. Pretend we are considering $\prod_{q_x} e^{iq_x \delta x}$ (with $e^{iL_x q_x}=1$) instead. When $\delta x$ is small, we will separate the resulting phase to an extensive in $L_x$ part and a non-extensive part, $\sum_{q_x} q_x\delta x=L_x(\pi\delta x)-\pi\delta x$. But when $\delta x=L_x$, it is not so reasonable to separate the two parts, as they really would add up together to ensure the phase is a multiple of $2\pi$. However, if we think of $\delta x$ as changing by $1$ each time, then we can clearly separate the extensive and the non-extensive part in the change $L_x \pi - \pi$ of the phase. In our actual problem, the ``extensive part'' itself depends non-linearly on $\delta x$ which will gradually grows towards $\delta x$, making the interpretation trickier.}
Moreover, it is unclear how an analysis along this line will be related to a lattice notion of ``trivialization of tangent bundle", in order to compare to the original understanding in the continuum \cite{Witten:1988hf}. So finding a good manifestation of the framing anomaly on the lattice {is an important task for a subsequent work} \cite{XuChen_preparation}.

\section{Wilson Loop Observable}
\label{Wilson_loop}
\subsection{Flux Attachment and Anyon Statistics}
The observables of a pure gauge theory are Wilson loops. In CS theory, a Wilson loop insertion corresponds to an anyon's worldloop, and (the phase of) the expectation value characterizes the anyon statistics---including mutual statistics and self statistics. The idea is, when we evaluate the Gaussian integral by finding the classical saddle, the saddle configuration is such that a gauge flux is ``attached'' to the vicinity of the anyon worldloop, so that the anyon's Aharonov-Bohm phase with the attached flux gives rise to the anyon statistics. Now we present this flux attachment process in our lattice theory.

In our lattice theory, Wilson loops are described by integers $W_l\in\mathbb{Z}$ on links, such that the lattice divergence on each vertex vanishes, $(\nabla\cdot W)_v=(\partial W)_v=0$. For example, $W_l$ can be $Q$ on a closed lattice loop $\gamma$ and $0$ elsewhere, so $\gamma$ is the anyon worldloop and $Q$ its charge; or $W_l$ can be $Q_1, Q_2$ on two closed lattice loops $\gamma_1, \gamma_2$ respectively, and zero elsewhere, so we have two anyon worldloops with two charges. The corresponding Wilson loop observable is
\begin{equation}
   e^{i\sum_l A_l W_l}.
\end{equation}
Note that $W_l\in\mathbb{Z}$ is needed for $A_l$ to be well-defined mod $2\pi$, and $(\nabla\cdot W)_v=0$ is needed for gauge invariance (if $\nabla\cdot W\neq 0$, then the expectation vanishes by Elitzur's theorem). When evaluating the expectation,
\begin{equation}
    \begin{split}
    &\left\langle e^{i\sum_l A_l W_l}\right\rangle \\
    =&\frac{1}{Z}\left[\prod_{\text{link} \: l} \int_{-\pi}^\pi \frac{dA_l}{2\pi}\right] \left[\prod_{\text{plaq.} \: p} \sum_{s_p\in\mathbb{Z}} \right]
    \left[\prod_{\text{cube} \: c} \int_{-\pi}^\pi \frac{d\lambda_c}{2\pi} \: e^{i\sum_c\lambda_c ds_c}\right] \ z_\chi[s]^k \\
    &\exp \left\{-\frac{1}{2e^2}\sum_p F_p^2+\frac{ik}{4\pi}\sum_c \left[(A\cup dA)_c - (A\cup 2\pi s)_c - (2\pi s\cup A)_c\right]\right\}\exp\left\{i\sum_{l} A_lW_l\right\} \ ,
\end{split}
\end{equation}
we can view $W_l$ as external source for the gauge field. (Here we used Euclidean signature. One may as well use Lorentzian signature, with some extra technical subtleties that we will comment on below.) 
In the below we will mostly focus on the local properties of the flux attachment, and take the spacetime lattice to be infinite, so that the $s_p$ can be dropped and $A_l$ turned into a real field, as we did in \cref{bulk_spectrum}. We will briefly explain the global aspects at the end of this subsection. 

Now that in infinite spacetime the path integral becomes Gaussian, it can be calculated by solving for its classical EoM
\begin{equation}
    (iK_E dA)_l= -W_l,
    \label{flux_attachment}
\end{equation}
where $iK_E$ is defined below \cref{M_E_def}. The solution for $dA$ is the flux attached to the Wilson loop, and we denote the solution for $A$ as $A^{\text{cl}}$. The expectation is then
\begin{equation}
    \left\langle e^{i\sum_l A_l W_l}\right\rangle=e^{(i/2) \sum_l A^\text{cl}_l W_l} \ .
\end{equation}
Note that since we are in the Euclidean signature, the solution $A^{\text{cl}}$ will contain some imaginary part, but this is not a problem because we really are just performing the Gaussian integral, upon substituting $A^{\text{cl}}$ back into the Gaussian. (In the Lorentzian signature, the solution $A^{\text{cl}}$ will be real, but due to the presence of the kernel of $iK$, i.e. the spectrum in \cref{bulk_spectrum}, the solution $A^{\text{cl}}$ will depend on the $i\epsilon$ prescription when inverting $iK$, which corresponds to the physical choice of initial and final conditions. This is the trade-off of technicalities between Euclidean and Lorentzian signature.)

\begin{figure}[htbp]
    \centering
    \begin{subfigure}[t]{0.44\textwidth}
        \centering
        \includegraphics[width=\textwidth]{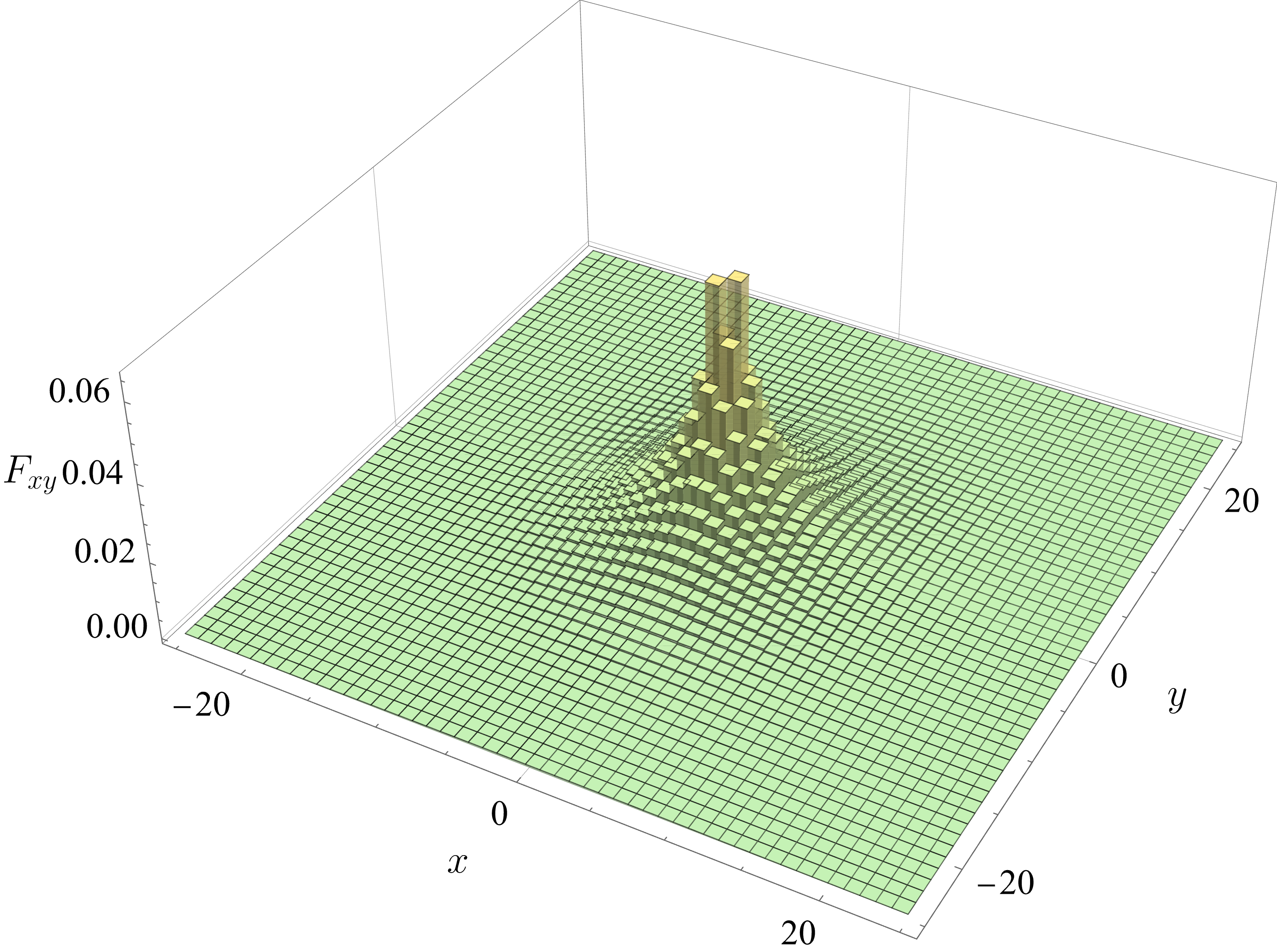}
        \caption{$e^2=1$}
    \end{subfigure}
    \begin{subfigure}[t]{0.44\textwidth}
        \centering
        \includegraphics[width=\textwidth]{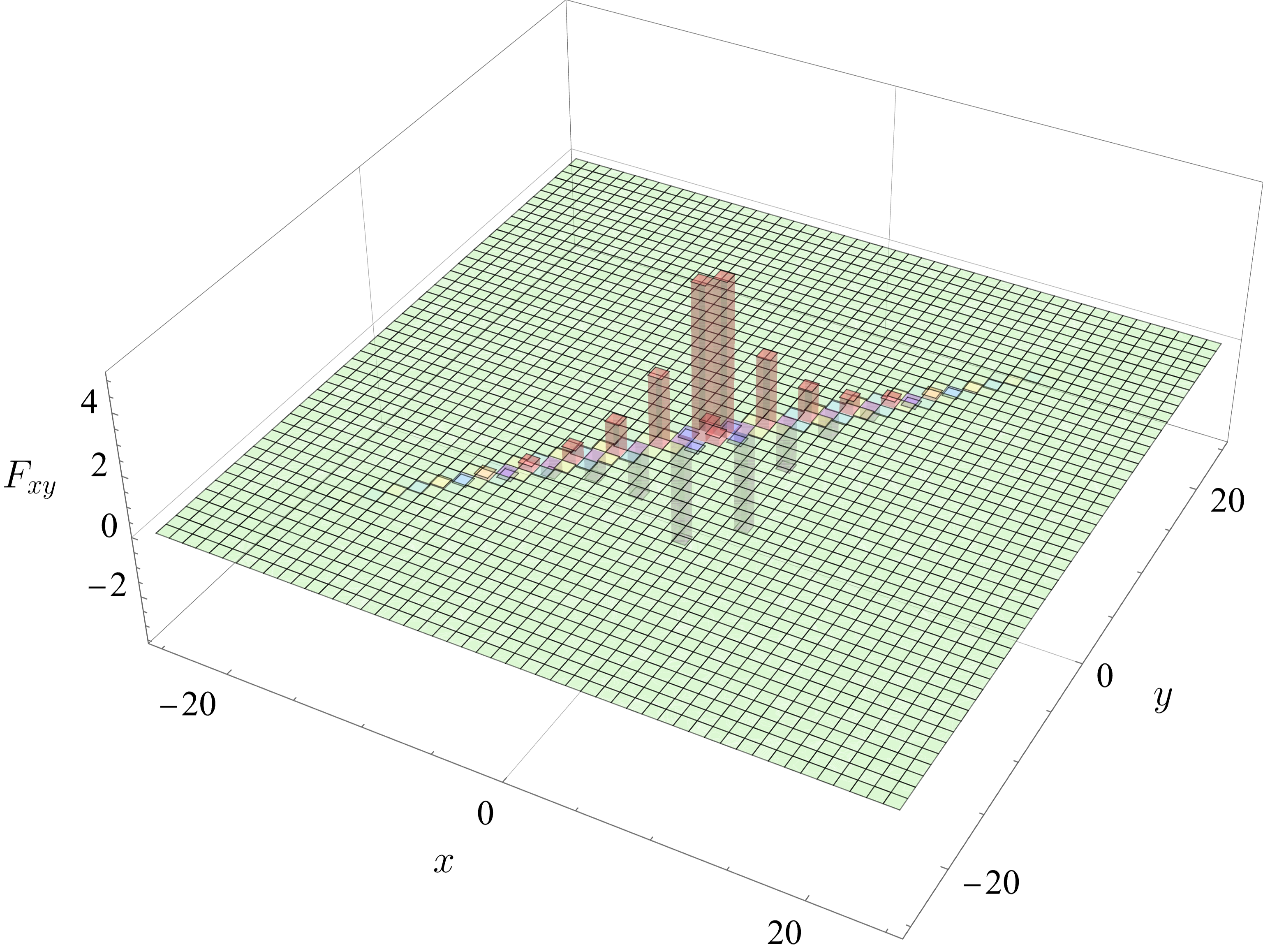}
        \caption{$e^2=64$}
    \end{subfigure}
    \caption{Magnetic flux attached to a single anyon static in space. We keep charge $Q=1$, $c=1$, $k=-1$ and tune $e^2$. When $k$ flips sign, the attached magnetic flux also flips sign. The total magnetic flux over the space sums up to $-2\pi Q/k$.}
    \label{flux_attachment_static}
\end{figure}

For Wilson loop(s) of generic shapes, the attached flux $dA^{\text{cl}}$ can always be found by inverting $iK_E$ in \cref{flux_attachment}. For the simplicity of presentation, in \cref{flux_attachment_static}, we consider a straight Wilson line running in the imaginary time direction, representing a single anyon staying at a fixed spatial position, and plot the strength of the magnetic flux---which takes real value---attached to its spatial vicinity. The total magnetic flux attached to the anyon sums up to $-2\pi Q/k$. In addition, there will be electric field (which we did not plot) circulating around the vicinity, with imaginary-valued field strength $\propto e^2$.

\begin{figure}[htbp]
    \centering
    \includegraphics[width=0.4\linewidth]{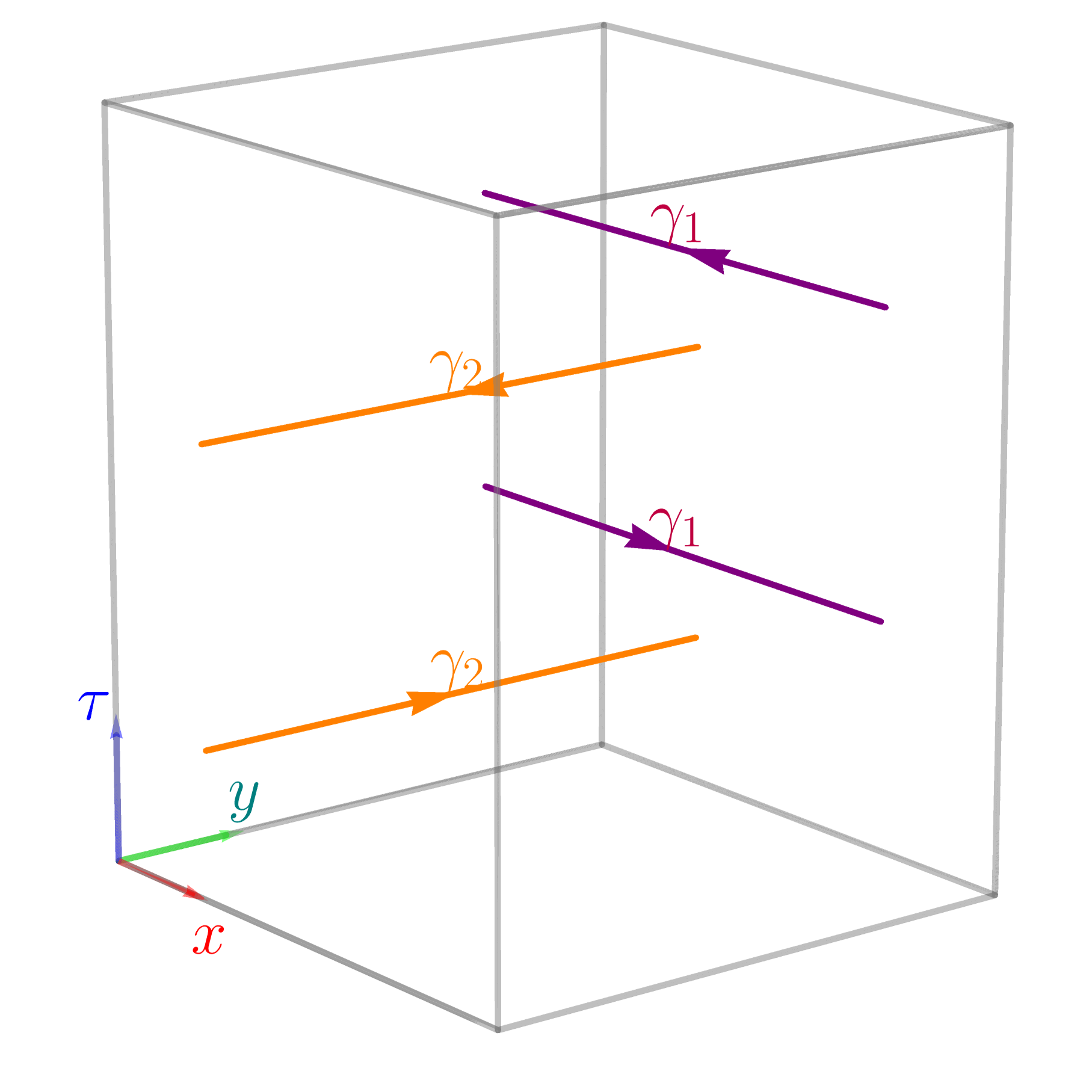}
    \caption{The setup for computing mutual statistics.}
    \label{mutual_stat}
\end{figure}

From the fact that the total magnetic flux attached to the anyon sums up to $-2\pi Q/k$, we can readily see the mutual statistical phase between well-separated Wilson loops (separated at a scale much larger than the smearing range of the attached flux) is indeed given by the Aharonov-Bohm phase
\begin{equation}
    e^{(i/2) \sum_l (A^\text{cl,(1)}_l W_l^{(2)} + A^\text{cl,(2)}_l W_l^{(1)})}=e^{(i/2) (-2\pi Q_1/k)Q_2+(i/2) (-2\pi Q_2/k)Q_1}=e^{i 2\pi Q_1Q_2/k}
\end{equation}
as expected, where we considered two Wilson loops $W=W^{(1)}+W^{(2)}$ along $\gamma_1, \gamma_2$ with charges $Q_1, Q_2$ as shown in \cref{mutual_stat}, and $A^\text{cl,(i)}$ is the flux attachment to $W^{(i)}$. The remaining contributions $e^{(i/2) \sum_l (A^\text{cl,(1)}_l W_l^{(1)} + A^\text{cl,(2)}_l W_l^{(2)})}$ are the self-statistics of each loop, which will we study below (and for \cref{mutual_stat} in particular the self-statistics would be trivial).

To compute the self-statistics on a single Wilson loop, we need to look into more details of the flux attachment. From \cref{flux_attachment_static}, we can see the interesting interpolation from Witten's point-split framing to Polyakov's geometrically framing as we decrease $e^2$, similar to what is known from the continuum \cite{Hansson:1989yu}. In the continuum, a ``framing'' of the Wilson loop is a regularization protocol to make the self-statistics phase well-defined (see below); on the lattice, since no further regularization should be needed, the ``framing'' process must arise automatically.

Let us begin with small $e^2$ for simplicity. The attached flux smears out with an exponential decay length of order $1/|m|c=2\pi c/|k|e^2$, as can be read-off from the behavior of \cref{def_G_0} at small $m^2 c^4$ and small $q$. Such smeared flux embodies Polyakov's geometrical framing \cite{Polyakov:1988md} as we compute $e^{(i/2) \sum_l A^\text{cl}_l W_l}$ for a single Wilson line. To be precise, in making this statement, we only looked at the real part of $A^\text{cl}$ associated with magnetic flux attached in the vicinity of the Wilson line; on the other hand, $A^\text{cl}$ also has an imaginary part associated with the imaginary electric flux in the vicinity of the Wilson line, and this part will contribute a suppression factor per unit length, which corresponds to a self-energy $\propto e^2$, that can be removed by a local counter term on the Wilson line if we want.

\begin{figure}[htbp]
    \centering
    \includegraphics[width=0.4\linewidth]{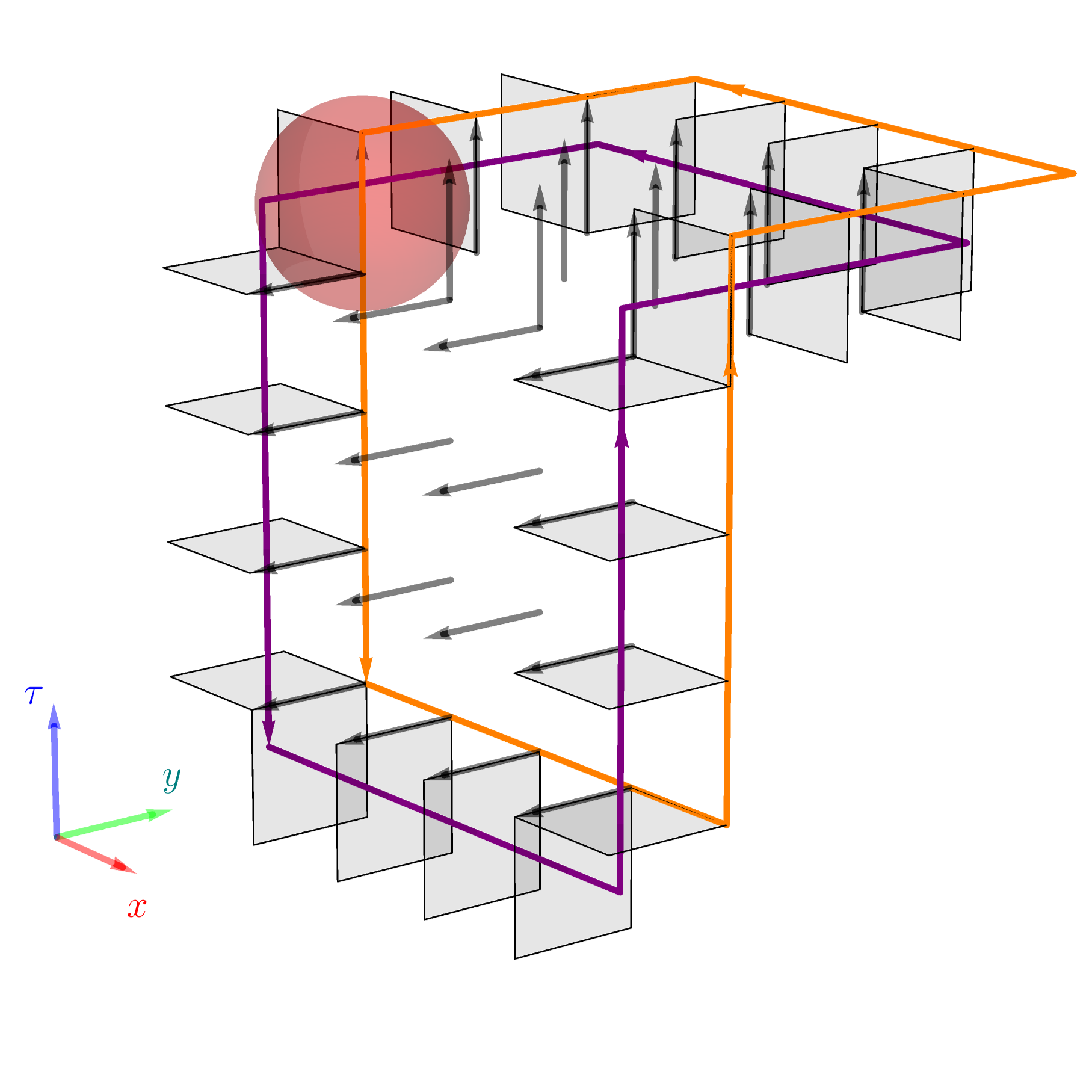}
    \caption{Linking number $\operatorname{link}(\tilde\gamma,\gamma)$ between the Wilson loop $\gamma$ (orange line on original lattice) and one of the attached flux tubes $\tilde\gamma$ (purple line on dual lattice); here we pictured a nearest flux tube. Gray plaquettes correspond to where $\sigma=1$, and gray arrows are the associated components in $A^{\text{cl}}$ (with fixed gauge choice). The circled part is where $A^{\text{cl}}_l W_l$ detects $\operatorname{link}(\tilde\gamma,\gamma)=-1$ in this example.}
    \label{self_linking}
\end{figure}

The flux attachment with large $e^2$ and its relation to Witten's point-split framing is more involved. In Witten's point-split framing regularization \cite{Witten:1988hf}, the $-2\pi Q/k$ attached flux is concentrated on a flux tube loop that is displaced slightly from the Wilson loop, and the detailed displacement can be chosen artificially as long as it is sufficiently small. On the lattice, when $e^2\gg 1$, the attached flux is not only splitted away from the anyon, moreover the $-2\pi Q/k$ total flux itself also further splits into many fluxes tubes, distributing along a displacement direction dictated by the cup product (which is the $\pm (\hat{x}+\hat{y}+\hat{\tau})$ direction for our choice of cup product, although the $\hat{\tau}$ part cannot be seen in \cref{flux_attachment_static}), so this is a more involved version of Witten's point-splitting framing. Let us denote this further split as (we focus on the real-valued magnetic flux part) 
\begin{equation}
    \operatorname{Re}(dA^{\text{cl}})_p=\frac{-2\pi Q}{k} \sum_{a} \lambda_a  (\sigma^a)_p
\end{equation}
where for each $a$, $(\sigma^a)_p$ traces out a flux tube $\tilde\gamma^a$ on the dual lattice, and $\lambda_a$ is the weight of this flux tube (subjected to $\sum_a \lambda_a=1$).
Note from \cref{flux_attachment_static} that as we move further away from the anyon, $|\lambda_a|$ exponentially decays over a length scale of $\delta x^2 |m|c= \delta x^2 |k|e^2/2\pi c$ (we have set the unit length $\delta x=1$) as can be read-off from the behavior of \cref{def_G_0} at large $m^2 c^4$ and around $q_x+q_y+q_\tau=\pi \mod 2\pi$; moreover the sign of $\lambda_a$ alternates. These features of $\lambda_a$ are reminiscence of the ``undesired zero mode'' that would have been there (recall from \cref{necessity_of_maxwell_term}) if the Maxwell coefficient $1/e^2=0$. Now, the self-statistics can be written as
\begin{equation}
    e^{(i/2) \sum_l \operatorname{Re} A^\text{cl}_l W_l} =e^{(i/2)(-2\pi Q^2/k)\sum_a \lambda_a\operatorname{link}(\tilde\gamma^a,\gamma)}\
\end{equation}
where the linking number between the Wilson loop $\gamma$ on the lattice and the flux tube loop $\tilde\gamma^a$ on the dual lattice as always well-defined, see \cref{self_linking}. When the shape of the Wilson loop is not changing rapidly over the length scale of $|m|c$, the linking numbers $\operatorname{link}(\tilde\gamma^a,\gamma)$ will be independent of $a$, so that we will retrive Witten's point-split framing result upon summing $\sum_a \lambda_a=1$. The self-statistics is $-\pi Q^2/k$ as expected. (And still, there is the imaginary part of $A^{\text{cl}}$ which contributes the self-energy suppression, but for this part there is no qualitative difference between large and small $e^2$.)

Finally we comment on what happens when the spacetime lattice has non-trivial topology (regardless of the signature). We can use the procedure demonstrated in \cref{ground_state_degeneracy} to make $A_l$ locally a real variable, meanwhile leaving the $\text{U(1)}$ flat holonomy $\delta A_l$ and some representative Dirac strings $[s]^{\text{rep}}_p$. The flat holonomy part $\delta A$ now appears as
\begin{equation}
     \exp \left\{\frac{ik}{4\pi}\sum_c \left[- (\delta A\cup 2\pi [s]^{\text{rep}})_c - (2\pi [s]^{\text{rep}}\cup \delta A)_c\right]+i\sum_l \delta A_l W_l\right\},
\end{equation}
playing the role of a Lagrange multiplier that enforces the constraint
\begin{equation}
     \sum_{p \text{ on any closed non-contractible surface}} \left(W_{l_p} - k [s]^{\text{rep}}_p \right) = 0
\end{equation}
where $l_p$ is the link associated with the plaquette $p$ via the cup product (which is, roughly speaking, ``perpendicular to'' the non-contractible surface). This constraint only has solution when the total number of anyons through the closed non-contractible surface is a multiple of $k$---a familiar physical conclusion. When the spacetime manifold has torsion (see \cref{torsion_in_homology_cohomology_theory} and \cref{a_more_rigorous_calculation_of_the_partition_function}), there is additional contribution to the phase of the expectation value from the linking number between the Wilson loop and the torsion loop, as has been known from the continuum. 

\subsection{Spin}
\label{spin}
The spin-statistics theorem applies to anyons, which means an anyon's spin must be equal to its self-statistics, up to a $1/2\pi$ conventional factor. In Witten's point-split framing \cite{Witten:1988hf}, this relation is manifested as the following. The linking number $\operatorname{link}(\tilde\gamma,\gamma)$ can change either due to a change of the shape of the Wilson loop $\gamma$ while fixing the displacement/framing direction between $\gamma$ and $\tilde\gamma$, or due to a local change of the displacement/framing direction of $\tilde\gamma$ from a given $\gamma$; the former process can create an anyon exchange, while the latter can create a local $2\pi$ rotation. Regardless of in which way the change of linking number occurs, the change of the phase of the Wilson loop expectation is always $-\pi Q^2/k$ times that change, hence manifesting the spin-statistics theorem.

On the lattice, as we have already seen, with large $e^2$, the phase of the Wilson loop expectation is again $-\pi Q^2/k$ times the point-split linking number. However, so far, the point-split displacement/framing convention is always fixed by the cup product, so it seems we can only discuss the first kind of process that corresponds to the self-statistics of anyon exchange, but hard to make sense of a local rotation process, hence hard to makes sense of the physical concept of spin.

\begin{figure}[htbp]
    \centering
    
    \begin{subfigure}[t]{0.3\textwidth}
    \centering
    \includegraphics[width=\linewidth]{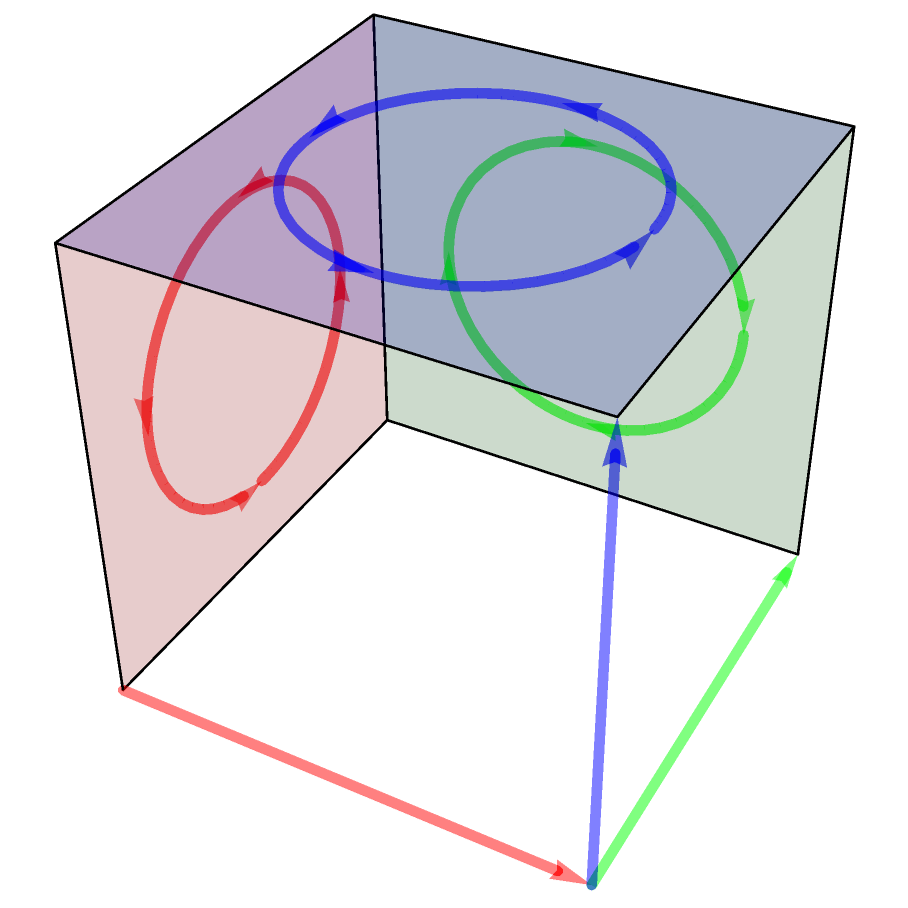}
    \caption{Rotated cup product}
    \label{trivial_rot_cup}
    \end{subfigure}
    \begin{subfigure}[t]{0.48\textwidth}
    \centering
    \includegraphics[width=\linewidth]{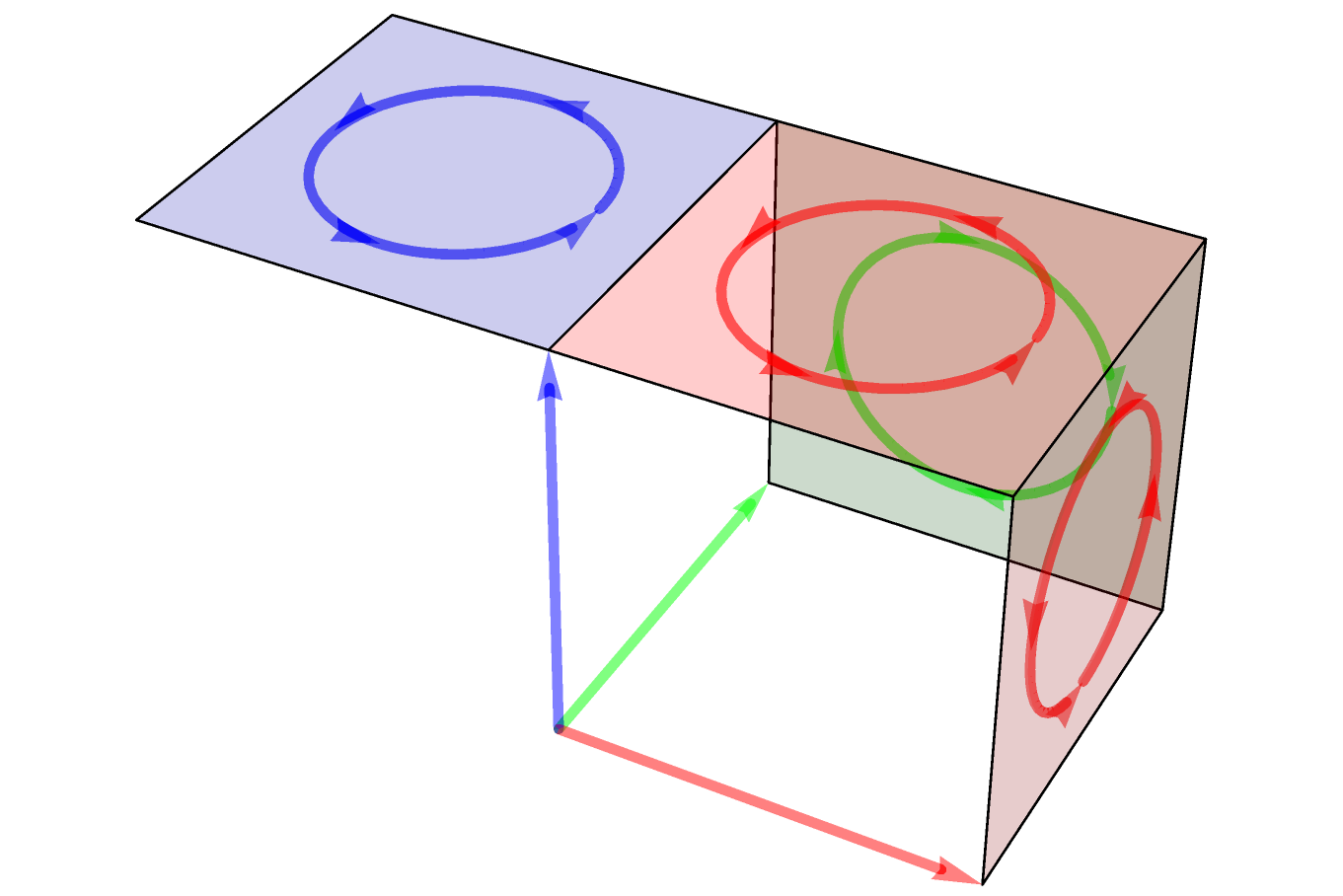}
    \caption{Intermediate layer}
    \label{rot_cup}
    \end{subfigure}
    \caption{Rotating the cup product}
\end{figure}

Now we show we can actually change the cup product convention from place to place on the lattice, so that a rotation of the framing can be generated. Originally our cup product convention is \cref{cup_product_on_cubic_lattice}, where the displacement is $(\hat{x}+\hat{y}+\hat{\tau})/2$. Suppose we use the original convention on cubes with $\tau<0$, but for cubes with $\tau>0$ we want to change the convention so that the displacement becomes $(-\hat{x}+\hat{y}+\hat{\tau})/2$, see \cref{trivial_rot_cup}. The problem is to retain the Leibniz rule of cup product, which is crucial for the gauge invariance of the theory. It turns out this is possible: all we need is that on the intermediate layer of cubes at $\tau=0$, we define the cup product as shown in \cref{rot_cup}. 

\begin{figure}[htbp]
    \centering
    \begin{subfigure}[t]{0.32\textwidth}
        \centering
        \includegraphics[width=\linewidth]{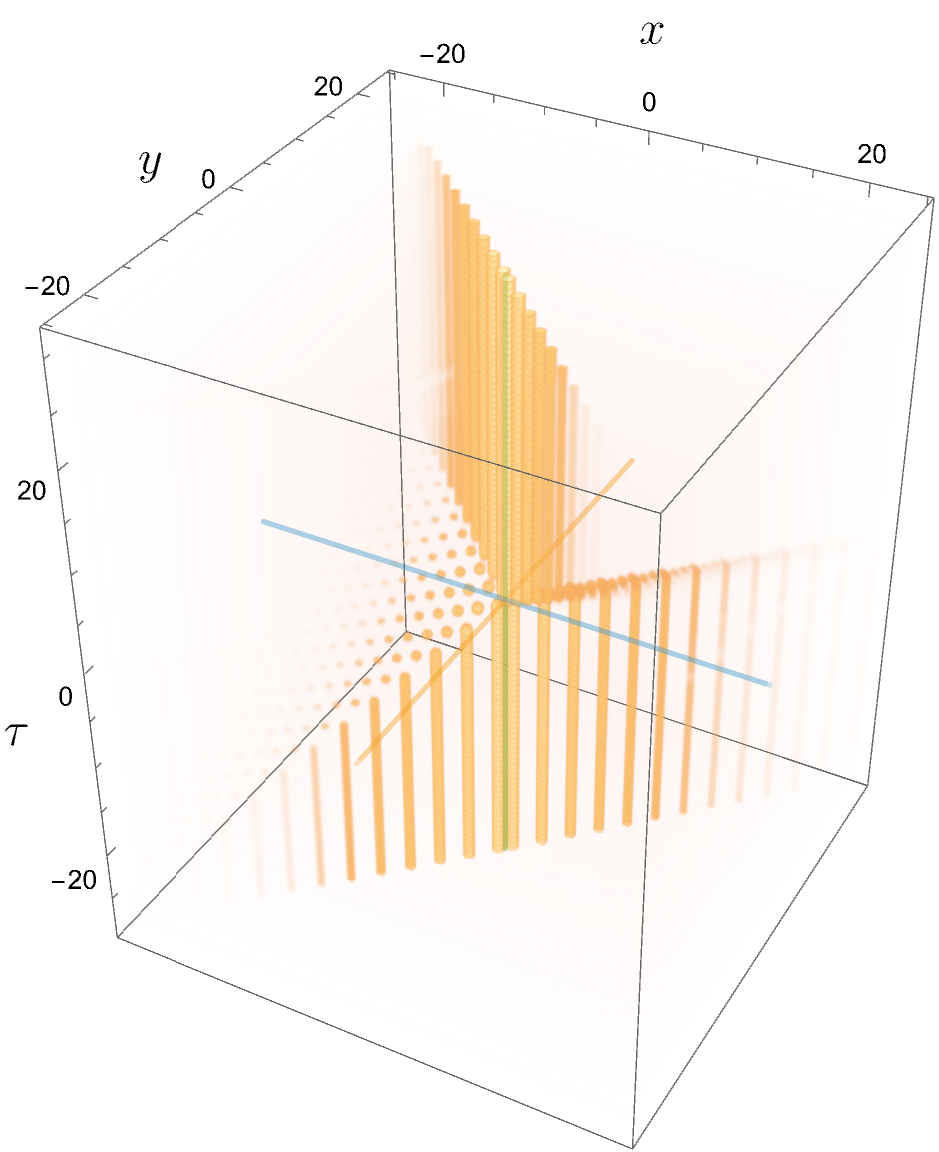}
        \caption{oblique view}
    \end{subfigure}
    \begin{subfigure}[t]{0.32\textwidth}
        \centering
        \includegraphics[width=\linewidth]{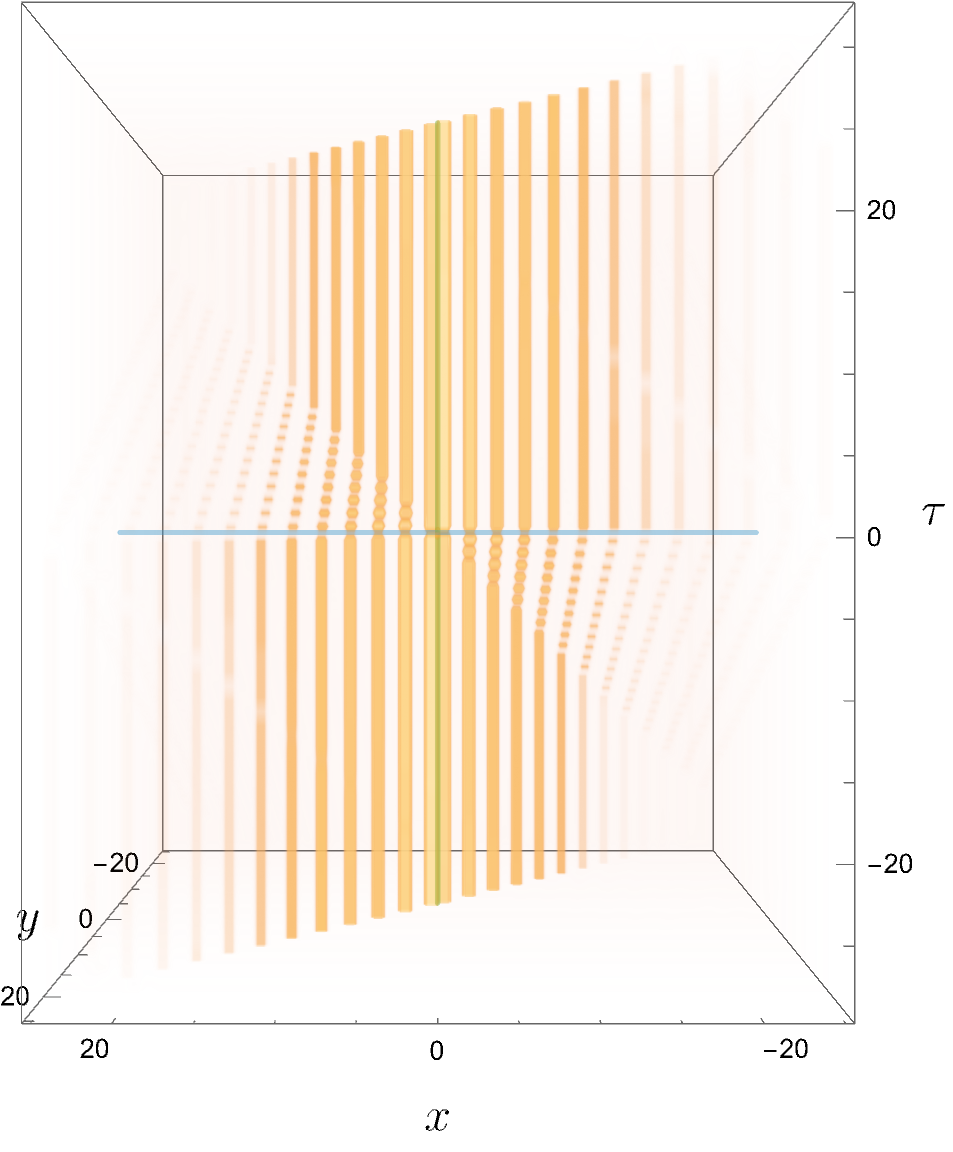}
        \caption{side view}
    \end{subfigure}
    \begin{subfigure}[t]{0.32\textwidth}
        \centering
        \includegraphics[width=\linewidth]{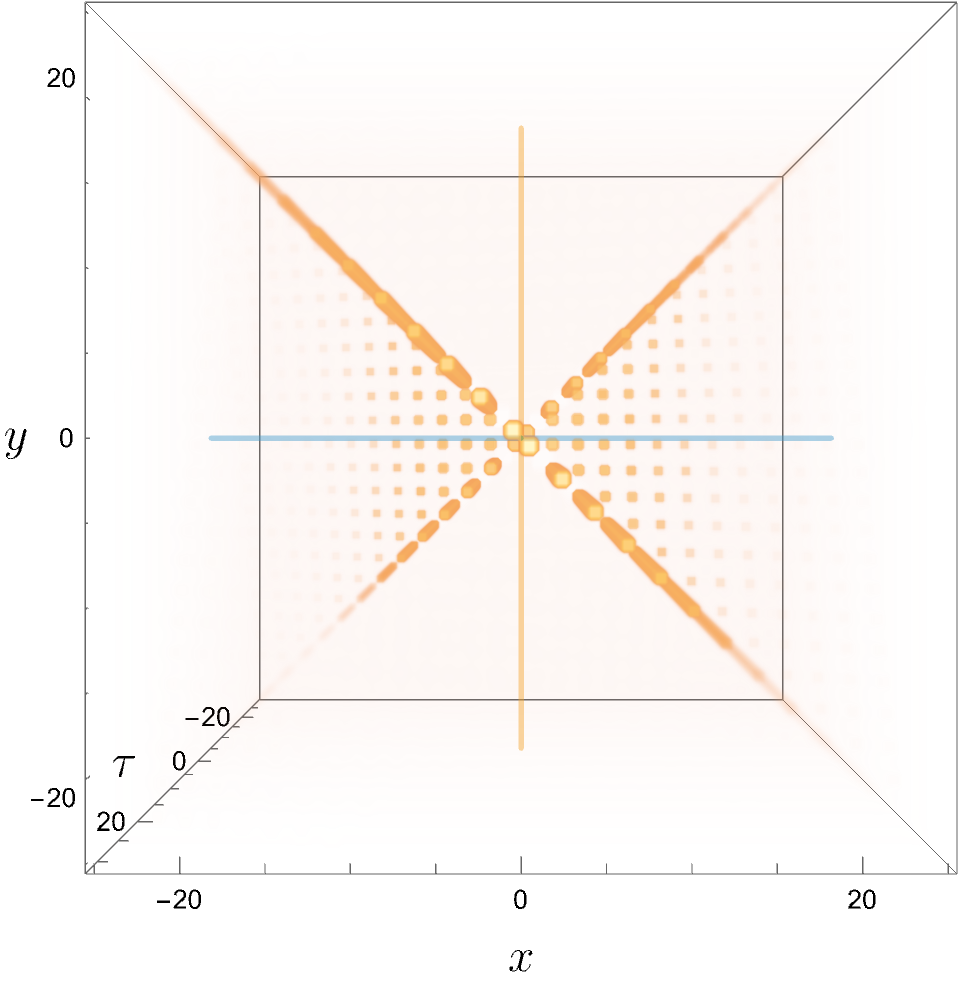}
        \caption{top view}
    \end{subfigure}
    \caption{Magnetic flux attached to a single anyon with $\tau$-dependent cup product.}
    \label{flux_attachment_spin}
\end{figure}

With this $\tau$-dependent cup product, we can numerically check that, for large $e^2$, the attached magnetic flux profile appears as \cref{flux_attachment_static} for very negative $\tau$, and gradually rotates counter-clockwise as $\tau$ increases, until a $\pi/2$ rotation is accumulated when $\tau$ becomes very positive. For the flux component located distance $r$ from the anyon, the rotation roughly takes a time span of $r/c$; and recall most of the flux is contained within $r\lesssim \Delta x^2 |m|c = \Delta x^2 |k| e^2/2\pi c$. See \cref{flux_attachment_spin} for the visualization. Repeating similar changes of the cup product for four times, we can accumulate a $2\pi$ rotation of the attached flux tubes around the Wilson line, hence generating a change of the linking number between the flux tubes and the Wilson line by $1$.

\section{Further Discussions}

We showed the Chern-Simons-Maxwell theory can be defined and solved on the lattice, and demonstrated the crucial properties of the theory. The manifestations of the chirality is particularly important, and for this purpose we presented a calculation, along with a physical interpretation, of the gravitational anomaly in relation to the chiral edge mode. The physical interpretation provided by the lattice is particularly illuminating for clarifying the subtleties in the usual continuum treatments.

Within the context of $\text{U(1)}$ Chern-Simons theory, now that almost all important properties expected from the continuum have been reproduced, the main remaining task is to find a good interpretation of framing anomaly on a finite lattice---even though we have already reproduced the closely related gravitational anomaly on an semi-infinite lattice. In the continuum \cite{Witten:1988hf}, the framing anomaly can be seen by, for instance, gluing the first and (the orientation reversal of) the third solid torus in \cref{twist} along their common boundary, which gives rise to the partition function on $S^2\times S^1$ but with a twisted trivialization of the tangent bundle. It is desirable to develop a similar lattice calculation which glues some twisted boundary to some untwisted boundary. One technical challenge is how to glue a twisted lattice grid (with $\delta x = L_x$ in \cref{anomaly_grid}) to an untwisted one, and some special treatment on the boundary similar to \cref{rot_cup} should be needed. {We will study this in a subsequent work} \cite{XuChen_preparation}. Another possible approach is to implement a procedure similar to that introduced in \cite{You:2015cga}. It will be conceptually important if such a calculation can be performed, accompanied with an explained connection to some suitable lattice notion of ``trivialization of tangent bundle".

Another task worth pursuing is, if we turn our spacetime lattice Lagrangian formalism to a spatial lattice Hamiltonian formalism (see \cref{hamiltonian_on_lattice}), the Hilbert space will not be a local product Hilbert space, but one with some 1-form $\mathbb{Z}$ gauge constraints and the familiar 0-form $\text{U(1)}$ gauge constraint. It will be interesting to see if such a setting can be relaxed to a local product Hilbert space (since a local product Hilbert space is usually desired for in-principle-physically-realizable microscopic models) with suitable energy penalties from which the desired constraint Hilbert space emerges at low energy. This turns out to be technically non-trivial but should be achievable, as has been demonstrated in doubled $\text{U(1)}$ Chern-Simons theory \cite{Han:2021wsx,Han:2022cnr}.

Finally, recall in the Introduction we said the Villainized Chern-Simons-Maxwell theory is a special solvable case within a much broader theme, i.e. that of refining lattice theories via category theory in order to better connect the lattice QFT to continuum QFT \cite{Chen:2024ddr}, especially at the topological level. A lattice construction for non-abelian Chern-Simons-Yang-Mills theory has been proposed through this approach \cite{Chen:2024ddr,Zhang:2024cjb}. While the non-abelian theory might not be solvable as what we have done for the abelian theory, it is still an important (and technically non-trivial) task to analyze the theory in suitable limits, and demonstrate that it has certain expected properties from the continuum theory.

\

\noindent \emph{Acknowledgement.} This work is supported by NSFC under Grants No.~12174213 and No.~12342501.

\

\appendix
\section{Hamiltonian Formalism on Lattice}
\label{hamiltonian_on_lattice}
We focus on the bosonic case, for which the partition function is
\begin{equation}
\begin{split}
    Z=&\left[\prod_{\text{link} \: l} \int_{-\pi}^\pi \frac{dA_l}{2\pi}\right] \left[\prod_{\text{plaq.} \: p} \sum_{s_p\in\mathbb{Z}} \right]
    \left[\prod_{\text{cube} \: c} \int_{-\pi}^\pi \frac{d\lambda_c}{2\pi}\right]
    e^{i\sum_c\lambda_c ds_c}\\
    &\exp \left\{-\frac{i}{2e^2}\sum_p \eta_p F_p^2+\frac{ik}{4\pi}\sum_c \left[(A\cup dA)_c - (A\cup 2\pi s)_c - (2\pi s\cup A)_c\right]\right\}
\end{split}
\end{equation}
with $k$ even. The treatment for the odd $k$ fermionic case is similar, except for involving an extra Majorana degree of freedom, in essentially the same way as in \cite{Han:2022cnr}, whose details will be omitted here.

We will use the coordinate representation $l=(r,\mu),p=(r,\mu\nu),c=(r,txy)$ of $l,p,c$ as in \cref{bulk_spectrum}. We will denote by $\Delta_\mu$ the forward lattice derivative, i.e. $\Delta_\mu X(r) = X(r+\hat{\mu})-X(r)$, while by $\nabla_\mu$ the backward lattice derivative, i.e. $\nabla_\mu X(r) = X(r)-X(r-\hat{\mu})$. We will often omit the $r$ label on the variables.

To arrive at the Hamiltonian formalism, it turns out we first need to restore the 1-form $\mathbb{Z}$ gauge discussed in \cref{lattice_model}, so that the $A_\mu$ will now be real variables; as we shall see, the 1-form $\mathbb{Z}$ gauge constraint will be recovered later. Next we Fourier transform the electric field $F_{ti}$ at each $p=(r, ti)$:
\begin{equation}
\begin{split}
    Z=&\left[\prod_{r,\mu} \int_{-\infty}^\infty \frac{dA_\mu}{2\pi}\right] \left[\prod_{r,\mu\nu} \sum_{s_{\mu\nu}\in\mathbb{Z}} \right]
    \left[\prod_{r} \int_{-\pi}^\pi \frac{d\lambda_{txy}}{2\pi}\right]
    e^{i\sum_r\lambda_{txy} (\Delta_t s_{xy}+\Delta_x s_{yt}+\Delta_y s_{tx})}\\
    &\left[\prod_{r,i} \sqrt{\frac{ie^2}{2\pi}}\int_{-\infty}^\infty d\Pi^i\right] e^{-i\sum_r {(\Pi^xF_{xt}+\Pi^yF_{yt})}}\\
    &\exp \bigg\{-i\sum_{r}
    \left[ \frac{e^2}{2} \left(\tilde{\Pi}^x\right)^2+ \frac{e^2}{2} \left(\tilde{\Pi}^y \right)^2 + \frac{c^2 F_{xy}^2}{2e^2} \right] + \frac{ik}{4\pi}\sum_r \left[A_t\cup F_{xy}- 2\pi s\cup A\right] \bigg\},
    \label{Canonical_Z}
\end{split}
\end{equation}
where $\Pi^i$ (living on $p=(r, ti)$ plaquettes) are the canonical momenta of $A_i$, and we defined the gauge invariant ``mechanical momenta'' as
\begin{equation}
\begin{aligned}
        \tilde\Pi^x(r)=\Pi^x(r)-\frac{k}{4\pi}A_y(r-\hat{y})\\
        \tilde\Pi^y(r)=\Pi^y(r)+\frac{k}{4\pi}A_x(r-\hat{x})
\end{aligned}
\end{equation}
whose classical equation of motion can be read-off to be
\begin{equation}
\begin{aligned}
        \left.\tilde\Pi^x(r) \right|_{\text{EoM}} &=\frac{1}{e^2}F_{tx}(r)\\
        \left.\tilde \Pi^y(r) \right|_{\text{EoM}} &=\frac{1}{e^2}F_{ty}(r) \ .
\end{aligned}
\end{equation}
\footnote{Note that as $e^2\rightarrow +\infty$, at EoM $\Pi^x$ approaches $(k/4\pi) A_y$ and $\Pi^y$ approaches $-(k/4\pi)A_x$. This recovers what is usually said in the continuum CS theory, that ``$A_x$ and $A_y$ are canonical variables''.}
Obviously, there is another pair of canonical variables in \cref{Canonical_Z}, namely the integer-valued $s_{xy}$ and the $\text{U(1)}$ valued $e^{i\lambda_{txy}}$. Moreover, $A_t$, $s_{yt}$, $s_{yt}$ appear linearly and without time derivative in the action; they will serve as Lagrange multipliers that give rise to constraints on the Hilbert space.

More explicitly, from \cref{Canonical_Z}, we can recognize that, as we pass on to spatial lattice (with spatial coordinates ${\bf r}$) and continuous time, the full Hilbert space consists of a real field on each spatial link $({\bf r}, i)$ and a rotor on each spatial plaquette $({\bf r}, xy)$:
\begin{align}
    [{A}_i({\bf r}),{\Pi}^j({\bf r}')]&=i\delta_{{\bf r},{\bf r}'}\delta_i^j\\
    [{s}_{xy}({\bf r}),e^{i{\lambda}_{txy}({\bf r}')}]&=\delta_{{\bf r},{\bf r}'} e^{i{\lambda}_{txy}({\bf r})} \ .
\end{align}
The full Hilbert space is subjected to some constraints. First is the 1-form $\mathbb{Z}$ gauge constraint on each spatial link, from summing out $s_{xt}$ and $s_{yt}$ in \cref{Canonical_Z}:
\begin{equation}
\begin{aligned}
    \exp i\left(2\pi\Pi^x({\bf r})+\nabla_y\lambda_{txy}({\bf r})+(k/2) {A}_y({\bf r}+\hat{x} \right) &=1\\[.2cm]
    \exp i\left(2\pi\Pi^y({\bf r})-\nabla_x\lambda_{txy}({\bf r})-(k/2) {A}_x({\bf r}+\hat{y})\right) &=1
\end{aligned}
\end{equation}
where the left-hand-side is the generator of the 1-form $\mathbb{Z}$ gauge transformation on each spatial link. Then there is also the familiar 0-form $\text{U(1)}$ gauge constraint on each spatial vertex, from integrating out $A_t$ in \cref{Canonical_Z}:
\begin{equation}
     \nabla_x \Pi^x({\bf r})+\nabla_y \Pi^y({\bf r})+\frac{k}{4\pi}( d{A}_{xy}({\bf r})-2\pi {s}_{xy}({\bf r}))-\frac{k}{2} {s}_{xy}({\bf r}-\hat{x}-\hat{y})=0
\end{equation}
where the left-hand-side is the generator of the 0-form $\text{U(1)}$ gauge transformation on each spatial vertex.
\footnote{This looks like an $\mathbb{R}$ constraint, but given the 1-form $\mathbb{Z}$ constraints above it reduces to a $\text{U(1)}$ constraint.}
Alternatively, this can be expressed in terms of the ``mechanical momenta" as
\begin{equation}
     \nabla_x {\tilde\Pi}^x({\bf r})+\nabla_y {\tilde\Pi}^y({\bf r})+\frac{k}{4\pi}{F}_{xy}({\bf r})+\frac{k}{4\pi}{F}_{xy}({\bf r}-\hat{x}-\hat{y})=0 \ .
\end{equation}
Finally, the Hamiltonian is
\begin{equation}
    {H}=\sum_{{\bf r}} \left[ \frac{e^2}{2} \left({\tilde{\Pi}}^x\right)^2+ \frac{e^2}{2} \left({\tilde{\Pi}}^y \right)^2 + \frac{c^2 {F}_{xy}^2}{2e^2} \right] \ .
\end{equation}
Note the gauge constraints commute with each other, and they all commute with the Hamiltonian.

In \cite{Han:2021wsx, Han:2022cnr}, it is shown that, in doubled $\text{U(1)}$ CS theory (where a Maxwell term is not needed), one can relax the strict gauge constraints into energy penalties accompanied with some mildly gauge non-invariant terms. It will be interesting to see if the same can be achieved here, for single chiral $\text{U(1)}$ CS-Maxwell theory. But this technical analysis is beyond the scope of the present work.

\

\emph{\textbf{Note}}: Ref.~\cite{Peng:2024xbl}, which appeared as this paper was being finalized, also had the same Lagrangian as ours and derived a Hamiltonian formulation. Compared to our Hamiltonian formulation above {which emphasizes manifest locality}, the Hamiltonian formulation in \cite{Peng:2024xbl} did not have a 1-form $\mathbb{Z}$ gauge constraint on each spatial link, but rather fixed this local 1-form $\mathbb{Z}$ gauge by setting the condition ${s}_{xy}=0$ on most except for one plaquette on each connected component of the spatial lattice. {Then they mainly focused on the topological sector where that last $s_{xy}$ is $0$ as well;  when including all topological sectors, the implementation will require non-local information of the space, i.e. its topology and/or global size.}

\section{Explicit Form of Operators and Structures}
\label{explicit_form_of_operators_and_structures}
For Lorentzian signature, after Fourier transformation, denoting 1-form $A=(A_t,A_x,A_y)^T$ and 2-form $d_1A=((d_1A)_{xy}, (d_1A)_{yt}, (d_1A)_{tx})^T=((d_1A)^t,(d_1A)^x,(d_1A)^y)^T$, we have
\begin{align}
    d_0(q) & =
    \begin{bmatrix}
        e^{iq_t}-1     \\
        e^{iq_x}-1 \\
        e^{iq_y}-1
    \end{bmatrix} \\[.2cm]
    d_1(q) & =
    \begin{bmatrix}
        0                                                      & -e^{iq_y}+1 & e^{iq_x}-1 \\
        e^{iq_y}-1  & 0                                                      & -e^{iq_t}+1      \\
        -e^{iq_x}+1 & e^{iq_t}-1     & 0
    \end{bmatrix} \\[.2cm]
    d_2(q) & =
    \begin{bmatrix}
        e^{iq_t}-1 & e^{iq_x}-1 & e^{iq_y}-1
    \end{bmatrix}
\end{align}
\begin{align}
    \eta            & =
    \begin{bmatrix}
        (1-i\epsilon)c^2 &    &    \\
            & -(1+i\epsilon) &    \\
            &    & -(1+i\epsilon)
    \end{bmatrix}                     \\[.2cm]
    \cup(q)            & =\begin{bmatrix}
        e^{iq_t} & 0                                                                                              & 0                                                                                              \\
        0                                                                                              & e^{iq_x} & 0                                                                                              \\
        0                                                                                              & 0                                                                                              & e^{iq_y}\\
    \end{bmatrix}.
\end{align}
Note that $q_t=-q^t=-\omega$. When calculating the specturum of classical EoM, the $i\epsilon$ prescription in $\eta$ will be neglected. For Euclidean signature, $q_t$ becomes $q_\tau=q^\tau$, and $\eta$ becomes
\begin{equation}
        \eta_{\text{E}} =
    \begin{bmatrix}
        c^2 &   &   \\
            & 1 &   \\
            &   & 1
    \end{bmatrix}                   
\end{equation}
(the $i\epsilon$ prescription in the Lorentzian $\eta$ is just $-i\epsilon\eta_{\text{E}}$) where we will mostly take $c^2=1$ when working with Euclidean signature.

In our Fourier transformation, if we choose the coordinate of other points on a link (such as its middle point) to label the link, the Fourier transformed link variable will differ by some matrix action $T_1$ on its left; similarly the Fourier transformed plaquette variable will change by some $T_2$. We have
\begin{equation}
    S=
    \frac{1}{V}\sum_{k}
    -\frac{1}{2e^2}(A^\dagger T_1^\dagger) (T_1 d^\dagger_1 T_2^\dagger) (T_2\eta T_2^\dagger)(T_2 d_1 T_1^\dagger )(T_1A)
    +\frac{k}{4\pi}(A^\dagger T_1^\dagger) (T_1\cup T_2^\dagger)(T_2d_1T_1^\dagger) (T_1 A).
\end{equation}
So $M$ will transform to $T_1^\dagger MT_1$. Since the transformation is unitary, the eigenvalues will not change.

\section{Proof of the Ansatz \cref{d_1A}}
\label{proof_of_ansatz}
From the picture \cref{diagrammatic_representation_of_EoMs} we can see the EoM $(iKd_1A)^{t,x}(\omega,q_x,y)=0$ only involves $d_1A$ at $y$ and $y-1$, while $(iK d_1A)^{y}(\omega,q_x,y)=0$ only involves $d_1A$ at $y$ and $y+1$ (recall that links and plaquettes are labeled by the point with smallest $y$). So both lines of \cref{eom_on_lattice_with_edge} only involve $d_1 A$ at $y$ and $y-1$, and therefore \cref{eom_on_lattice_with_edge} can be recognized as a first-order linear recurrence relation, i.e. transfer matrix equation,
\begin{align}
    R(\omega, q_x) (d_1A)(\omega,q_x,y)-T(\omega, q_x) (d_1A)(\omega,q_x,y-1) = 0 \ \ \ \ \forall y>0
\end{align}
where $R,T$ are some $3\times 3$ matrices independent of $y$. Therefore
\begin{align}
(d_1A)(\omega,q_x,y)=\left[(R^{-1}T)(\omega, q_x)\right]^y \: (d_1A)(\omega,q_x,0)  \ \ \ \ \forall y\geq 0 \ .
\end{align}
(Here the $y$ superscript is the power to the integer $y$, not the $y$ component index.) This means we can diagonalize $R^{-1}T$ and find solutions that decompose $d_1 A$ into eigenmodes
\begin{equation}
    (d_1A)(\omega,q_x,y)=\sum_{q_y}e^{iq_y y}(d_1A)(\omega,q_x, q_y)
\end{equation}
where, naively, the summation is over all three eigenvalues $e^{iq_y}$ of the $3\times 3$ matrix $(R^{-1}T)(\omega, q_x)$. However, because $d_2 d_1 A=0$, we can say $d_1 A$ really only has two rather than three independent components (just like in the continuum, given $E_x$ and $E_y$, we have $B=(q_x E_y-q_y E_x)/\omega$), and the matrix equation is essentially $2\times 2$. Moreover, according to the discussion below \cref{solving_q}, out of the two remaining modes, we have two possibilities: Depending on $\omega, q_x$, either we have two modes both with $|e^{iq_y}|=1$, which are the bulk modes that we have already seen in \cref{bulk_spectrum}; or we have a decaying mode with $|e^{iq_y}|<1$, which is the edge mode we are interested in, and a diverging mode with $|e^{iq_y}|>1$, which we will discard.

After having this transfer matrix picture in mind and confirming the validity of the ansatz \cref{d_1A}, it turns out we do not need to actually write out the details of $R$ and $T$. We can just work with $K$ as in the main text to solve for the decaying mode $e^{iq_y}$ in the ansatz \cref{d_1A}.

\section{Torsion in Homology and Cohomology, and Reidemeister Torsion}
\label{torsion_in_homology_cohomology_theory}
In this appendix we review the torsion in homology and cohomology. Although the three-torus we consider in the main text has trivial torsion, we believe it is still helpful to have the more general cases in mind, especially for the purpose of understanding the Jacobians of the exterior derivatives and the Reidemeister torsion.

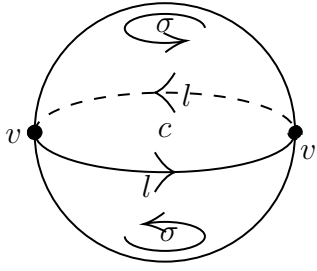
\begin{figure}[htbp]
    \centering

\tikzset{every picture/.style={line width=0.75pt}} 

\begin{tikzpicture}[x=0.75pt,y=0.75pt,yscale=-1,xscale=1]

\draw   (50,175) .. controls (50,139.1) and (79.1,110) .. (115,110) .. controls (150.9,110) and (180,139.1) .. (180,175) .. controls (180,210.9) and (150.9,240) .. (115,240) .. controls (79.1,240) and (50,210.9) .. (50,175) -- cycle ;
\draw   (110,185) .. controls (113.33,190.56) and (116.67,193.89) .. (120,195) .. controls (116.67,196.11) and (113.33,199.44) .. (110,205) ;
\draw   (120,165) .. controls (116.67,159.44) and (113.33,156.11) .. (110,155) .. controls (113.33,153.89) and (116.67,150.56) .. (120,145) ;
\draw  [draw opacity=0][dash pattern={on 4.5pt off 4.5pt}] (50,175) .. controls (50,175) and (50,175) .. (50,175) .. controls (50,163.95) and (79.1,155) .. (115,155) .. controls (150.9,155) and (180,163.95) .. (180,175) -- (115,175) -- cycle ; \draw [dash pattern={on 4.5pt off 4.5pt}] [dash pattern={on 4.5pt off 4.5pt}]  (50,175) .. controls (50,175) and (50,175) .. (50,175) .. controls (50,163.95) and (79.1,155) .. (115,155) .. controls (150.9,155) and (180,163.95) .. (180,175) ; \draw [shift={(180,175)}, rotate = 72.42] [color={rgb, 255:red, 0; green, 0; blue, 0 }  ][fill={rgb, 255:red, 0; green, 0; blue, 0 }  ][dash pattern={on 4.5pt off 4.5pt}][line width=0.75]      (0, 0) circle [x radius= 3.35, y radius= 3.35]   ; \draw [shift={(50,175)}, rotate = 270] [color={rgb, 255:red, 0; green, 0; blue, 0 }  ][fill={rgb, 255:red, 0; green, 0; blue, 0 }  ][dash pattern={on 4.5pt off 4.5pt}][line width=0.75]      (0, 0) circle [x radius= 3.35, y radius= 3.35]   ;
\draw  [draw opacity=0] (180,175) .. controls (180,175) and (180,175) .. (180,175) .. controls (180,175) and (180,175) .. (180,175) .. controls (180,186.05) and (150.9,195) .. (115,195) .. controls (79.1,195) and (50,186.05) .. (50,175) -- (115,175) -- cycle ; \draw    (180,175) .. controls (180,175) and (180,175) .. (180,175) .. controls (180,186.05) and (150.9,195) .. (115,195) .. controls (79.1,195) and (50,186.05) .. (50,175) ; \draw [shift={(50,175)}, rotate = 252.42] [color={rgb, 255:red, 0; green, 0; blue, 0 }  ][fill={rgb, 255:red, 0; green, 0; blue, 0 }  ][line width=0.75]      (0, 0) circle [x radius= 3.35, y radius= 3.35]   ; 
\draw  [draw opacity=0] (125.69,128.87) .. controls (122.6,129.59) and (118.93,130) .. (115,130) .. controls (103.95,130) and (95,126.73) .. (95,122.7) .. controls (95,118.67) and (103.95,115.4) .. (115,115.4) .. controls (125.8,115.4) and (134.6,118.52) .. (134.99,122.43) -- (115,122.7) -- cycle ; \draw    (123.72,129.27) .. controls (121.08,129.74) and (118.13,130) .. (115,130) .. controls (103.95,130) and (95,126.73) .. (95,122.7) .. controls (95,118.67) and (103.95,115.4) .. (115,115.4) .. controls (125.8,115.4) and (134.6,118.52) .. (134.99,122.43) ;  \draw [shift={(125.69,128.87)}, rotate = 174.94] [color={rgb, 255:red, 0; green, 0; blue, 0 }  ][line width=0.75]    (10.93,-4.9) .. controls (6.95,-2.3) and (3.31,-0.67) .. (0,0) .. controls (3.31,0.67) and (6.95,2.3) .. (10.93,4.9)   ;
\draw  [draw opacity=0] (104.31,221.13) .. controls (107.4,220.41) and (111.07,220) .. (115,220) .. controls (126.05,220) and (135,223.27) .. (135,227.3) .. controls (135,231.33) and (126.05,234.6) .. (115,234.6) .. controls (103.95,234.6) and (95,231.33) .. (95,227.3) -- (115,227.3) -- cycle ; \draw    (106.28,220.73) .. controls (108.92,220.26) and (111.87,220) .. (115,220) .. controls (126.05,220) and (135,223.27) .. (135,227.3) .. controls (135,231.33) and (126.05,234.6) .. (115,234.6) .. controls (103.95,234.6) and (95,231.33) .. (95,227.3) ;  \draw [shift={(104.31,221.13)}, rotate = 354.94] [color={rgb, 255:red, 0; green, 0; blue, 0 }  ][line width=0.75]    (10.93,-4.9) .. controls (6.95,-2.3) and (3.31,-0.67) .. (0,0) .. controls (3.31,0.67) and (6.95,2.3) .. (10.93,4.9)   ;

\draw (34,172) node [anchor=north west][inner sep=0.75pt]   [align=left] {$\displaystyle v$$ $};
\draw (181,180) node [anchor=north west][inner sep=0.75pt]   [align=left] {$\displaystyle v$$ $};
\draw (122,149) node [anchor=north west][inner sep=0.75pt]   [align=left] {$\displaystyle l$$ $};
\draw (102,195) node [anchor=north west][inner sep=0.75pt]   [align=left] {$\displaystyle l$$ $};
\draw (109,116) node [anchor=north west][inner sep=0.75pt]   [align=left] {$\displaystyle \sigma $};
\draw (111,221) node [anchor=north west][inner sep=0.75pt]   [align=left] {$\displaystyle \sigma $};
\draw (110,169) node [anchor=north west][inner sep=0.75pt]   [align=left] {$\displaystyle c$};

\end{tikzpicture}
    \caption{Triangulation of $\mathbb{R}P^3$}
    \label{triangulation_of_RP3}
\end{figure}

When calculating homology and cohomology groups of a topological space, we can use different kinds of coefficient. $\mathbb{Z}$ and $\mathbb{R}$ are the most commonly used choices. Take the homology case for example. One may naively expect they are essentially the same, and the difference is just changing from some $\mathbb{Z}^{b_i}$ to $\mathbb{R}^{b_i}$ where $b_i$ is the $i$th Betti number. However, (co)homology groups in coefficient $\mathbb{Z}$ contain something more. They are the torsion part of (co)homology groups. A basic example is $\mathbb{R}P^3$ (the topological space of $SO(3)$), where each point represents a straight line through $0$ in $\mathbb{R}^4$; more intuitively, we can picture it as a 3$d$ ball $D^3$ but with antipodal points on the $S^2$ surface identified. This topological space can be triangulated as \cref{triangulation_of_RP3}. It turns out the first homology group $H_1(\mathbb{R}P^3; \mathbb{Z})$ contains torsion: we can calculate it as $H_1(\mathbb{R}P^3)=\mathbb{Z}(l)/\mathbb{Z}(\partial \sigma=2l)=\mathbb{Z}/2\mathbb{Z}$. While if we use $\mathbb{R}$ as coefficient, we get $H_1(\mathbb{R}P^2; \mathbb{R})=\mathbb{R}(l)/\mathbb{R}(\partial \sigma=2l)=0$ because $2$ has an inverse in $\mathbb{R}$ (but not in $\mathbb{Z}$). In general, a $\mathbb{Z}$ coefficient homology group is a direct sum of a ``torsion part'' and a ``free part'', where the ``torsion part''  is some finite abelian group, while the ``free part'' is some infinite abelian group of the form $\mathbb{Z}^{b_i}$. The $\mathbb{R}$ coefficient homology only sees the free part $\mathbb{R}^{b_i}$.

Torsion will also show up in the cohomology, through the universal coefficient theorem, which says there is a short exact sequence
\begin{equation}
\begin{tikzcd}
	0 & {\operatorname{Ext}^1(H_{i-1}(M;\mathbb{Z}),A)} & {H^i(M;A)} & {\operatorname{Hom}(H_i(M;\mathbb{Z}),A)} & 0
	\arrow[from=1-1, to=1-2]
	\arrow[from=1-2, to=1-3]
	\arrow[from=1-3, to=1-4]
	\arrow[from=1-4, to=1-5]
\end{tikzcd},
\end{equation}
where $A$ is any coefficient group. For $A=\mathbb{Z}$, the $\operatorname{Ext}$ part basically counts the torsion part of $H_{i-1}$, while the $\operatorname{Hom}$ part counts the free part of $H_{i}$. For $A=\mathbb{R}$, the $\operatorname{Ext}$ part becomes trivial.

Now, let us see in general where the torsion may show up for an orientable $3$d manifold $M$. For orientable manifold, we have a further constraint $H^{3-i}(M; \mathbb{Z})\cong H_i(M; \mathbb{Z})$ from the Poincare duality. Now we consider $i=0, 1, 2, 3$. We know $H_0(M; \mathbb{Z})$ just counts how many connected components $M$ has, so it is torsion free, and hence so is $H^3(M; \mathbb{Z})$. On the other hand, $\cong H^2(M; \mathbb{Z})$, and we have already seen through an example that here might contain non-trivial torsion. And for $H_2(M;\mathbb{Z})\cong H^1(M;\mathbb{Z})$, using the universal coefficient theorem, we have a short exact sequence
\begin{equation}
\begin{tikzcd}
	0 & {\operatorname{Ext}^1(H_0(M;\mathbb{Z}),\mathbb{Z})} & {H^1(M;\mathbb{Z})} & {\operatorname{Hom}(H_1(M;\mathbb{Z}),\mathbb{Z})} & 0
	\arrow[from=1-1, to=1-2]
	\arrow[from=1-2, to=1-3]
	\arrow[from=1-3, to=1-4]
	\arrow[from=1-4, to=1-5]
\end{tikzcd},
\end{equation}
which implies
\begin{equation}
    H^1(M;\mathbb{Z})\cong \operatorname{Hom}(H_1(M;\mathbb{Z}),\mathbb{Z}),
\end{equation}
because $H_0(M; \mathbb{Z})$ contains no torsion. The torsion part in $H_1(M)$ will not contribute to $\operatorname{Hom}(H_1(M;\mathbb{Z}),\mathbb{Z})$, so $H_2(M;\mathbb{Z})\cong H^1(M;\mathbb{Z})$ is also torsion free.  As for $H_3(M,\mathbb{Z})\cong H^0(M;\mathbb{Z})$, using the universal coefficient theorem, again we get
\begin{equation}
    H^0(M;\mathbb{Z})\cong \operatorname{Hom}(H_0(M;\mathbb{Z}),\mathbb{Z}),
\end{equation}
which is torsion free. So the only possible place any torsion may show up is in $H_1(M;\mathbb{Z})\cong H^2(M;\mathbb{Z})$, and we call the torsion part $\mathcal{T}$.

It seems with $\mathbb{R}$ coefficient cochains, we will just loss the information about torsion. But the information about torsion will show up in a more subtle way, through the Jacobians of the (continuum or lattice) exterior derivatives. The Reidemeister torsion for a $3$d lattice $\mathcal{M}$ (the generalization to other dimensions is obvious) is
\begin{equation}
    R=
    \frac{d_2(V_{C^2/Z^2})\wedge V_{H^3}}{V_{C^3}}
    \frac{V_{C^2}}{V_{C^2/Z^2}\wedge d_1(V_{C^1/Z^1})\wedge V_{H^2}}
    \frac{V_{C^1/Z^1}\wedge d_0(V_{C^0/H^0})\wedge V_{H^1}}{V_{C^1}}
    \frac{V_{C^0}}{V_{C^0/H^0}\wedge V_{H^0}},
\end{equation}
(an analogous continuum definition is called the Ray-Singer torsion, which turns out equal to the Reidemeister torsion \cite{muller1978analytic,cheeger1979analytic}) where $V_{W}$ denotes a volume form of the vector space $W$, all the cohomology here are evaluated with coefficient $\mathbb{R}$ and we drop the $(\mathcal{M};\mathbb{R})$ label for all the vector spaces. The notation $d_0(V_{C^0/H^0})$ is the volume form of its image $B^1$ obtained by applying $d_0$ to a volume form of its coimage $C^0/H^0$. As we can see, the value of $R$ depends on the choice of $V_{H^i},V_{C^i}$, but the dependence on the choice of $V_{C^i/Z^i}$ is cancelled out---the crucial reason why $R$ is defined in such a way. On a lattice, we do have a nature choice of basis for $C^i$, and for $V_{H^i}$, we can choose the basis corresponding to the embedding of the basis for free part in $\mathbb{Z}$-valued cohomology groups.

Consider $\mathbb{R}P^3$ as an example. Using the triangulation in \cref{triangulation_of_RP3}, a straight forward calculation tells us the Reidemeister torsion of $\mathbb{R}P^3$ is
\begin{equation}
    R=\frac{e_{c}}{e_{c}}\frac{e_{\sigma}}{2 e_{\sigma}}\frac{e_l}{e_l}\frac{e_v}{e_v}=\frac{1}{2}=\frac{1}{\left|\mathbb{Z}_2\right|},
\end{equation}
where we have $e_{c,\sigma,l,v}$ as the chosen basis. We find $R$ equals to the reciprocal of the torsion size. In fact this is not a coincidence.

Recall that for \cref{summation_of_discrete_dofs} we have claimed that if we choose the basis for coimage $C^i/Z^i$ and image $B^i$ of $d_i$ to be those embedded by the $\mathbb{Z}$-valued spaces, there is no Jacobian, which means under this choice of $V_{C^i/Z^i}$ and $V_{B^i}$ we have $d_i(V_{C^i/Z^i})=V_{B^{i+1}}$. The Reidemeister torsion becomes
\begin{equation}
    R=
    \frac{V_{B^3}\wedge  V_{H^3}}{V_{C^3}}
    \frac{V_{C^2}}{V_{C^2/Z^2}\wedge V_{B^2}\wedge V_{H^2}}
    \frac{V_{C^1/Z^1}\wedge V_{B^1}\wedge V_{H^1}}{V_{C^1}}
    \frac{V_{C^0}}{V_{C^0/H^0}\wedge V_{H^0}}.
\end{equation}
Now the only difference between denominator and numerator of each factor is the torsion part of $H^i(\mathcal{M};\mathbb{Z})$, so in the end we get the Reidemeister torsion
\begin{equation}
    |R|=\frac{1}{|\mathcal{T}|}
\end{equation}
measuring the size of the torsion.

The Reidemeister torsion and the continuum Ray-Singer torision can also be defined for covariant derivatives with non-trivial flat gauge fields, and the result is still a topological invariant in a suitable sense. We will not encounter these more general cases in the present work.

\section{A More Rigorous Calculation of the Partition Function}
\label{a_more_rigorous_calculation_of_the_partition_function}
For a general spacetime $M$, each $[s]\in H^2(M;\mathbb{Z})\cong H_1(M;\mathbb{Z})$ corresponds to a (topological class of) non-contractible loop on the dual lattice, and we have learnt in \cref{torsion_in_homology_cohomology_theory} that they fall into two different categories: those that become contractible after going around the loop a certain number of times, and those that never become contractible this way. We call the first kind the torsion part (isomorphic to some $\bigoplus_i \mathbb{Z}_{p_i}$, which is finite) and the second kind the free part (isomorphic $\mathbb{Z}^{b_1}$ for some integer $b_1$, called the first Betti number, which is intuitively ``the number of holes''). The integral over $A'$ involves flat fluctuations, which take value in $H^1(M;\mathbb{R})$. From the universal coefficient theorem, we know that $H^1(M;\mathbb{R})$ is isomorphic to $\operatorname{Hom}(H_1(M;\mathbb{Z}),\mathbb{R})$, in which only the free part of $H_1$ gets mapped non-trivially to $\mathbb{R}$. Thus, the flat fluctuation of $A'$ serves as a Lagrange multiplier for the free part of $[s']$ through their cup product of in the action. As a result, the contribution of $[s']$ vanishes unless the free part of $[s']$ is trivial. Therefore only the torsion part contributes:
\begin{equation}
\begin{split}
     Z&=\frac{\left[\prod_{l} \int_{-\infty}^{\infty} \frac{dA_l'}{2\pi}\right]}{\left[\prod_{v} \int_{-\infty}^{\infty}  \frac{d\phi_v'}{2\pi}\right]}\frac{\sum_{[\kappa]}}{\sum_{[n']}} \sum_{[s']\in\mathcal{T}}\\ &(z_\chi[[s']^{\text{rep}}])^k\exp \left\{-\frac{1}{2e^2}\sum_p F_p^2+\frac{ik}{4\pi}\sum_c \left[(A'\cup dA')_c - (A'\cup 2\pi [s']^{\text{rep}})_c - (2\pi [s']^{\text{rep}}\cup A')_c\right]\right\}.
\end{split}
\end{equation}

To eliminate $[s']$ in $F=dA'-2\pi [s']^{\text{rep}}$, note that when we map $H^2(M;\mathbb{Z})$ to $H^2(M;\mathbb{R})$, the kernel is the torsion part, since $H^2(M;\mathbb{R})$ has no torsion. This means for $[s']\in\mathcal{T}$ that satisfies $p[s']=[0]\in\mathcal{T}$ for some integer $p$, exists some $A'_{[s']^{\text{rep}}}\in C^1(M; 2\pi\mathbb{Z}/p)$ such that $dA'_{[s']^{\text{rep}}}=2\pi [s']^{\text{rep}}$ (but  $A'_{[s']^{\text{rep}}}\notin C^1(M; 2\pi\mathbb{Z})$ unless $[s']=[0]$), so after a redefinition of $A''=A'-A'_{[s']^{\text{rep}}}$, we can absorb these $[s']^{\text{rep}}$, and write $F=dA''$. The action becomes
\begin{equation}
    S=-\frac{1}{2e^2}\sum_p(dA'')_p^2+\frac{ik}{4\pi}\sum_c[(A''\cup dA'')_c-(2\pi[s']^{\text{rep}}\cup A'_{[s']^{\text{rep}}} )_c],
\end{equation}
which consists of two independent parts: a free theory for $A''_l\in\mathbb{R}$, and $-i(k/2)\sum_c([s']^{\text{rep}}\cup A'_{[s']^{\text{rep}}})_c$ which is proportional to the self-linking number $\operatorname{link}([s']^{\text{rep}},[s']^{\text{rep}})=\sum_c([s']^{\text{rep}}\cup A'_{[s']^{\text{rep}}}/2\pi)_c$ of loops in torsion part.

Note the self-linking number of a torsion loop is fractional---suppose $[s']\in\mathcal{T}$ is such that $p[s']$ is contractible, then $A'_{[s']^{\text{rep}}}$ is a multiple of $2\pi/p$, hence the self-linking number is a multiple of $1/p$. For even $k$, only the fractional part of the self-linking number contributes to the partition function; under change of branching structure (which changes loop framing), the self-linking number only changes by an integer, and thus its contribution to the partition function is unchanged. For odd $k$, under change of branching structure, the change of the self-linking number may contribute an extra $e^{i\pi}$ to the partition function, but this change will always be compensated by the change of $z_\chi$.

The free theory part can be calculated by Faddeev–Popov method. We have
\begin{equation}
\begin{split}
    Z=&\left|\frac{1}{\sqrt{[(2\pi)^{-1}M ](V_{C^1/Z^1},V_{C^1/Z^1})}}\frac{V_{C^1/Z^1}\wedge d_0(V_{C^0/H^0})\wedge V'_{H^1}}{V'_{C^1}}
    \frac{V'_{C^0}}{V_{C^0/H^0}\wedge V'_{H^0}}\right|\\
    &\left[\sum_{[s']\in\mathcal{T}}(z_\chi[[s']^{\text{rep}}])^ke^{-{i\pi k}\operatorname{link}([s']^{\text{rep}},[s']^{\text{rep}})}\right] \ .
\end{split}
\end{equation}
Here, since $M$ is a bilinear form with null space $Z^1$, we use $[(2\pi)^{-1}M ](V_{C^1/Z^1},V_{C^1/Z^1})$ to denote the product of its non-zero eigenvalues $\det'[(2\pi)^{-1}M ]$ once a basis is chosen on $C^1/Z^1$. More rigorously, $M$ should be regarded as a linear operator from $C^1/Z^1$ to the dual linear space $(C^1/Z^1)^*$. If we pick a basis of $C^1/Z^1$, we also get its dual basis. Comparing $[(2\pi)^{-1}M ](V_{C^1/Z^1})$ with the volume form ${V_{C^1/Z^1}}^*$ induced by the dual basis, we have
\begin{equation}
    [(2\pi)^{-1}M ](V_{C^1/Z^1},V_{C^1/Z^1})=\operatorname{det}'[(2\pi)^{-1}M ]=\frac{[(2\pi)^{-1}M ](V_{C^1/Z^1})}{{V_{C^1/Z^1}}^*},
\end{equation}
and as the notation suggests, $[(2\pi)^{-1}M ](\lambda V_{C^1/Z^1},\lambda V_{C^1/Z^1})=\lambda^2[(2\pi)^{-1}M ](V_{C^1/Z^1},V_{C^1/Z^1})$ when scaling the volume form.

Recall that we already have canonical basis for $C^i,H^i$ as defined in \cref{torsion_in_homology_cohomology_theory}. The difference between the canonical volume form $V_{\bullet}$ (here $\bullet$ can only be filled by $C^i,H^i$) and the volume form $V'_{\bullet}$ in the path integral measure (i.e. $\prod_l \int (dA_l)/(2\pi)$ and other integrals or summations) is just a $(2\pi)^{\operatorname{dim}C^i}$ or $(2\pi)^{\operatorname{dim}H^i}$ factor. More precisely, the factor is $(2\pi)^{-\dim C^1+\dim C^0+\dim H^1-\dim H^0}=(2\pi)^{-\dim C^1/Z^1}$, which can be absorbed by changing $\det'[(2\pi)^{-1}M]^{-1/2}$ to $\det'[(2\pi)M]^{-1/2}$.

We continue to define $\det'$ for other linear operators. Take $d_0$ as an example:
\begin{equation}
    \operatorname{det}'d_0=\frac{d_0(V_{C^0/H^0})}{V_{B^1}} \ .
\end{equation}
Note that $\det'$ depends on a certain choice of volume form of image and coimage space. Now we can write
\begin{equation}
\begin{split}
    Z=\left|\frac{\det' d_0}{\sqrt{\det'(2\pi M )}}
    \frac{V_{C^1/Z^1}\wedge V_{B^1}\wedge V_{H^1}}{V_{C^1}}
    \frac{V_{C^0}}{V_{C^0/H^0}\wedge V_{H^0}}\right|
    \left[\sum_{[s']\in\mathcal{T}}(z_\chi[[s']^{\text{rep}}])^ke^{-{i\pi k}\operatorname{link}([s']^{\text{rep}},[s']^{\text{rep}})}\right]
    \ .
\end{split}
\label{Z_formal}
\end{equation}
We would like to calculate $Z$ in two ways: one is to use the Fourier basis and give a precise meaning of \cref{gsd}; the other is to use the Poincare duality and extract a factor of $\sqrt{|R|}$ to arrive at \cref{Z_K}.

Consider the $\mathbb{T}^3$ lattice $\mathbb{Z}_{L_\tau}\times\mathbb{Z}_{L_x}\times\mathbb{Z}_{L_y}$. For $C^0$, we use the Fourier basis $\phi(r)=e^{iqr}/\mathcal{V} $ (where $\mathcal{V}=L_\tau L_x L_y$) with $q\neq 0$ for $C^0/H^0$ and an additional basis vector $\phi(r)=1$ for $H^0$. One may wonder why we do not simply use $\phi(r)=1/\mathcal{V}$ for $q=0$ as well, and this can be traced back to how we introduced these $q=0$ trivial gauge transformations in the first place---we introduced them as the embedding of $[\kappa]\in H^0(\mathbb{T}^3,\mathbb{Z})$ into $H^0(\mathbb{T}^3,\mathbb{R})$ (which is what we call $H^0$ now; recall the $2\pi$'s has been absorbed before). 

Similarly, for $C^1$, we use the basis $A(r)=(1,0,0)^Te^{iqr}/\mathcal{V} $, $A(r)=(0,1,0)^Te^{iqr}/\mathcal{V}$ and $A(r)= (0,0,1)^Te^{iqr}/\mathcal{V}$ with $q\neq 0$ for $C^1/H^1$, and, instead of the usual $q=0$ basis, we use $A(r)=(1,0,0)^T\delta_{\tau,0}$, $A(r)= (0,1,0)^T\delta_{x,0}$ or $A(r)=(0,0,1)^T\delta_{y,0}$ for $H^1$, which are the embedding of $[n]\in H^1(\mathbb{T}^3,\mathbb{Z})$ into $H^0(\mathbb{T}^3,\mathbb{R})$ (which is what we call $H^1$ now).

Let us first sort out the Jacobians carefully. For $C^0$, if we have used the Fourier basis not only for $q\neq 0$ but also for $q=0$, then we would have the Jacobians
\begin{align}
    \frac{V_{C^0}}{V_{C^0,\text{Fourier}}}&=\mathcal{V}^{\frac{1}{2}\dim C^0} \ .
\end{align}
(If the orthonormal Fourier basis---that with $\sqrt{\mathcal{V}}$ in the denominator---is used instead, this Jacobian would become $1$.) However, instead of the $q=0$ Fourier basis, we actually use the embedded basis for $H^0$, so there is an extra Jacobian
\begin{equation}
    \frac{V_{C^0}}{V_{C^0/H^0,\text{Fourier}}\wedge V_{H^0}} =\mathcal{V}^{\frac{1}{2}\dim C^0 - 1} \ .
\end{equation}
(If for $C^0/H^0$ the orthonormal Fourier basis is used instead, this Jacobian will become $\mathcal{V}^{-1/2}$.)

Similarly, for $C^1$, if we have used the Fourier basis not only for $q\neq 0$ but also for $q=0$, then we have the Jacobian
\begin{equation}
    \frac{V_{C^1}}{V_{C^1,\text{Fourier}}}=\mathcal{V}^{\frac{1}{2}\dim C^1} \ .
\end{equation}
(If the orthonormal Fourier basis is used instead, the Jacobian will be $1$.)
But our modified basis has the embedded basis replacing the $q=0$ components of the Fourier basis. To find the extra Jacobian, we decompose the embedded basis as
\begin{equation}
    \begin{bmatrix}
        1\\0\\0
    \end{bmatrix}\delta_{\tau,0}=
    \sum_{q=0}\begin{bmatrix}
        1\\0\\0
    \end{bmatrix}L_xL_y \frac{1}{\mathcal{V}}e^{i q r} 
    +\sum_{q \neq 0}\begin{bmatrix}
        1\\0\\0
    \end{bmatrix}L_xL_y\delta_{q_x,0}\delta_{q_y,0} \frac{1}{\mathcal{V}}e^{i qr} \ ,
\end{equation}
and the extra Jacobian is extracted from the $q=0$ coefficient, which is $L_xL_y$. (If the orthonormal Fourier basis is used instead, the factor will be $L_xL_y/\sqrt{\mathcal{V}}$.) Similar for other two directions. Therefore
\begin{equation}
    \frac{V_{C^1}}{V_{C^1/H^1,\text{Fourier}}\wedge V_{H^1}} =\mathcal{V}^{\frac{1}{2}\dim C^1 - 2} \ .
\end{equation}
(If for $C^1/H^1$ the orthonormal Fourier basis is used instead, this Jacobian will become $\mathcal{V}^{-1/2}$.)

Under the basis transformation $\det'(2\pi M)$ also transforms:
\begin{equation}
    \operatorname{det}'(2\pi M)^{-1/2}=\prod_{q\neq 0}\operatorname{det}'(2\pi M(q)/\mathcal{V})^{-1/2} = \mathcal{V}^{\frac{1}{2}\dim C^1-\frac{1}{2}\dim Z^1}\prod_{q\neq 0}\operatorname{det}'(2\pi M(q))^{-1/2},
\end{equation}
where the $1/\mathcal{V}$ factor in $M$ comes from the fact
\begin{equation}
    \sum_x\sum_y A(x)M(x-y)A(y)=\frac{1}{\mathcal{V}}\sum_q A(-q)M(q)A(q).
\end{equation}
(If the orthonormal Fourier basis is used instead, no $\mathcal{V}$ factor will appear here.)

So we get
\begin{equation}
    \begin{split}
    Z=&
    \left|\prod_{q\neq 0}\frac{\det' d_0(q)}{\sqrt{\det'(2\pi M (q))}}
    \frac{V_{(C^1/Z^1)(q)}\wedge V_{B^1(q)}}{V_{(C^1/H^1)(q),\text{Fourier}}}\right|
    \mathcal{V}^{\frac{1}{2}\dim C^1-\frac{1}{2}\dim Z^1-\frac{1}{2}\dim C^1+2+\frac{1}{2}\dim C^0-1}\\
    &\left[\sum_{[s']\in\mathcal{T}}(z_\chi[[s']^{\text{rep}}])^ke^{-{i\pi k}\operatorname{link}([s']^{\text{rep}},[s']^{\text{rep}})}\right]\\
    =&
    \left|\prod_{q\neq 0}\frac{\det' d_0(q)}{\sqrt{\det'(2\pi M (q))}}
    \frac{V_{(C^1/Z^1)(q)}\wedge V_{B^1(q)}}{V_{(C^1/H^1)(q),\text{Fourier}}}\right|
    \left[\sum_{[s']\in\mathcal{T}}(z_\chi[[s']^{\text{rep}}])^ke^{-{i\pi k}\operatorname{link}([s']^{\text{rep}},[s']^{\text{rep}})}\right],
\end{split}
\end{equation}
where $V_{(C^1/Z^1)(q)}$ and $V_{(B^1)(q)}$ are volume form for $(C^1/Z^1)(q)$, $B^1(q)$ which are spanned by vectors in $C^1/Z^1$, $B^1$ with momentum $q$ (since $d_i$ can be decomposed into direct sum of $d_i(q)$, this decomposition of $C^1/Z^1$, $B^1$ is valid) and $M(q)$ is a $3\times3$ matrix while $d_0(q)$ is a $3\times 1$ row vector. Notice that all the $\mathcal{V}$ factors cancel out as expected.

We are now very close to the expression \cref{gsd}. Observe that $M(q)$ is a hermitian matrix since $M$ is symmetric and real. So we can use a unitary transformation to diagonalize it. We also know that $d_0(q)/\sqrt{(d_0(q))^\dagger d_0(q)}$ is one of its normalized eigenvectors, which actually belongs to $M(q)$'s kernel. Now we may choose $V_{B^1(q)}$ basis to be $d_0(q)/\sqrt{(d_0(q))^\dagger d_0(q)}$, and the other two normalized eigenvectors of $M(q)$ as basis of $C^1/Z^1(q)$. Thus in the end we have
\begin{equation}
    Z=\left|\prod_{q\neq 0}\frac{\sqrt{(d_0(q))^\dagger d_0(q)}}{\sqrt{\det'(2\pi M (q))}}\right|
    \left[\sum_{[s']\in\mathcal{T}}(z_\chi[[s']^{\text{rep}}])^ke^{-{i\pi k}\operatorname{link}([s']^{\text{rep}},[s']^{\text{rep}})}\right],
\end{equation}
where $\det'(2\pi M (q))$ just means the product of $M(q)$'s non-zero eigenvalue. This gives the precise meaning of \cref{gsd}: since $M(q)$ is the matrix elements under orthonormal Fourier basis for $M$, when using the canonical basis, no Jacobian shows up and the product of all non-zero eigenvalues of $2\pi M$ matrix in canonical basis is $\det'(2\pi M)=\prod_{q\neq 0}\det'(2\pi M (q))$ while $\prod_{q\neq 0}\sqrt{(d_0(q))^\dagger d_0(q)}=\sqrt{\det'(d_0^Td_0)}$.

Now we show how to extract a factor of $\sqrt{|R|}$ from $Z$ to arrive at \cref{Z_K}. Going back to \cref{Z_formal}, we can identify dual spaces $(C^i)^*$,$(B^i)^*$,$(H^i)^*$ with $C_{i}$,$C_{i}/Z_{i}$,$H_{i}$ respectively. Using the corresponding dual volume form, we have
\begin{equation}
    \operatorname{det}'d_0=\frac{d_0(V_{C^0/H^0})}{V_{B^1}}=\frac{V_{B^1}^*}{(d_0(V_{C^0/H^0}))^*}=\frac{d_0^T(V_{B^1}^*)}{V_{C^0/H^0}^*}=\frac{(\partial_1)(V_{C_1/Z_1})}{V_{B_0}}=\operatorname{det}'(\partial_1),
\end{equation}
where we have the corresponding dual volume form denoted as $V^*_{\bullet}$.
We also have
\begin{equation}
\begin{split}
    &\frac{V_{C^1/Z^1}\wedge V_{B^1}\wedge V_{H^1}}{V_{C^1}}
    \frac{V_{C^0}}{V_{C^0/H^0}\wedge V_{H^0}}\\
    =&\frac{V_{C^1}^*}{V_{C^1/Z^1}^*\wedge V_{B^1}^*\wedge V_{H^1}^*}
    \frac{V_{C^0/H^0}^*\wedge V_{H^0}^*}{V_{C^0}^*}\\
    =&\frac{V_{C_1}}{V_{B_1}\wedge V_{C_1/Z_1}\wedge V_{H_1}}
    \frac{V_{B_0}\wedge  V_{H_0}}{V_{C_0}} \ .
\end{split}
\end{equation}
The partition function can be written as
\begin{equation}
\begin{split}
    &Z=\sum_{[s']\in\mathcal{T}}(z_\chi[[s']^{\text{rep}}])^ke^{-S[A_{[s']^{\text{rep}}}]}\\
    &\sqrt{\left|\frac{\det' d_0\det' \partial_1}{\det'(2\pi M )}
    \frac{V_{C_1}}{V_{B_1}\wedge V_{C_1/Z_1}\wedge V_{H_1}}
    \frac{V_{B_0}\wedge  V_{H_0}}{V_{C_0}}
    \frac{V_{C^1/Z^1}\wedge V_{B^1}\wedge V_{H^1}}{V_{C^1}}
    \frac{V_{C^0}}{V_{C^0/H^0}\wedge V_{H^0}}\right|} \ .
\end{split}
\end{equation}
On a cubic lattice, we know the dual lattice is isomophic to the origin lattice. So we have Poincare duality: $C_i\cong C^{3-i}$, $B_i\cong B^{3-i}$,$Z_i\cong Z^{3-i}$,$H_i\cong H^{3-i}$. The partition function can be further simplified as
\begin{equation}
    \begin{split}
    &Z=
    \sum_{[s']\in\mathcal{T}}\left|\frac{1}{\sqrt{\det'(2\pi K)}}\right|(z_\chi[[s']^{\text{rep}}])^ke^{-S[A_{[s']^{\text{rep}}}]}\\
    &\sqrt{\left|\frac{\det' (-d_2)\det' d_0}{\det'd_1}\frac{V_{B^3}\wedge  V_{H^3}}{V_{C^3}}
    \frac{V_{C^2}}{V_{C^2/Z^2}\wedge V_{B^2}\wedge V_{H^2}}
    \frac{V_{C^1/Z^1}\wedge V_{B^1}\wedge V_{H^1}}{V_{C^1}}
    \frac{V_{C^0}}{V_{C^0/H^0}\wedge V_{H^0}}\right|}\\
    &=\sum_{[s']\in\mathcal{T}}
    \left|\frac{\sqrt{R}}{\sqrt{\det'(2\pi K)}}\right|(z_\chi[[s']^{\text{rep}}])^ke^{-S[A_{[s']^{\text{rep}}}]},
\end{split}
\end{equation}
where we have factorized $M=Kd_1$ to extract the Reidemeister torsion $R$. Here $\det'(2\pi K)=(2\pi K)(V_{C^1/Z^1},V_{B^2})$, and since we have used the Poincare duality, if we have chosen a basis for $B^2$, then it induces a basis for $C_2/Z_2\cong C^1/Z^1$. And notice that when changing the basis
\begin{equation}
    \frac{V'_{B^2}}{V_{B^2}}=\frac{V_{C_2/Z_2}}{V'_{C_2/Z_2}}=\frac{V_{C^1/Z^1}}{V'_{C^1/Z^1}},
\end{equation}
which indicates that the $(2\pi K)(V'_{C^1/Z^1},V'_{B^2})=(2\pi K)(V_{C^1/Z^1},V_{B^2})$ is a deterministic number regardless of the choice of basis.


\bibliographystyle{JHEP}
\bibliography{biblio.bib}






\end{document}